\documentclass[
aip,
jcp,
 amsmath,amssymb,
 reprint,%
]{revtex4-1}

\usepackage{graphicx}
\usepackage{dcolumn}
\usepackage{bm}

\usepackage[version=3]{mhchem} 
\usepackage[utf8]{inputenc}
\usepackage[T1]{fontenc}
\usepackage{etoolbox}

\usepackage{url}
\usepackage{xcolor}
\usepackage[normalem]{ulem}
\usepackage{xr}
\usepackage{filecontents}

\newcommand{\lDinf}{\lambda_{\mathrm{D}}}
\newcommand{\kBT}{k_{\mathrm{B}} T} 
\newcommand{\eps}{\epsilon \epsilon _0}

\newcommand{\Dp}{D_+ + D_-}

\newcommand{\cinf}{c_{\mathrm{\infty}}}
\newcommand{\bcs}{c_{\mathrm{s}_1}}

\newcommand{\lDuinf}{l_{\mathrm{Du}} ^{\infty}}
\newcommand{\lDu}{l_{\mathrm{Du}}}
\newcommand{\lGC}{l_{\mathrm{GC}}}
\newcommand{\appropto}{\mathrel{\vcenter{
  \offinterlineskip\halign{\hfil$##$\cr
    \propto\cr\noalign{\kern2pt}\sim\cr\noalign{\kern-2pt}}}}}

 \newcommand{\addition}[1]{#1}              
 \newcommand{\deletion}[1]{}                 

\graphicspath{{./figures_paper/}}

\makeatletter
\def\@email#1#2{%
 \endgroup
 \patchcmd{\titleblock@produce}
  {\frontmatter@RRAPformat}
  {\frontmatter@RRAPformat{\produce@RRAP{*#1\href{mailto:#2}{#2}}}\frontmatter@RRAPformat}
  {}{}
}%
\makeatother

\begin{document}


\title{Revisiting the access conductance of a nanopore in a charged membrane}
\author{Holly C. M. Baldock}
\author{David M. Huang}
\email{david.huang@adelaide.edu.au}
\affiliation{School of Physics, Chemistry and Earth Sciences, The University of Adelaide, SA 5005, Australia}

\date{\today}

\begin{abstract}
Electric-field-driven electrolyte transport through nanoporous membranes is important for applications including osmotic power generation, sensing and iontronics. We derive an analytical equation in the Debye--H\"uckel regime and a semi-analytical equation for arbitrary surface potentials for the electric-field-driven electric current through a pore in an ultrathin membrane, which predict scaling with fractional powers of the pore size and Debye length. We show that our theory for arbitrary electric potentials accurately quantifies the ionic conductance through an ultrathin membrane in finite-element method numerical simulations for a wide range of parameters, and generalizes a widely used theory for the access electrical conductance of a membrane nanopore to a broader range of conditions. Our theory predicts that fractional scaling of the ionic conductance with electrolyte concentration at low concentrations is a general an intrinsic property of charged ultrathin membranes and also occurs for thicker membranes for which the access contribution to the conductance dominates, which could help to explain experimental observations of this widely debated phenomenon.
\end{abstract}


\maketitle 

\section{Introduction}
\label{sec:introduction}

Electric-field-driven transport through porous membranes underpins numerous applications of micro- and nanofluidics, including osmotic energy harvesting,\cite{aluruConfinementSingleDigitNanpore2023, zhangNanofluidicsOsmoticEnergy2021a, siriaNewAvenuesLargescale2017, rankinEffectHydrodynamicSlip2016} nanopore-based sensing and sequencing,\cite{siwyNanporesDNASequencingFiltration2023, yingBeyondDNASequencing2022} and emerging iontronic systems for signal processing and energy storage.\cite{robinMemorySynapseDynamicsChannels2023, robinNanofluidicsCrossroads2023, kamsmaBrainIontronicNanochannels2024, gkoupidenisOrganicBioinspiredElectrotronics2024} Such technologies take advantage of the ability of electric fields to drive selective ion flow through confined geometries,\cite{aluruConfinementSingleDigitNanpore2023} offering pathways to convert ionic gradients into power via reverse electrodialysis,\cite{aluruConfinementSingleDigitNanpore2023, zhangNanofluidicsOsmoticEnergy2021a, siriaNewAvenuesLargescale2017, rankinEffectHydrodynamicSlip2016} discriminate between analytes at the molecular scale,\cite{siwyNanporesDNASequencingFiltration2023, yingBeyondDNASequencing2022} and build biologically inspired information-processing devices.\cite{robinMemorySynapseDynamicsChannels2023, robinNanofluidicsCrossroads2023, kamsmaBrainIontronicNanochannels2024, gkoupidenisOrganicBioinspiredElectrotronics2024}  

Two-dimensional (2D) membranes, such as those made from graphene,\cite{rollingsIonSelectivityGraphene2016} molybdenum disulfide (MoS$_2$),\cite{fengSinglelayerMoS2Nanopores2016} and hexagonal boron nitride (hBN),\cite{liu2024hbn} offer exceptional promise for these applications due to their atomic thickness and large surface-area-to-volume ratios.\cite{sahuColloquiumIonicPhenomena2019} The extreme confinement of 2D membranes results in large transmembrane gradients that can be orders of magnitudes greater than those achieved in conventional membranes,\cite{macha2DMaterialsEmerging2019} with exceedingly high osmotic power densities measured in single-layer MoS$_2$ membranes,\cite{fengSinglelayerMoS2Nanopores2016} single-layer hBN membranes,\cite{liu2024hbn} and a 2D nanoporous polymer membrane.\cite{chengGateableOsmoticPowerPolymer2023} Moreover, single-layer graphene membranes,\cite{cantley2019} as well as single- and few-layer MoS$_2$ membranes,\cite{Liu2014} have demonstrated remarkable sensitivity in detecting single-molecule translocation events, highlighting their potential for highly selective and low-concentration sensing applications.\cite{cantley2019}

Understanding fluid transport through 2D membranes has wider implications, as models of 2D membranes have been shown to accurately describe the influence of the pore ends on fluid transport through thicker membranes,\cite{sisanEndNanochannels2011, sahuColloquiumIonicPhenomena2019} known as “entrance effects”.\cite{rankinEntranceEffectsConcentrationgradientdriven2019} Entrance effects can become significant as the membrane thickness approaches the pore size\cite{maoElectroosmoticFlowNanopore2014, melnikovElectrosmoticFlowNanopore2017} and when the solid--fluid friction flow is low,\cite{sisanEndNanochannels2011} such as in carbon nanotubes.\cite{Manghi2021CNTs, rankinEffectHydrodynamicSlip2016} Access resistance plays a central role in biological ion transport, where nanoscale confinement near membrane pores significantly limits electrical current and governs the rate of ion transport under certain conditions.\cite{sahu2018bio, alcaraz2017bio, Aguilella2020}

While theories have been derived to quantify the electric-field-driven access resistance in uncharged\cite{hallAccessResistance1975} and charged\cite{leeLargeElectricSizeSurfaceCondudction2012} nanopores, many experimental phenomena observed in 2D membranes are not fully understood theoretically.\cite{macha2DMaterialsEmerging2019} For example, single-layer MoS$_2$\cite{fengSinglelayerMoS2Nanopores2016} and single-layer graphene \cite{rollingsIonSelectivityGraphene2016} membranes appear to have similar electric-field-driven ionic conductances at similar pore sizes and electrolyte concentrations, while the MoS$_2$ membrane exhibits a concentration-gradient-driven ionic conductance that is an order of magnitude greater than that of the graphene membrane.\cite{fengSinglelayerMoS2Nanopores2016, rollingsIonSelectivityGraphene2016, macha2DMaterialsEmerging2019} Furthermore, scaling of the ionic conductance with fractional powers of the electrolyte concentration have been observed in 2D and thin membranes at low electrolyte concentrations,\cite{liu2024hbn, chengGateableOsmoticPowerPolymer2023, shan2013graphene, VenkatesanGraphene2012} which differs to the saturation at low electrolyte concentration predicted by the existing theory of the access resistance of a charged membrane. \cite{leeLargeElectricSizeSurfaceCondudction2012} 

Scaling of the ionic conductance with fractional powers of the electrolyte concentration at low concentrations has also been observed in longer nanopores, including carbon nanotubes,\cite{secchiScalingTransportCarbonNanotubes2016, Uematsu2018CNTs} biomimetic boron nitride nanotube porins,\cite{zhongwuIonTransportBNNT2024} and biological nanochannels,\cite{Martin2018} which deviates from theoretical predictions and typical experimental behavior for long nanopores.\cite{bocquetNanofluidicsBulkInterfaces2010, siriaGiantOsmoticEnergy2013} 
Various mechanisms have been proposed to explain this phenomenon, including a salinity-dependent surface charge,\cite{Secchi2016, Biesheuvel2016, Manghi2018}  interfacial slip, \cite{Manghi2018} variable degrees of counter-ion binding at the surface,\cite{Uematsu2018CNTs} and electric potential “leakage” from the pore.\cite{Noh2020} Recently, concentration-gradient-driven ion fluxes through a 2D membrane have been shown by theory and simulation to scale with fractional powers of the pore size and average electrolyte concentration, which implies that similar scaling relationships could apply to the electrical access resistance given the similarities between concentration-gradient- and electric-field-driven transport.\cite{baldockCDF2025} Thus, entrance effects could be responsible for some of the aforementioned experimental observations, although this hypothesis had been unverified until now.

Here we derive an analytical equation in the Debye–Hückel regime and a semi-analytical equation for arbitrary surface potentials to describe the electric-field-driven electric current through a circular aperture in an infinitesimally thin planar (i.e. ultrathin) membrane, which predict scaling of the ionic conductance with fractional powers of the pore size and Debye screening length. The latter equation accurately quantifies the ionic conductance in finite-element method numerical simulations over a wide range of system parameters, and captures scaling of the ionic conductance with fractional powers of the electrolyte concentration observed in simulations at low bulk electrolyte concentrations.
We thereby generalize an existing theory for the access electrical conductance of a membrane nanopore\cite{leeLargeElectricSizeSurfaceCondudction2012}---which has been widely used to infer the surface charge densities of 2D and thicker membranes from experimental measurements \cite{liu2024hbn, fengSinglelayerMoS2Nanopores2016, yazda2021hBn,LinSiliconNitride2021, leeLargeElectricSizeSurfaceCondudction2012}---to make accurate predictions over a wider range of conditions without any adjustable parameters. 
This leads to an expression for the ionic conductance of a membrane nanopore that predicts scaling of the ionic conductance with fractional powers of the electrolyte concentration in ion-selective membranes at low electrolyte concentrations, including those for which the membrane thickness exceeds the pore size.

\section{Theory}
\label{sec:theo}

\begin{figure} 
    \centering
    \includegraphics{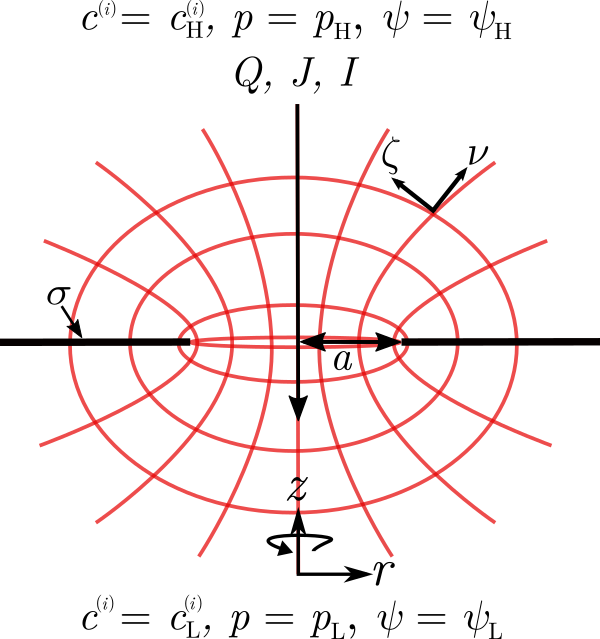}
    \caption{Schematic of flow of an electrolyte solution through a circular aperture of radius $a$ in an infinitesimally thin planar wall with surface charge density $\sigma$. $c_{\mathrm{H}}^{(i)}$ is the concentration of species~$i$, $p_{\mathrm{H}}$ is the pressure, and $\psi_{\mathrm{H}}$ is electric potential far from the membrane in the upper half plane. $c_{\mathrm{L}}^{(i)}$ is the concentration of species~$i$, $p_{\mathrm{L}}$ is the pressure, and $\psi_{\mathrm{L}}$ is electric potential far from the membrane in the lower half plane. $Q$, $J$ and $I$ are the flow rate, solute flux and electric current, respectively. $\zeta$ and $\nu$ are oblate–-spheroidal coordinates, while $r = a \sqrt{(1 + \nu^2)(1 - \zeta^2 )}$ and $z  = a\nu\zeta$  are cylindrical coordinates. Contours of constant $\zeta$ and $\nu$ are shown as solid red lines and unit vectors are shown at one point in space. The system has rotational symmetry about $z$.}
    \label{fig:schematic-2d_membrane}
\end{figure}

We consider transport of an electrolyte solution through the circular aperture shown in Fig.~\ref{fig:schematic-2d_membrane} induced by a potential difference $\Delta \psi = \psi_\mathrm{H} - \psi_{\mathrm{L}}$ between the two sides of an infinitesimally thin planar membrane, where $\psi_\mathrm{H}$ and $\psi_{\mathrm{L}}$ are the electric potential at the upper and lower reservoir boundaries, respectively.  The concentration of the solution and pressure far from the pore are the same on both sides, i.e.,  $c_{\mathrm{H}}^{(i)} = c_{\mathrm{L}}^{(i)}$ and $p_{\mathrm{H}} = p_{\mathrm{L}}$. We represent the system using the same model as in Ref.~\citenum{baldockCDF2025} for concentration-gradient-driven electrolyte transport in the same geometry, which involves solving the Poisson--Nernst--Plank--Stokes equations, in which the fluid is treated as an incompressible continuous medium and the ions as point particles that only interact electrostatically. Without loss of generality, we write the concentration of ions of type $i$ as $c^{(i)} = c_{\mathrm{s}}^{(i)} \exp[-Z_ie\psi/(\kBT)]$, where the virtual variable $c_{\mathrm{s}}^{(i)}$ represents the ion concentration at which the electric potential is zero, \cite{baldockCDF2025} $Z_i$ is the valence of species~$i$, $e$ is the elementary charge, $k_{\mathrm{B}}$ is the Boltzmann constant, and $T$ is the temperature.

Analogously to Ref.~\citenum{baldockCDF2025}, we assume that each system variable can be represented by a perturbation expansion, where $\varepsilon \ll 1$ is a dimensionless parameter that characterizes the magnitude of the applied field. Thus, $\boldsymbol{u} = \varepsilon \boldsymbol{u}_1$, $c_{\mathrm{s}}^{(i)} = \cinf^{(i)} + \varepsilon \bcs^{(i)}$ and $\psi = \psi _0 + \varepsilon \psi _1$ represent the non-equilibrium velocity, $c_{\mathrm{s}}^{(i)}$ and electric potential, respectively. In the linear-response regime, we can write the ion flux density of species $i$ as 
\begin{equation}
        \label{eq:ion_flux_dens-lin}
        \boldsymbol{j} _i =  \left[ (\varepsilon\boldsymbol{u}_1) \cinf^{(i)}  - D_i  \nabla (\varepsilon c_{\mathrm{s}_1} ^{(i)}) \right]\exp{\left(-\frac{Z_i e \psi_0}{\kBT}\right)},
\end{equation}
where $\psi_0$ is the electric potential at equilibrium, and $\boldsymbol{u}_1$, $c_{\mathrm{s}_1}^{(i)}$ and $\psi_1$ are the first-order perturbations to the fluid velocity, $c_{\mathrm{s}}^{(i)}$ and electric potential, respectively.\cite{baldockCDF2025} Furthermore, $\cinf^{(i)}$ and $D_i$ are the bulk concentration and diffusivity  of species~$i$, respectively (model parameters and variables are defined in Table~\ref{stab:parameters} in the supplementary material). Note that $Z_ie \psi_0$ is the equilibrium electric potential energy of an ion of type $i$ and $k_{\mathrm{B}}T$ is the thermal energy. We assume no slip of the fluid and no penetration of the ions at the surface boundary. Thus, $\boldsymbol{u}_1=0$ and $\boldsymbol{\hat{n}} \cdot \nabla c_{\mathrm{s}_1} =0$ at this boundary, where $\boldsymbol{\hat{n}}$ is the unit normal to the surface.

Taking the divergence of Eq.~\eqref{eq:ion_flux_dens-lin}, we can write the conservation equation for ion species~$i$ ($\nabla \cdot \boldsymbol{j} _i = 0$) as
\begin{equation}
    \label{eq:ion_conservation-lin}
    \nabla^2 \left(\frac{\varepsilon c_{\mathrm{s}_1}^{(i)}}{\cinf^{(i)}} \right) = \nabla \left(\frac{Z_i e \psi_0}{\kBT} \right) \cdot \left[\frac{\varepsilon \boldsymbol{u}_1}{D_i} + \nabla \left(\frac{\varepsilon c_{\mathrm{s}_1}^{(i)}}{\cinf^{(i)}} \right) \right].
\end{equation}
For electric-field-driven flow through the pore geometry in Fig.~\ref{fig:schematic-2d_membrane}, the boundary conditions on the variables are $\psi \rightarrow \pm \Delta \psi/2$ in the upper and lower half planes, respectively, and $c^{(i)} \rightarrow \cinf$ at both boundaries. Given that $\psi_0 \rightarrow 0$ at both boundaries, $\varepsilon \psi_1 \rightarrow \pm \Delta \psi/2$ in the upper and lower half planes, respectively, where the condition $Ze |\varepsilon \psi_1|/2 \ll \kBT$ characterizes the linear--response regime. Substituting the boundary conditions for $c^{(i)}$ and $\psi$ into the expression for the ion concentration and assuming linear response of fluxes to the applied potential difference gives $c_{\mathrm{s}} \rightarrow \cinf[ 1 \pm Ze \Delta \psi /(2 \kBT)]$ in the upper and lower half planes, respectively. Full details of the model, including further details on the derivation of the boundary conditions for $c_{\mathrm{s}}^{(i)}$, can be found in Sec.~\ref{ssec:theo-full_model} of the supplementary material. 

The electric current $I$ across the membrane is
\begin{align}
    \label{eq:elec_curr-int}
    I & = \iint _S \mathrm{d} S \,  \boldsymbol{\hat{n}} \cdot \boldsymbol{j}_{\mathrm{e}},
\end{align}
where $\boldsymbol{j}_{\mathrm{e}} = e\sum _i Z_i \boldsymbol{j}_i$ is the electric current density, $\boldsymbol{\hat{n}} = \boldsymbol{\hat{\nu}}$ is the unit vector normal to the pore mouth, and the surface integral is across the pore aperture.  The electric-field-driven ionic conductance is $G = -I/\Delta \psi$.\cite{leeLargeElectricSizeSurfaceCondudction2012} 

\subsection{Weak ion--membrane interactions (Debye--H\"uckel regime)} 
\label{sec:theo-DH}

Assuming that the right-hand-side of Eq.~\eqref{eq:ion_conservation-lin} is small, which is the case if the gradient of the velocity near the surface is small relative to the diffusivity of species~$i$ and the equilibrium electric potential energy is small relative to the thermal energy, Eq.~\eqref{eq:ion_conservation-lin} reduces to $\nabla^2 ({\varepsilon c_{\mathrm{s}_1}^{(i)}} ) = 0$. Solving this expression subject to the boundary conditions imposed on $\varepsilon c_{\mathrm{s}_1}^{(i)}$ gives \cite{morse1953b}
\begin{align}
    \label{eq:cs-DH}
    \varepsilon c_{\mathrm{s}_1}^{(i)} = \frac{\cinf ^{(i)} Z_i e \Delta \psi}{\pi \kBT} \tan^{-1}{\nu}.
\end{align}
Analogous simple analytical expressions in terms of oblate--spheroidal coordinates have been derived for $\psi_1$ in electric-field-driven flow\cite{maoElectroosmoticFlowNanopore2014} and for $\bcs ^{(i)}$ in concentration-gradient-driven flow\cite{baldockCDF2025} of an electrolyte in the same membrane geometry in the linear-response and Debye--H\"uckel regimes. Substituting Eqs.~\eqref{eq:ion_flux_dens-lin} and~\eqref{eq:cs-DH} into Eq.~\eqref{eq:elec_curr-int} gives
\begin{align}
    \label{eq:elec_curr-DH-full} 
    I = -\frac{a \eps (\Dp) \Delta \psi}{(\lDinf)^2} \left[ 1 + \frac{1}{2}\int ^1 _0 \mathrm{d} \zeta \left(\frac{Ze \psi_0|_{\nu=0}}{\kBT}\right)^2 \right]
\end{align}
as the electric-field-driven electric current of a $Z$:$Z$ electrolyte in the Debye--H\"uckel regime ($Ze|\psi_0| \ll \kBT$) when $D_+ \approx D_-$, where
\begin{align}
\label{eq:debye_length}
    \lDinf = \sqrt{\frac{\eps \kBT}{2(Ze)^2 \cinf}}
\end{align}
is the Debye screening length for a $Z$:$Z$ electrolyte with bulk electrolyte concentration $\cinf$, $\epsilon_0$ is the vacuum permittivity, and $\epsilon$ is the dielectric constant of the medium (taken to be water here). The first and second terms in Eq.~\eqref{eq:elec_curr-DH-full} represent the bulk and surface contributions to the conductance, respectively.   A full analytical expression for the equilibrium electric potential $\psi_0$ in the Debye--H\"uckel regime in this geometry can be found in Ref.~\citenum{maoElectroosmoticFlowNanopore2014}, and is given in Eq.~\eqref{seq:epot-DH} in the supplementary material. More details of the derivation of Eq.~\eqref{eq:elec_curr-DH-full}, including an analogous expression to Eq.~\eqref{eq:elec_curr-DH-full} when $D_+ \neq D_-$ and a derivation of the electric-field-driven solute flux, can be found in Sec.~\ref{ssec:theo-DH} of the supplementary material. 

\subsubsection{Thick electric double layer}
\label{sec:theo-DH-thickEDL}

When the electric double layer (EDL) is much wider than the pore radius ($\lDinf \gg a$), $\psi_0 \approx \sigma \lDinf/(\eps)$ in the Debye--H\"uckel regime.\cite{baldockCDF2025, maoElectroosmoticFlowNanopore2014} In this limit and assuming $D_+ \approx D_-$, Eq.~\eqref{eq:elec_curr-DH-full} reduces to
\begin{align}
     I \approx -\frac{a \eps (\Dp) \Delta \psi}{(\lDinf)^2}\left[1 + 2\left(\frac{\lDinf}{l_{\mathrm{GC}}} \right)^2 \right],
  \label{eq:elec_curr-DH-thickEDL}
\end{align}
where
\begin{align}
\label{eq:GC_length}
    \lGC = \frac{2 \eps \kBT}{Ze |\sigma|}
\end{align}
is the Gouy--Chapman length for a $Z$:$Z$ electrolyte. We also derive expressions for the electric-field-driven electric current and solute flux for $D_= \neq D_-$ in this regime in Sec.~\ref{ssec:theo-DH-thickEDL} of the supplementary material.

\subsubsection{Thin electric double layer}
\label{sec:theo-DH-thinEDL}
 
When the EDL is much narrower than the pore radius ($\lDinf \ll a$), we can use the results in Ref.~\citenum{baldockCDF2025} to show that 
\begin{equation}
    \label{eq:int_epot-sq-DH}
    \int^1 _0 \mathrm{d}\zeta \, \left( \frac{Ze \psi_0|_{\nu=0}}{k_{\mathrm{B}} T} \right) ^2 \propto \left(\frac{\lDinf}{a} \right)^{\frac{1}{2}} \left(\frac{\lDinf}{\lGC} \right)^2 .
\end{equation}
We have found the constant of proportionality in Eq.~\eqref{eq:int_epot-sq-DH} to be $\approx 2/3$ by numerically integrating the left-hand side using the analytical expression for $\psi_0$ in the Debye--H\"uckel regime \cite{maoElectroosmoticFlowNanopore2014} for various values of $\lDinf/a \ll 1$ and fitting the results to the relationship on the right-hand side (see Sec.~\ref{ssec:theo-DH-thinEDL} of the supplementary material for details). Substituting Eq.~\eqref{eq:int_epot-sq-DH} with this proportionality constant into Eq.~\eqref{eq:elec_curr-DH-full} gives 
\begin{equation}
    I \approx -\frac{a\eps (\Dp) \Delta \psi}{(\lDinf)^2} \left[1 + \frac{1}{3} \left(\frac{\lDinf}{a} \right)^{\frac{1}{2}} \left(\frac{\lDinf}{\lGC} \right)^2 \right]
    \label{eq:elec_curr-DH-thinEDL}
\end{equation}
as the electric-field-driven electric current in the Debye--H\"uckel regime for a thin EDL and $D_+ \approx D_-$. In Sec.~\ref{ssec:theo-DH-thinEDL} of the supplementary material, we also determine expressions for the electric current and solute flux in this regime for $D_+ \neq D_-$ by fitting to the finite-element method (FEM) simulations described in Sec.~\ref{sec:results}. Equation~\eqref{eq:elec_curr-DH-thinEDL} indicates scaling of the current with fractional powers of the pore radius and Debye length, while the theory for a thick EDL in the Debye--H\"uckel regime (Eq.~\eqref{eq:elec_curr-DH-thickEDL}) indicates scaling with integer powers. 

\subsection{Arbitrary ion--membrane interactions and a thin electric double layer}
\label{sec:theo-thinEDL}

As the bulk conductance dominates for a thin EDL in the Debye--H\"uckel regime, the utility of Eq.~\eqref{eq:elec_curr-DH-thinEDL} is limited. Nevertheless, the results in Eq.~\eqref{eq:elec_curr-DH-thinEDL} indicate unusual scaling of the surface conductance with fractional power laws of the pore size and Debye length for an ultrathin membrane that is not seen in thicker membranes, e.g. for long cylindrical pores or slit pores.\cite{bocquetNanofluidicsBulkInterfaces2010, rankinEffectHydrodynamicSlip2016} Although scaling with fractional powers of the pore radius and Debye length were not previously predicted in Ref.~\citenum{leeLargeElectricSizeSurfaceCondudction2012} for the access contribution to the ionic conductance of a nanopore, similar fractional scaling relationships appear in the theory for concentration-gradient-driven electrolyte transport in the same ultrathin-membrane geometry studied here.\cite{baldockCDF2025} Assuming that these or similar relationships are valid outside the Debye--H\"uckel regime, this unusual scaling could be significant at larger magnitudes of the electric potential energy when the surface conductance is large. To test this hypothesis, we first assume that the electric-field-driven electric current through an ultrathin membrane can be approximated as the sum of bulk and surface contributions via
\begin{equation}
\label{eq:elec_curr-gen}
    I \approx -\left(2a \kappa_{\mathrm{b}} + \kappa_{\mathrm{s}} \right) \Delta \psi ,
\end{equation}
where
\begin{align}
    \label{eq:bulk_conduct}
    \kappa_{\mathrm{b}} = \frac{\eps (\Dp)}{2(\lDinf)^2}
\end{align}
is the bulk conductivity ($2a \kappa_{\mathrm{b}}$ is the bulk conductance\cite{hallAccessResistance1975}) and $\kappa_{\mathrm{s}}$ is the surface conductance of an ultrathin membrane. While Eq.~\eqref{eq:elec_curr-gen} is similar to the theory of the access conductance of a pore in Ref.~\citenum{leeLargeElectricSizeSurfaceCondudction2012}, we do not assume $\kappa_{\mathrm{s}}$ to be directly proportional to the surface conductivity of the interior of a long pore. 

Comparing Eq.~\eqref{eq:elec_curr-gen} to Eq.~\eqref{eq:elec_curr-DH-thinEDL}, we arrive at 
\begin{equation}
    \kappa_{\mathrm{s}} \approx \frac{(a \lDinf)^{\frac{1}{2}} \eps (\Dp)}{3 (\lGC)^2}
       \label{eq:surf_conduct-DH} 
\end{equation}
in the Debye--H\"uckel regime for a thin EDL and $D_+ \approx D_-$. The surface conductivity of a planar wall \cite{leeLargeElectricSizeSurfaceCondudction2012} (which is a good approximation for the surface conductivity of a long planar channel or cylindrical pore for a thin EDL) for $D_+ \approx D_-$ is
\begin{equation}
    \label{eq:surf_conduct-plane}
    \kappa_\mathrm{s}^{\infty} = 
    \frac{\eps(\Dp)}{\lGC} \left[-\frac{\lGC}{\lDinf} + \sqrt{\left( \frac{\lGC}{\lDinf} \right)^2 + 1 } \right].
\end{equation}
In Ref.~\citenum{leeLargeElectricSizeSurfaceCondudction2012}, the access contribution to the surface conductance from each pore end was taken as $\beta \kappa_\mathrm{s}^{\infty}$ for a thin EDL, where $\beta$ is a numerical constant. In the Debye--H\"uckel regime, Eq.~\eqref{eq:surf_conduct-plane} reduces to
\begin{equation}
\label{eq:surf_conduct-plane-DH}
    \kappa_{\mathrm{s}}^{\infty} \approx \frac{ \lDinf \eps (\Dp)}{2 (\lGC)^2} ,
\end{equation}
and we see that $\kappa_{\mathrm{s}} / \kappa_{\mathrm{s}}^{\infty} \approx 2 (a/\lDinf)^{\frac{1}{2}}/3$ in this regime. Similarly to the case of concentration-gradient-driven electrolyte transport through an ultrathin membrane,\cite{baldockCDF2025} the scaling of $\kappa_{\mathrm{s}}$ with fractional powers of $a$ and $\lDinf$ arises from the geometrical interplay between the variables that describe the properties of the electrolyte solution and the membrane (see Fig.~\ref{fig:schematic-2d_membrane}), which is due to the non-uniformity of the electric-field lines as the ions enter and exit the bulk at the pore ends. While the Debye length $\lDinf$ (which is $\propto (\cinf)^{-\frac{1}{2}}$) and the Gouy--Chapman length $\lGC$ (which is $\propto 1/\sigma$) describe the properties of an electrolyte solution and a charged planar surface, respectively, the pore radius $a$ governs the pore geometry. 

For a charged planar surface, at which the potential is proportional to $\lDinf/\lGC$ in the Debye--H\"uckel regime, $\lDinf/\lGC$ is a dimensionless parameter that describes the interactions of the ions with the surface. As $\kappa_{\mathrm{s}} / \kappa_{\mathrm{s}}^{\infty} \approx 2 (a/\lDinf)^{\frac{1}{2}}/3$ for $\lDinf \ll a$ and $Ze|\psi_0| \ll \kBT$---where $\kappa_{\mathrm{s}}$ and $\kappa_{\mathrm{s}}^{\infty}$ are both proportional to $1/(\lGC)^2$---we infer that, in this regime, $[2 (a/\lDinf)^{\frac{1}{2}}(\lDinf/\lGC)^{2}/3]^{\frac{1}{2}} = (2/3)^{\frac{1}{2}} a^{\frac{1}{4}}(\lDinf)^{\frac{3}{4}}/\lGC$ describes the influence of ion--surface interactions on the surface conductance of an ultrathin membrane. Thus, for the case of a thin EDL in an ultrathin membrane, we deduce that $(2/3)^{\frac{1}{2}} a^{\frac{1}{4}}(\lDinf)^{\frac{3}{4}}/\lGC$ plays an analogous role to $\lDinf/\lGC$ in the theory of the surface conductivity near a planar wall outside the Debye--H\"uckel regime. 
Thus, we substitute $(2/3)^{\frac{1}{2}} (a/\lDinf)^{\frac{1}{4}}/ \lGC$ for all instances of $1/\lGC$ in Eq.~\eqref{eq:surf_conduct-plane} to yield
\begin{equation}
    \label{eq:surf_conduct}
    \kappa_\mathrm{s} \approx \frac{\eps(\Dp)}{\lGC} \left[-\frac{\lGC}{\lDinf} + \sqrt{\left( \frac{\lGC}{\lDinf} \right)^2 + \frac{2}{3} \left(\frac{a}{\lDinf} \right)^{\frac{1}{2}}} \right]
\end{equation}
for a thin EDL and all magnitudes of the electric potential. Equation~\eqref{eq:surf_conduct} is the main result of this work; further details of this derivation and a more in-depth comparison with the case of electric-field-driven electrolyte transport in a planar channel can be found in Sec.~\ref{ssec:thinEDL-gen} of the supplementary material. We also give an expression for the surface charge density with respect to the electric-field-driven electric current derived from this expression in Eq.~\eqref{seq:surface-charge} in the supplementary material. When $ \lGC \ll (2/3)^{\frac{1}{2}} a^{\frac{1}{4}}(\lDinf)^{\frac{3}{4}}$---where $\lGC$ is small at high surface charge---$\kappa_\mathrm{s} \propto (a/\lDinf)^{\frac{1}{4}}$. Thus, scaling of the electric current with fractional powers of the pore radius and Debye length is seen at high magnitudes of the electric potential energy as well as in the Debye--H\"uckel regime. 

\subsection{Limits of validity of the theory}
\label{sec:theo-thinEDL-utility}

For a counter-ion-only (i.e. ion-selective) membrane, which the system will approximate at very high surface charge densities, the Gouy--Chapman length quantifies the width of the EDL.\cite{Herrero2024BoltzmannFormulae} Moreover, a pore can be considered ion selective when the surface conductance dominates over the bulk conductance,\cite{poggioli2019, rankinEffectHydrodynamicSlip2016} which corresponds to a Dukhin length $\lDu = \kappa_{\mathrm{s}}/\kappa_{\mathrm{b}}$ that is much larger than the pore size $2a$ (see Eq.~\eqref{eq:elec_curr-gen}). For a thin EDL and $D_+ \approx D_-$, we can calculate the Dukhin length of an ultrathin membrane using Eqs.~\eqref{eq:bulk_conduct} and \eqref{eq:surf_conduct}. In this case, the condition $\lDu \gg 2a$ corresponds to
\begin{equation}
\label{eq:selec_pore-cond}
    \sqrt{\frac{3}{2}} \left(\frac{a}{\lDinf} \right)^{\frac{1}{4}} \left( {\frac{a}{\lDinf} + 2 } \right)^{\frac{1}{2}} \ll \frac{\lDinf}{\lGC},
\end{equation}
which reduces to $\sqrt{3/2} (a/\lDinf)^{\frac{3}{4}} \ll {\lDinf}/{\lGC}$ when $\lDinf \ll a/2$ and $\sqrt{3} (a/\lDinf)^{\frac{1}{4}} \ll {\lDinf}/{\lGC}$ when $\lDinf \gg a/2$. When Eq.~\eqref{eq:selec_pore-cond} holds, Eq.~\eqref{eq:surf_conduct} is expected to be accurate when $\lGC \ll a$ rather than being restricted to systems with $\lDinf \ll a$ (as $\lDinf \gtrsim \lGC$ in this case). As $\lGC \propto 1/\sigma$, the Gouy--Chapman length will be smaller than or comparable to the pore radius at moderate to high surface charge densities for a wide range of pore sizes. For example, a Gouy--Chapman length of $1$~nm corresponds to a surface charge density magnitude of $35$~mC~m$^{-2}$ for a monovalent $Z$:$Z$ aqueous electrolyte under ambient conditions, while surface charge densities of $-240$~mC m$^{-2}$ and $-600$~mC~m$^{-2}$ have been reported for surface-modified graphene at pH~7 and graphene at pH~8, respectively.\cite{rollingsIonSelectivityGraphene2016, shan2013graphene} Moreover, surface charge densities with magnitudes of $24$--$88$~mC~m$^{-2}$ have been reported for MoS$_2$ at pH~$5$, which could be increased to $300$--$800$~mC~m~$^{-2}$ by increasing the pH.\cite{fengSinglelayerMoS2Nanopores2016} 

When the condition in Eq.~\eqref{eq:selec_pore-cond} is not met, the thin EDL regime applies when $\lDinf \ll a$. Thus, we expect Eq.~\eqref{eq:surf_conduct} to break down for small pore sizes and low surface charge densities. At very low surface charge densities, we can apply the theory derived in the Debye--H\"uckel regime, for which the bulk conductance dominates. In this case, $\lDu \ll 2a$ and $I \approx -2a \kappa_{\mathrm{b}}\Delta \psi$ for arbitrary widths of the EDL, which is identical to the electrical current obtained from Hall's theory for the electrical access resistance of an uncharged  pore.\cite{hallAccessResistance1975} As the surface conductance of an ultrathin membrane in Eq.~\eqref{eq:surf_conduct} also diminishes as $a/\lDinf \rightarrow 0$ at fixed surface charge density, we expect our theory for an ultrathin membrane to make reasonable predictions about the electric-field-driven electric current over a wide range of conditions.

For electrolyte transport through a planar channel with a non-overlapping EDL, the contribution to the electric current due to the difference in ion diffusivities never has a higher order dependence on the surface charge density than the term involving the sum (see Eq.~\eqref{seq:elec_curr-plane-thinEDL} of the supplementary material for details). In this case, the assumption $D_+ \approx D_-$ can applied to electrolytes outside the Debye--H\"uckel regime even when $D_+ \neq D_-$. Similarly to electrolyte transport through a planar channel, we expect the contribution due to the difference in ion diffusivities to be small for an ultrathin membrane at surface charge densities that are of experimental interest and leave further investigation for future work. 

\subsection{Revisiting access electric-field-driven ionic conductance}
\label{sec:theo-acc_resist}

In the theory of the access contribution to the electric-field-driven ionic conductance of a nanopore of Lee et al.,\cite{leeLargeElectricSizeSurfaceCondudction2012} the surface contribution to the access conductance was assumed to be proportional to the surface conductivity of the pore interior, which was in turn approximated as the surface conductivity $\kappa_{\mathrm{s}}^{\infty}$ of a planar wall in the thin EDL limit. Including a bulk contribution proportional to the bulk conductivity, the contribution of each end of the nanopore to the conductance, which we will call the ``access conductance'',  was thus taken to be 
\begin{equation}
    \label{eq:lee-access_conductance}
    G_{\mathrm{a}}^{\mathrm{LB}} = 2 \alpha a \kappa_{\mathrm{b}} + \beta \kappa_{\mathrm{s}}^{\infty}
\end{equation}
where $\alpha$ and $\beta$ are numerical constants.\cite{leeLargeElectricSizeSurfaceCondudction2012}
The ionic conductance of a nanopore was then approximated as $G = 1/(R_{\mathrm{p}} + 2R_{\mathrm{a}})$, where $R_{\mathrm{p}}$
is the electrical resistance of the pore interior and $R_{\mathrm{a}}$ is that of each of the pore ends,\cite{leeLargeElectricSizeSurfaceCondudction2012} with the access resistance $R_{\mathrm{a}}$ the inverse of the access conductance and the pore resistance $R_{\mathrm{p}}$ the inverse of the conductance of the pore interior. As shown in Ref.~\citenum{hallAccessResistance1975}, the bulk contribution to the access resistance is $1/(4a \kappa_{\mathrm{b}})$; thus, Lee et al. assigned $\alpha =2$ and obtained $\beta = 2$ by fitting to continuum simulations of nanopores.

As the ionic conductance of the pore interior diminishes as the membrane thickness $L$ approaches the pore size $2a$,\cite{leeLargeElectricSizeSurfaceCondudction2012} the ionic conductance of an ultrathin membrane $G_{L/(2a) \rightarrow 0}$ is equivalent to half the access conductance $G_{\mathrm{a}}$.\cite{sahuColloquiumIonicPhenomena2019} Using Eq.~\eqref{eq:elec_curr-gen},  we can write in our theory
\begin{equation}
    \label{eq:acc_resist}
    G_{\mathrm{a}} = \frac{1}{R_{\mathrm{a}}} \approx 2 G_{L/(2a) \rightarrow 0} = {4a \kappa_{\mathrm{b}} + 2\kappa _{\mathrm{s}}}
\end{equation}
for a thin EDL, where the bulk conductivity and the surface conductance of an ultrathin membrane are given by Eqs.~\eqref{eq:bulk_conduct} and~\eqref{eq:surf_conduct}, respectively. Using Eq.~\eqref{eq:acc_resist}, we can write the ionic conductance of a finite-length pore as
\begin{equation}
\label{eq:ion_conduct-finite_len}
    G = \left[\frac{L}{\pi a (a\kappa_{\mathrm{b}} + 2 \kappa_{\mathrm{s}}^{\infty})} + \frac{1}{2a \kappa_{\mathrm{b}} + \kappa_{\mathrm{s}}} \right]^{-1} ,
\end{equation} 
where $G_{\mathrm{p}} = 1/R_{\mathrm{p}} = \pi a (a\kappa_{\mathrm{b}} + 2 \kappa_{\mathrm{s}}^{\infty})/L$ is the contribution to the ionic conductance due to the pore interior for a thin EDL as previously given in Ref.~\citenum{leeLargeElectricSizeSurfaceCondudction2012}. 

\begin{figure*}
\centering
    \includegraphics{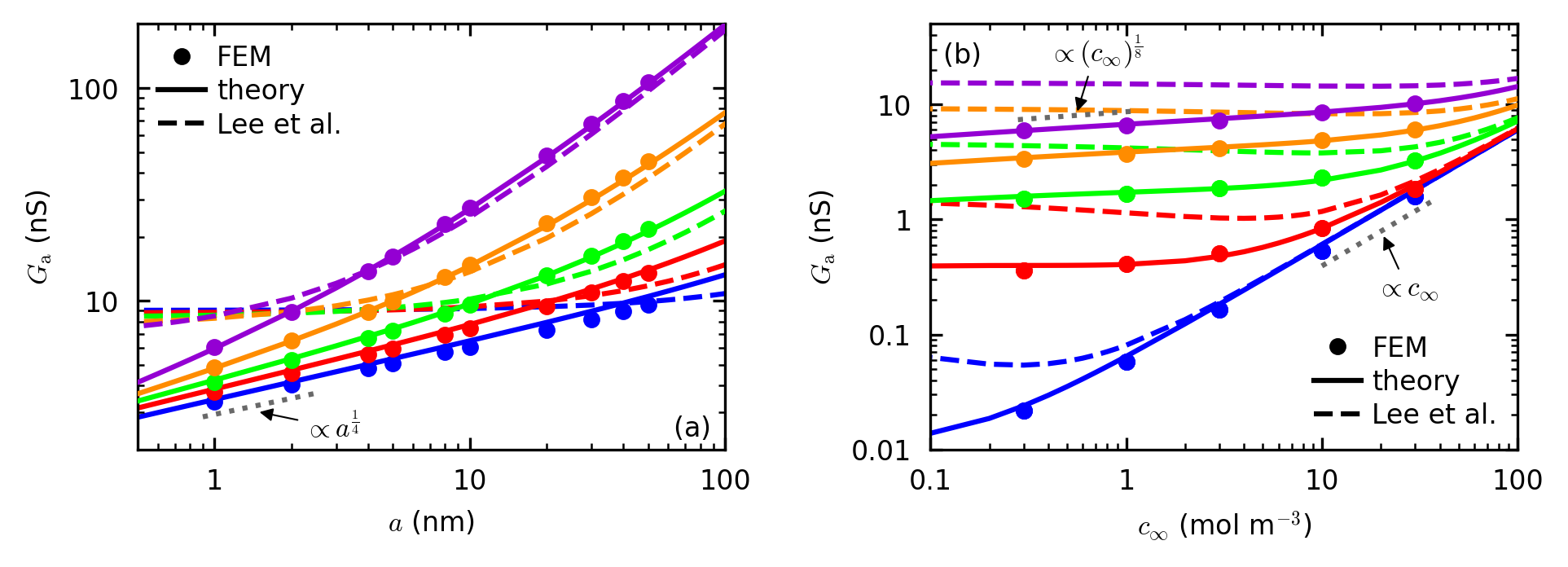}
    \caption{\label{fig:acc_resist-vs-radius-cave-thinEDL} Access conductance $G_\mathrm{a}$ vs (a) pore radius $a$ for surface charge density $\sigma = -60$~mC~m$^{-2}$ and bulk electrolyte concentration $\cinf = 0.3$ (blue), $1$ (red), $3$ (green), $10$ (orange), and $30$ (purple)~mol~m$^{-3}$ and (b) bulk electrolyte concentration $\cinf$ for pore radius $a = 1$~nm and surface charge density $\sigma = -1$ (blue), $-10$ (red), $-30$ (green), $-60$ (orange), and $-100$ (purple)~mC~m$^{-2}$. Symbols are FEM simulations of an ultrathin membrane, solid lines are the theory in this work (Eqs.~\eqref{eq:surf_conduct} and \eqref{eq:acc_resist}), dashed lines are the theory in Ref.~\citenum{leeLargeElectricSizeSurfaceCondudction2012} (Eqs.~\eqref{eq:surf_conduct-plane} and \eqref{eq:lee-access_conductance}) for $\alpha = \beta = 2$, and dotted lines indicate scaling relationships.}
\end{figure*}

Analogously to Lee et al., \cite{leeLargeElectricSizeSurfaceCondudction2012} we can interpret our results for the access conductance in terms of the increased apparent electric size (or radius) of a pore in an ultrathin membrane relative to the access conductance of an uncharged pore described by the theory of Hall.\cite{hallAccessResistance1975} Using our theory, the apparent electric pore size (diameter) is
\begin{align}
    2a_{\mathrm{app}} = \frac{G_{\mathrm{a}}}{2\kappa_{\mathrm{b}}} = 2a + \lDu,
    \label{eq:Dapp}
\end{align}
where $\lDu = \kappa_{\mathrm{s}}/\kappa_{\mathrm{b}}$ is the Dukhin length of an ultrathin membrane. On the other hand, $2a^{\mathrm{LB}}_{\mathrm{app}} = 2a+\beta\lDuinf/\alpha$ is the apparent electric pore size in Ref.~\citenum{leeLargeElectricSizeSurfaceCondudction2012}, where $\lDuinf = \kappa_{\mathrm{s}}^{\infty}/\kappa_{\mathrm{b}}$ is the Dukhin length 
at a planar surface. The theory in Ref.~\citenum{leeLargeElectricSizeSurfaceCondudction2012} predicts that the length scale over which  surface conduction effects within the pore extend outside the pore---referred to as the “healing length”---depends on the surface charge density and electrolyte concentration.\cite{leeLargeElectricSizeSurfaceCondudction2012} As the Dukhin length of an ultrathin membrane scales with fractional powers of the pore radius, Eq.~\eqref{eq:Dapp} indicates that the healing length also depends non-trivially on the pore geometry.  

\section{Results and discussion}
\label{sec:results}

To validate the theory in Sec.~\ref{sec:theo}, finite-element-method (FEM) simulations of electric-field-driven transport of potassium chloride (KCl) solutions ($0.3 \leq \cinf \leq 30$~mol~m$^{-3}$) through $0.2$-nm-thick membranes containing pores of various radii ($1 \leq a \leq 50$ nm) and surface charge densities ($|\sigma| \leq 100$ mC nm$^{-2}$) were carried out with COMSOL Multiphysics (version 4.3a).\cite{comsol4.3a} The no-slip boundary condition was applied at the solid--liquid boundary and potential differences up to $70$~mV, which are all within the linear-response regime,\cite{maoElectroosmoticFlowNanopore2014} were applied across the membrane. We verified that the measured fluxes did not depend significantly on the mesh, discretization order, membrane thickness or reservoir size for parameter variations around the chosen values. Further details of the simulations are given in Sec.~\ref{ssec:FEM} of the supplementary material. 

We have verified that the advective contribution to the electric current was negligibly small in all the simulations: Figure~\ref{sfig:Iadv-contr} in the supplementary material shows that the ratio of the advective to the electro-diffusive contributions to the electric current was $< 0.1$ inside of the pore mouth. We have also determined approximate expressions for the advective electric current in the thick and thin electric-double-layer limits for weak--ion membrane interactions in Sec.~\ref{ssec:adv_elec_curr} of the supplementary material. In the regimes for which they were derived, these expressions accurately capture the scaling of the advective electric current with pore radius, Debye length and surface charge density in the simulations. 

\subsection{Theory vs finite-element method simulations}
\label{sec:results-acc_resist}

We have compared the expressions for the access conductance derived in this work (Eqs.~\eqref{eq:surf_conduct} and~\eqref{eq:acc_resist}) and by Lee et al. in Ref.~\citenum{leeLargeElectricSizeSurfaceCondudction2012} (Eqs.~\eqref{eq:surf_conduct-plane} and \eqref{eq:lee-access_conductance}) for $\alpha = \beta = 2$ with our FEM simulation results, taking $G_{\mathrm{a}} = -2I/\Delta\psi$ to be twice the measured conductance of the ultrathin membrane. Figure~\ref{fig:acc_resist-vs-radius-cave-thinEDL} depicts the comparison for a range of pore radii and bulk electrolyte concentrations at a fixed surface charge density of $-60$~mC~m$^{-2}$, and for a range of bulk electrolyte concentrations and surface charge densities at a fixed pore radius of $1$~nm. At a fixed surface charge density, the theory in Ref.~\citenum{leeLargeElectricSizeSurfaceCondudction2012} captures the ionic conductance in the FEM simulations accurately for sufficiently large pore radii or bulk electrolyte concentrations such that $\lDinf \lesssim a$. Thus, for parameters for which the Debye length $\lDinf$ quantifies the EDL width (i.e. not a counter-ion only system), Eq.~\eqref{eq:lee-access_conductance} is accurate in the thin EDL regime. As shown in Fig.~\ref{fig:acc_resist-vs-radius-cave-thinEDL}, modifying this equation by replacing the surface conductivity near a planar wall in Eq.~\eqref{eq:surf_conduct-plane} with the surface conductance of an ultrathin membrane in Eq.~\eqref{eq:surf_conduct} to give Eq.~\eqref{eq:acc_resist} extends the validity of the theory to smaller pore sizes and lower electrolyte concentrations. \addition{In particular, when the pore is ion selective (i.e. when Eq.~\eqref{eq:selec_pore-cond} holds), the Gouy–Chapman length $\lGC$ rather than the Debye length $\lDinf$ quantifies the EDL width,\cite{Herrero2024BoltzmannFormulae} and so the theory is accurate for moderate to high surface charge densities even when $\lDinf > a$, as explained in Sec.~\ref{sec:theo-thinEDL-utility}. We further verify in Sec.~\ref{ssec:EDL_width} of the supplementary material that the Gouy–Chapman length quantifies the EDL width for an ion-selective membrane in the FEM simulations.} While the simulations at low electrolyte concentrations and small surface charge densities in Fig.~\ref{fig:acc_resist-vs-radius-cave-thinEDL}(b) are outside both the $\lDinf \ll a$ and $\lDu \gg 2a$ regimes, the theory derived in this work makes reasonable predictions about the ionic conductance even in these regimes, \deletion{because the EDL width is controlled by the Gouy--Chapman length when the pore is ion selective} \addition{because the bulk conductance dominates in these cases and the thin EDL approximation for the surface conductance becomes irrelevant to the accuracy of the theory}, as explained in Sec.~\ref{sec:theo-thinEDL-utility}.

In particular, Eq.~\eqref{eq:surf_conduct} captures the scaling of the ionic conductance with fractional powers of the pore radius and electrolyte concentration seen in simulations of small, highly charged pores at low electrolyte concentrations. Although the scaling law $G_{\mathrm{a}} \propto (\cinf)^{\frac{1}{8}}$ indicates a weak dependence of the ionic conductance on electrolyte concentration under these conditions, Fig.~\ref{fig:acc_resist-vs-radius-cave-thinEDL}(b) shows that assuming saturation of the access conductance at low electrolyte concentrations could lead to noticeably large discrepancies at low electrolyte concentration in ultrathin membranes with small pores and high surface charge. For a surface charge density of $-60$~mC~m$^{-2}$, pore radius of $1$~nm and electrolyte concentration of $0.1$~mol~m$^{-3}$, Fig.~\ref{fig:acc_resist-vs-radius-cave-thinEDL}(b) shows that the two theories deviate by a factor of $\approx 3$. Our theory predicts that the access electrical resistance of a nanopore continues to increase as the bulk concentration of the electrolyte solution is reduced, which 
could explain the increasing contribution from the access resistance observed in nanoscale biological ion channels as the electrolyte concentration is reduced.\cite{Aguilella2020, Martin2018}

\begin{figure}
    \centering
    \includegraphics{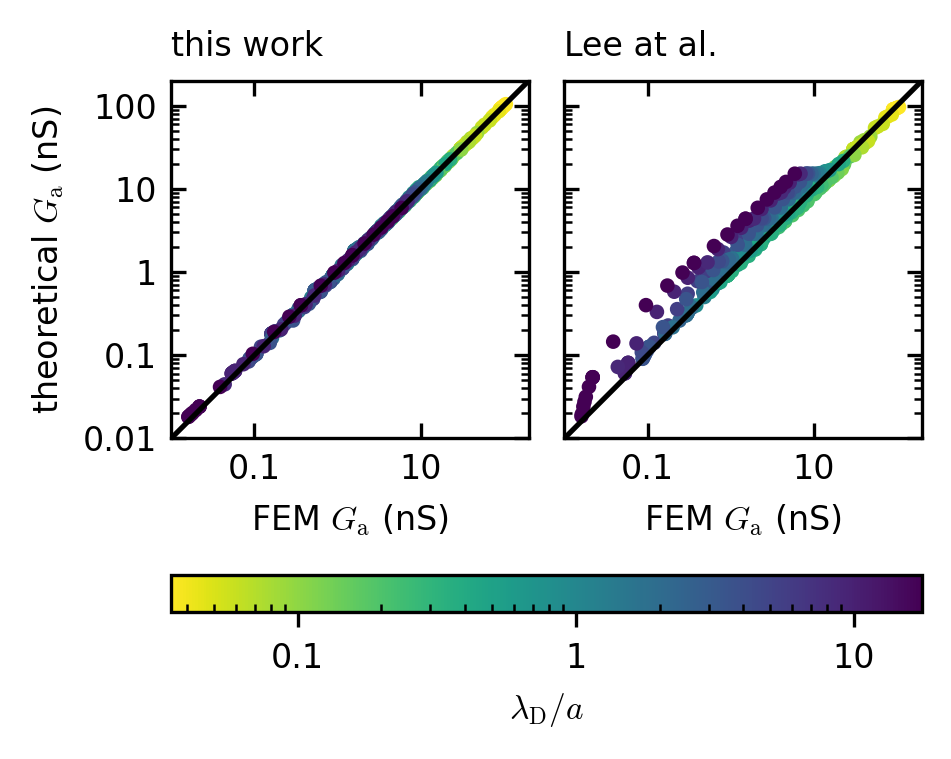}
    \caption{\label{fig:agreement_lines} Access conductance $G_\mathrm{a}$ from the theory derived in this work (Eqs.~\eqref{eq:surf_conduct} and~\eqref{eq:acc_resist}) and from the theory in Ref.~\citenum{leeLargeElectricSizeSurfaceCondudction2012} (Eqs.~\eqref{eq:surf_conduct-plane} and~\eqref{eq:lee-access_conductance}) for $\alpha = \beta = 2$, vs corresponding conductance from FEM simulations (symbols). The color map depicts variations in the ratio of the Debye length $\lDinf$ to the pore radius $a$ and the solid line indicates perfect agreement between the theory and simulations.}
\end{figure}

Figure~\ref{fig:agreement_lines} compares the access conductance derived in this work and that from the theory of Lee et al. \cite{leeLargeElectricSizeSurfaceCondudction2012} (for $\alpha = \beta = 2$) with the FEM simulations over a wide range of conditions. A line of perfect agreement is shown in black, where variations in the ratio of the Debye length to the pore radius, $\lDinf/a$, is indicated by the colors. As discussed in Sec.~\ref{sec:results-acc_resist} for the simulations shown in Fig.~\ref{fig:acc_resist-vs-radius-cave-thinEDL}, both theories can predict the ionic conductance from FEM simulations accurately when $\lDinf \lesssim a$. While the theory in Ref.~\citenum{leeLargeElectricSizeSurfaceCondudction2012} begins to deviate from the line of perfect agreement as $\lDinf/a$ increases, we have generalized this theory to remain accurate for a wide range of $\lDinf/a$.  

\subsection{Generalizing predictions of access conductance}
\label{sec:generalisation}

\begin{figure}
 \centering
 \includegraphics{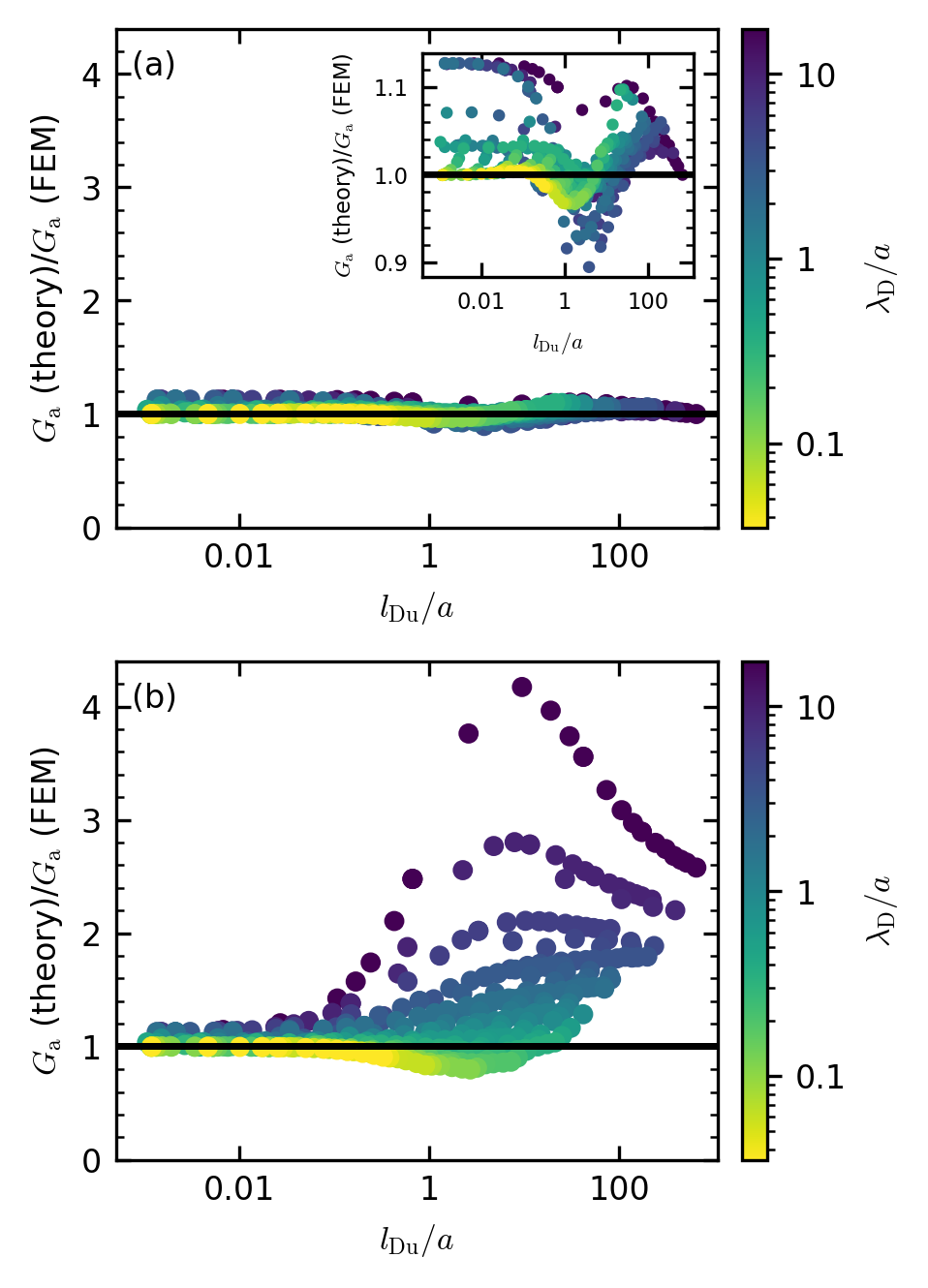}
    \caption{\label{sfig:theory-vs-FEM}
    Ratio of access conductance $G_\mathrm{a}$ from theory to that from FEM simulations (symbols) for the theory (a) in this work (Eqs.~\eqref{eq:surf_conduct} and~\eqref{eq:acc_resist}) and (b) Ref.~\citenum{leeLargeElectricSizeSurfaceCondudction2012} (Eqs.~\eqref{eq:surf_conduct-plane} and~\eqref{eq:lee-access_conductance}) for $\alpha = \beta = 2$, vs $\lDu/a$. $\lDu$ is the Dukhin length (of an ultrathin membrane) and $a$ is the pore radius. The color map depicts variations $\lDinf/a$, where $\lDinf$ is the Debye length and $a$ is the pore radius. The solid line indicates perfect agreement between the theory and simulations.
    }
 \end{figure}

Figure~\ref{sfig:theory-vs-FEM} shows the ratio of the access conductance from theory to that from simulations for the theory derived in this work and that in Ref.~\citenum{leeLargeElectricSizeSurfaceCondudction2012} for $\alpha = \beta = 2$ as the ratio of the Dukhin length $\lDu/a$ is varied. Analogous to Fig.~\ref{fig:agreement_lines}, a line of perfect agreement is shown in black, while the color map indicates variations in $\lDinf/a$. The largest ratio of the Dukhin length to the pore size considered in the simulations is $\approx 300$ (i.e. $\lDu/a \approx 600$), for which Fig.~\ref{sfig:theory-vs-FEM} shows good agreement between the theory derived in this work and the simulations. 

For the parameters used in the simulations, the maximum discrepancy between Eq.~\eqref{eq:acc_resist} (using Eq.~\eqref{eq:surf_conduct}) and the simulations is $\approx 14 \%$, which occurs at small $\lDu/a$ and large $\lDinf/a$. These conditions correspond to a thick EDL in the Debye--H\"uckel regime, for which Eq.~\eqref{eq:elec_curr-DH-thickEDL} or Hall's theory\cite{hallAccessResistance1975} can be used instead. The discrepancy of $\approx 10 \%$ seen for $\lDinf < a$ and $10 \lesssim \lDu/a \lesssim 100$ reflects the small deviation of the theory from the simulations with increasing pore size at an electrolyte concentration of $0.3$~mol~m$^{-3}$ and surface charge density of $-60$~mC~m$^{-2}$ (see Fig.~\ref{fig:acc_resist-vs-radius-cave-thinEDL}(a)), which we attribute tentatively to the effect of advection on the ion concentration distributions, which is not taken into account in the theory. On the other hand, the peak in the discrepancy between the theory in Ref.~\citenum{leeLargeElectricSizeSurfaceCondudction2012} and the simulations occurs for intermediate surface charge densities (intermediate $\lDu/a$) for which the bulk conductance does not dominate and the EDL width as quantified by the Debye length is sufficiently large that the approximation of the surface conductivity of a planar surface breaks down. As the surface charge density ($\lDu/a$) is further increased and the pore becomes more ion selective, the EDL width is instead quantified by the Gouy--Chapman length, which becomes smaller and thus the discrepancy is reduced.  

We have calculated the access electrical resistance from the ionic conductances in our simulations and from the access conductances in Eq.~\eqref{eq:lee-access_conductance} (using Eq.~\eqref{eq:surf_conduct-plane}) and Eq.~\eqref{eq:acc_resist} (using Eq.~\eqref{eq:surf_conduct}) to make similar comparisons to those in Fig.~\ref{fig:acc_resist-vs-radius-cave-thinEDL} for a range of pore radii, surface charge densities and bulk electrolyte concentrations in Figs.~\ref{sfig:acc_resist-vs-radius-thinEDL}--\ref{sfig:acc_resist-vs-cave-thinEDL} of the supplementary material. Moreover, we have calculated the access electrical resistance from the expressions for the electric current derived in the Debye--H\"uckel regime and compare these results with the simulations as the pore radius and surface charge density are varied in Sec.~\ref{ssec:acc_resist-DH} of the supplementary material. An analogous comparison to that in Fig.~\ref{fig:agreement_lines}, except with variations in $\lDu/a$ colored, is in Fig.~\ref{sfig:agreement_lines} of the supplementary material, which depicts the same trend with $\lDu/a$ as shown in Fig.~\ref{sfig:theory-vs-FEM}.

The parameter $\beta$ in the theory in Ref.~\citenum{leeLargeElectricSizeSurfaceCondudction2012} could be tuned to give better agreement for the access conductance when $\lDinf \lesssim a$ than shown in Figs.~\ref{fig:acc_resist-vs-radius-cave-thinEDL}--\ref{sfig:theory-vs-FEM}, particularly when Eq.~\eqref{eq:surf_conduct-plane} can be linearized. At high surface charge densities, $\kappa_{\mathrm{s}}^{\infty} \approx \eps(\Dp)/\lGC$ and $\kappa_{\mathrm{s}} \approx \sqrt{2/3}(a/\lDinf)^{\frac{1}{4}}\eps(\Dp)/\lGC$, for which $\beta \approx 2\sqrt{2/3}(a/\lDinf)^{\frac{1}{4}}$ upon a direct comparison between the theory for the access conductance derived in this work and that in Ref.~\citenum{leeLargeElectricSizeSurfaceCondudction2012}. However, our theory has the advantage that it does not contain any free parameters. In addition, given the dependence of the surface contribution to the access conductance on $\lDinf/a$ predicted by our theory in Eq.~\eqref{eq:surf_conduct}, the value of $\beta$ optimized for a particular $\lDinf/a$ in the theory in Ref.~\citenum{leeLargeElectricSizeSurfaceCondudction2012} will not necessarily give the correct conductance at other values of $\lDinf/a$, which could be problematic for interpreting experimental data. For example, the surface charge density of nanoporous membranes is often measured by fitting experimental conductance data to a theory over a range of electrolyte concentrations (i.e. varying $\lDinf/a$),\cite{leeLargeElectricSizeSurfaceCondudction2012, liu2024hbn, chengGateableOsmoticPowerPolymer2023, fengSinglelayerMoS2Nanopores2016, shan2013graphene, siriaGiantOsmoticEnergy2013, yazda2021hBn, LinSiliconNitride2021} which could result in an inaccurate surface charge density if the theory does not accurately capture the  $\lDinf/a$ dependence. 

As an example, we have fitted the simulations results in Fig.~\ref{fig:acc_resist-vs-radius-cave-thinEDL}(b) to the theory derived in this work and to that in Ref.~\citenum{leeLargeElectricSizeSurfaceCondudction2012}
for $\alpha=2$ and $\beta = 1/2$, $1$, $2$ and $4$.\cite{leeLargeElectricSizeSurfaceCondudction2012} For simulations of pores with surface charge densities of $-1$, $-10$, $-30$, $-60$ and $-100$~mC~m$^{-2}$ (where $a=1$~nm), we fitted surface charge densities with magnitudes of $0.72$, $9.8$, $30.$, $60.$ and $99$~mC~m$^{-2}$, respectively, using our theory (Eqs.~\eqref{eq:surf_conduct} and \eqref{eq:acc_resist}). We note that although 1~nm is a small pore radius, continuum hydrodynamic models  have been shown to be accurate down to the nanometer scale,\cite{bocquetNanofluidicsBulkInterfaces2010} including for electrolytes, when compared with molecular dynamics simulations.\cite{HuangHydrophobicNanochannels2007,HuangSuperhydrophobicSurfaces2008} Moreover, similar continuum models to that derived in this work have previously been used to fit surface charge densities from conductance measurements of nanopores on this scale.\cite{chengGateableOsmoticPowerPolymer2023, fengSinglelayerMoS2Nanopores2016} The most accurate fitted surface charge densities obtained from fits to the theory in Ref.~\citenum{leeLargeElectricSizeSurfaceCondudction2012} (Eqs.~\eqref{eq:surf_conduct-plane} and~\eqref{eq:lee-access_conductance}) had magnitudes of $0.61$~mC~m$^{-2}$ ($\beta = 1/2$), $12$~mC~m$^{-2}$ ($\beta = 1/2$), $26$~mC~m$^{-2}$ ($\beta = 1$), $54$~mC~m$^{-2}$ ($\beta = 1$) and $94$~mC~m$^{-2}$ ($\beta = 1$), respectively. The fits with Eqs.~\eqref{eq:surf_conduct} and \eqref{eq:acc_resist} gave the lowest relative error and typically the most favorable coefficient of determination $R^2$. The fitted surface charge densities ($\hat{\sigma}$), as well as the relative errors, standard errors of the fits, and $R^2$ of the ionic conductances, are given in Table~\ref{stab:fitted-surface_charge} of the supplementary material. Note that the $R^2$ was close to $1$ for all fits where surface charge densities of $-1$~mC~m$^{-2}$ and $-10$~mC~m$^{-2}$ were used in the simulations, even when the fitted surface charge density was inaccurate by a factor of $~4$. In this case, a poor fit was shielded by a relatively large bulk conductance. Furthermore, fits using Eq.~\eqref{eq:lee-access_conductance} with $\beta = 1/2$ always gave a favorable $R^2$ despite significantly overestimating the surface charge densities in simulations where $\sigma = -30$, $-60$, and $-100$~mC~m$^{-2}$; $|\hat{\sigma}| = 46$, $103$, and $183$~mC~m$^{-2}$ were obtained from these fits. These examples demonstrate that statistical measures of precision do not always indicate accurate results, particularly when there are tuneable parameters. Note that the values of $\beta$ that corresponded to the best fits were smaller than the optimal value of $\beta =2 $ found in Ref.~\citenum{leeLargeElectricSizeSurfaceCondudction2012} as the simulation data here are for smaller pores than considered in Ref.~\citenum{leeLargeElectricSizeSurfaceCondudction2012}.

\subsection{Case of a charged pore in an uncharged membrane}
\label{sec:uncharged-membrane}

It has been shown experimentally and using molecular simulations that 2D materials can acquire surface charge via chemisorption or physisorption of ions, such that charge is often distributed over the entire membrane surface, as assumed in our model.\cite{WanghBn2025, Advinculagraphene2025, McCaffreygraphene2017} For example, using surface-specific vibrational spectroscopy, atomic force microscopy, and machine-learning-based molecular dynamics, it has been shown that hBN in water at neutral pH acquires a significant negative charge due to \ce{OH-} adsorption.\cite{WanghBn2025}  Similar ion accumulation has been observed at graphene interfaces in both molecular simulations\cite{Advinculagraphene2025} and experiments.\cite{McCaffreygraphene2017} In order to match experimental data, modeling of ion transport through 2D membranes using FEM or molecular dynamics simulations has generally had to assume a non-zero surface charge density across the entire membrane and not just at the pore edge,
\cite{fengSinglelayerMoS2Nanopores2016, GrafMoS22019, rollingsIonSelectivityGraphene2016, yazda2021hBn} suggesting that in many circumstances such surfaces are effectively charged. Furthermore, membrane materials often have a surface charge over the entire surface under typical experimental conditions, which is necessary to take into account when modeling entrance effects.\cite{leeLargeElectricSizeSurfaceCondudction2012} For instance, silicon nitride, which commonly used in solid-state nanopores, typically has a surface charge density magnitude on the order of $10$ to $100$~mC~m$^{-2}$ at $\mathrm{pH} > 6$.\cite{LinSiliconNitride2021}

Nonetheless, we have tested the effect of applying surface charge only to the pore edge in our FEM model. In this case, the access resistances calculated from our simulations of ultrathin membranes with various pore radii ($1 \leq a \leq 50$~nm), bulk electrolyte concentrations ($0.3 \leq \cinf \leq 30$~mol~m$^{-3}$) and surface charge densities ($|\sigma| \leq 100$~mC~m$^{-2}$) are adequately described by Hall's theory for an uncharged pore, despite the high surface charge density magnitudes considered (see Sec.~\ref{ssec:access-resistance-uncharged-membrane} of the supplementary material for further details).\cite{hallAccessResistance1975} This result is in line with those of Ref.~\citenum{rollingsIonSelectivityGraphene2016}, in which FEM simulations showed that applying a very large surface charge of $-1$~C~m$^{-2}$ only at the pore edge of an ultrathin membrane yields a conductance close to the bulk value for concentrations between 30 and 3000~mol~m$^{-3}$. Consequently, matching experimental conductance data for a graphene membrane required a surface charge of $-0.6$~C~m$^{-2}$ distributed over the rest of the membrane surface.\cite{rollingsIonSelectivityGraphene2016} These findings differ markedly to those in Fig.~\ref{fig:acc_resist-vs-radius-cave-thinEDL} for a uniformly charged membrane, which indicate that the access electrical resistance can be reduced by orders of magnitudes by increasing the magnitude of the surface charge density.

\subsection{Fractional power-law scaling of conductance with concentration}
\label{sec:fractional_scaling}

Experimental data for the ionic conductance through a nanopore as a function of electrolyte concentration are often fitted to a power-law relation of the form\cite{zhongwuIonTransportBNNT2024, secchiScalingTransportCarbonNanotubes2016, Uematsu2018CNTs, Martin2018} 
\begin{equation}
\label{eq:arbitrary-scaling}
    G \propto (\cinf)^{\chi}.
\end{equation}
Conventional wisdom predicts linear scaling ($\chi = 1$) in the high-concentration (bulk-transport-dominated) limit and saturation of the conductance ($\chi = 0$) in the low-concentration (surface-transport-dominated) limit,\cite{bocquetNanofluidicsBulkInterfaces2010} which has been verified experimentally in many cases.\cite{stein2004, siriaGiantOsmoticEnergy2013} However, scaling with fractional exponents ($0< \chi < 1$) has also been observed in various experiments\cite{liu2024hbn, zhongwuIonTransportBNNT2024, chengGateableOsmoticPowerPolymer2023, secchiScalingTransportCarbonNanotubes2016, Uematsu2018CNTs, Martin2018, shan2013graphene, VenkatesanGraphene2012} and explained by various mechanisms.\cite{Noh2020, Uematsu2018CNTs, Secchi2016, Biesheuvel2016, Manghi2018}

Our theory of the ionic conductance of a finite-length pore (Eq.~\eqref{eq:ion_conduct-finite_len} with Eq.~\eqref{eq:surf_conduct}) and simulations (Fig.~\ref{fig:acc_resist-vs-radius-cave-thinEDL}(b)) indicates that fractional power-law scaling of the ionic conductance at low electrolyte concentrations is a general property of ultrathin membranes and thus of any system dominated by entrance effects, with $\chi$ approaching $1/8$ at low concentrations for an ultrathin ($L \ll 2a$) membrane. On the other hand, in line with conventional expectations $\chi \rightarrow 0$ at low concentrations when conductance inside the nanopore (i.e. $L \gg 2a$) dominates, and $\chi \sim 1$ for bulk-dominated electrolyte transport regardless of the length of the pore.\cite{hallAccessResistance1975, leeLargeElectricSizeSurfaceCondudction2012} Outside of these regimes, Eq.~\eqref{eq:ion_conduct-finite_len} predicts that $\chi$ varies continuously with the pore size, membrane thickness, electrolyte concentration, and surface charge density. 

\begin{figure}
    \centering
    \includegraphics{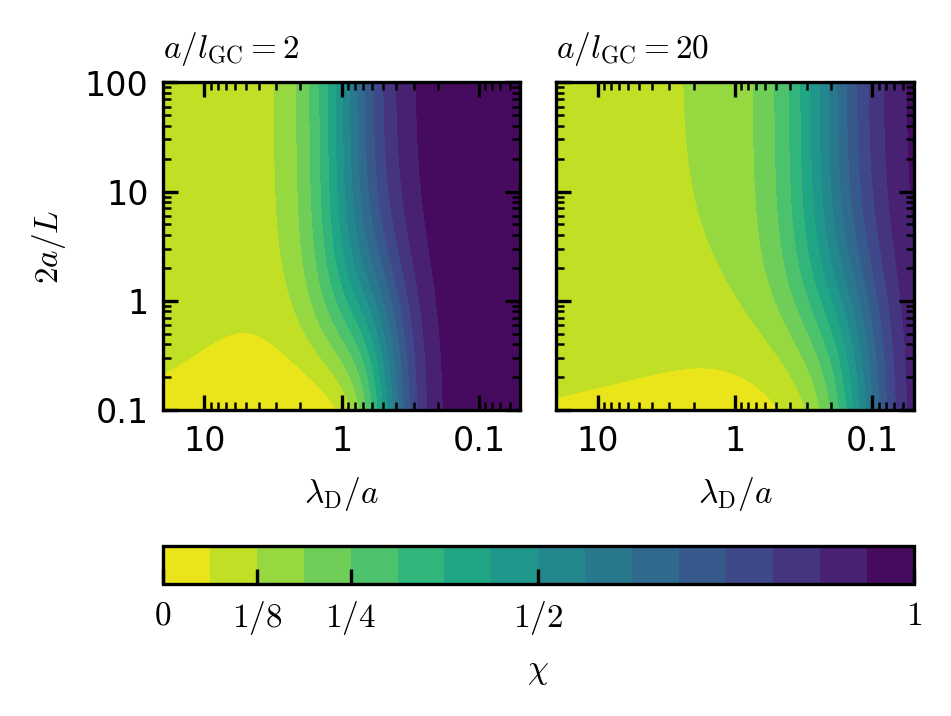}
    \caption{\label{fig:exponents} Apparent power-law scaling exponent of the ionic conductance of a membrane pore vs electrolyte concentration from the theory in Eq.~\eqref{eq:ion_conduct-finite_len} (using Eq.~\eqref{eq:surf_conduct-plane} and Eq.~\eqref{eq:surf_conduct} in the pore and access contributions, respectively) as a function of $2a/L$ and $\lDinf/a$, at $a/\lGC =$ $2$ and $20$, where $a$ is the pore radius, $L$ is the membrane thickness, $\lDinf$ is the Debye length and $\lGC$ is the Gouy--Chapman length.}
\end{figure}

To determine the value of $\chi$ as a function of electrolyte concentration from our theory, we have implemented the non-dimensionalization $G = \eps(\Dp)\hat{G}/a$ into Eq.~\eqref{eq:ion_conduct-finite_len}, which enables the dimensionless ionic conductance $\hat{G}$ to be written in terms of the length ratios $L/a$, $\lDinf/a$ and $\lGC/a$. As $\lDinf/a$ is the only parameter in $\hat{G}$ that depends on electrolyte concentration, Eq.~\eqref{eq:arbitrary-scaling} is equivalent to the relationship $\hat{G} \propto (a/\lDinf)^{2\alpha}$. Taking the logarithm of this expression, we can evaluate $\chi$ via the derivative
\begin{equation}
\label{eq:scaling_derivative}
    \chi = \frac{1}{2} \frac{\partial \log(\hat{G})}{\partial \log(a/\lDinf)}.
\end{equation}
We have evaluated $\chi$ over a wide range of conditions using central finite differences and present these results in Fig.~\ref{fig:exponents}. Details of the derivation of Eq.~\eqref{eq:scaling_derivative} and evaluation of $\chi$ are in Sec.~\ref{ssec:chi} of the supplementary material. Since $\chi$ defined in this way is not constant except in certain limiting regimes, we refer to it as the apparent power-law scaling exponent; however, it can appear constant when fitting experimental data if it varies sufficently slowly with concentration.

For the ratios of the length scales depicted in Fig.~\ref{fig:exponents}, $\chi$ converges to a constant value at very low electrolyte concentrations as predicted, and notably to a non-zero value of $\chi = 1/8$ for $L \ll 2a$. Since $\lDinf \propto (\cinf)^{-\frac{1}{2}}$, Fig.~\ref{fig:exponents} also indicates that, for an ultrathin membrane, the transitionary regime between $\chi = 1/8$ and bulk ($\chi = 1$)  electrolyte transport occurs over several orders of magnitudes of electrolyte concentration, particularly when the membrane is highly charged and the pores are small (i.e. $\lGC \ll a$). This transitionary regime is similarly broad for large-aspect-ratio ($L < 2a$) pores, in which transport is dominated by entrance effects. Thus, any power law between $1/8$ and $1$ could be obtained from fits of the ionic conductance to the electrolyte concentration when electrolyte concentrations within this transitionary regime are considered. Scaling of the ionic conductance with fractional powers of the electrolyte concentration is observed at very low electrolyte concentrations in experimental data of membranes with large aspect ratios (i.e. $L < 2a$), including single-layer hexagonal boron nitride,\cite{liu2024hbn} a two-dimensional nanoporous polymer membrane,\cite{chengGateableOsmoticPowerPolymer2023} and multilayered graphene and graphene-Al$_2$O$_3$ membranes.\cite{shan2013graphene, VenkatesanGraphene2012}

Electrical “imperfections” in the pore have previously been proposed to explain the scaling of the ionic conductance in a nanopore with a varying fractional powers of electrolyte concentration, where a fractional scaling exponent that represents electric potential “leakage” out of the pore, quantified by the difference between the Donnan potential and the potential measured in molecular dynamics simulations, was implemented into a theory for the surface conductance.\cite{Noh2020} Although the concept of electric potential “leakage” in Ref.~\citenum{Noh2020} is related to the non-uniformity of the electric field outside the pore described in our theory, the theory in Ref.~\citenum{Noh2020} models a change in the concentration of ions inside the pore, whereas the theory in this work predicts that the effects of surface conduction outside a nanopore contributes to the overall ionic conductance. In turn, the same exponent was assigned to the pore and access conductances in Ref.~\citenum{Noh2020}, while Eq.~\eqref{eq:ion_conduct-finite_len} predicts different scaling of these conductances with electrolyte concentration. Thus, the theory in Ref.~\citenum{Noh2020} predicts linear scaling of the ionic conductance in an ultrathin membrane with the electrolyte concentration at all concentrations, which differs from the fractional scaling predicted by our theory and observed in our FEM simulations (Fig.~\ref{fig:acc_resist-vs-radius-cave-thinEDL}).

As the surface conductance of an ultrathin membrane (Eq.~\eqref{eq:surf_conduct}) is not proportional to that near a planar wall (Eq.~\eqref{eq:surf_conduct-plane}) and in turn that predicted for electrolyte transport inside a long pore,\cite{leeLargeElectricSizeSurfaceCondudction2012} Fig.~\ref{fig:exponents} depicts a non-monotonic relationship between the scaling exponent and electrolyte concentration at low concentrations for intermediate aspect ratios of the pore, $2a/L$. For a counter-ion selective pore,
Eq.~\eqref{eq:ion_conduct-finite_len} reduces to
\begin{equation}
    \label{eq:another_condition}
    G \approx \frac{\eps (\Dp)}{\lGC} \left[ \frac{L}{2 \pi a} + \sqrt{\frac{3}{2}}\left( \frac{\lDinf}{a} \right)^{\frac{1}{4}} \right]^{-1}
\end{equation}
at very low electrolyte concentrations, where the first and second terms correspond to the pore conductance and access conductance, respectively. At low concentrations,  $\chi \rightarrow 0$ when the pore conductance dominates, but a further decrease in concentration can lead to the access conductance dominating due to an increase in $\lDinf$, resulting in $\chi \rightarrow 1/8$. Thus, our theory indicates that access surface conduction can be significant for ion-selective nanopores at low electrolyte concentrations, even if the pore length is much larger than its diameter. This leads to similar scaling behavior to that seen in ultrathin membranes at low electrolyte concentration, including a dependence of the ionic conductance on concentration. Thus, entrance effects could help to explain the electrolyte-concentration-dependent ionic conductance observed in $\approx$$38$-nm-long, $\approx$$2$-nm-diameter boron nitride nanotube porins at low electrolyte concentrations\cite{zhongwuIonTransportBNNT2024} (which is not seen in micrometer-length boron nitride nanotubes with larger diameters\cite{siriaGiantOsmoticEnergy2013}).

It should be noted that, to isolate effects of the pore geometry on ion transport, we have employed the simplest model that captures the effects of the unique pore geometry of an ultrathin membrane without the confounding effects of other phenomena. Although our model may not explain the complexities of real systems in quantitative detail, we have demonstrated that the pore geometry alone can explain the scaling of ionic conductance with fractional powers of electrolyte concentration. Noting that the surface charge of ultrathin and other membranes can vary with pH\cite{fengSinglelayerMoS2Nanopores2016, rollingsIonSelectivityGraphene2016, WanghBn2025} and electrolyte concentration,\cite{LinSiliconNitride2021} our theory can be readily adapted to use pH- and salinity-dependent surface charge densities calculated from a model of the charge regulation equilibrium at the surface.\cite{Uematsu2018CNTs} Interfacial slip, which is particularly relevant in carbon nanotubes,\cite{rankinEffectHydrodynamicSlip2016, Manghi2018, Manghi2021CNTs} is not included in our model. However, it can be captured within the Poisson--Nernst--Planck--Stokes framework using an empirical slip length.\cite{bocquetboundaryconditions2007} The high slip inside carbon nanotubes is expected to reduce the pore resistance and thus could make entrance effects on the ionic conductance prominent even for long nanotubes. A more quantitative model would also account for inhomogeneities in interfacial fluid properties,\cite{BonthuisBeyondContinuum2013} although including such inhomogeneities would likely not allow the model to be solved analytically. While continuum theory has been shown to be accurate down to the nanometer scale,\cite{bocquetNanofluidicsBulkInterfaces2010} molecular effects are significant at the sub-nanometer scale, and thus our theory may not be applicable in this regime. We leave such considerations for future work.

\section{Conclusions} 
\label{sec:conclusions}

We have derived an analytical equation in the Debye--H\"uckel regime and a semi-analytical equation for arbitrary surface potentials for the electric-field-driven electric current through an ultrathin membrane, which predict scaling with fractional powers of the pore size and Debye length. This unusual scaling behavior is not seen in theories of slit pores or thick membranes,\cite{bocquetNanofluidicsBulkInterfaces2010, rankinEffectHydrodynamicSlip2016} but was previously predicted for concentration-gradient-driven solute and electrolyte transport in the same geometry.\cite{rankinEntranceEffectsConcentrationgradientdriven2019, baldockCDF2025} 
The theory derived for arbitrary surface potentials accurately quantifies the ionic conductance through an ultrathin membrane in our finite-element method numerical simulations, which were carried out over a wide range of conditions. Thus, we have generalized the theory in Ref.~\citenum{leeLargeElectricSizeSurfaceCondudction2012} for the access electrical conductance of a membrane nanopore to make accurate predictions for an ultrathin membrane with small pores and moderate to high surface charge over a range of electrolyte concentrations.

Consequently, we have obtained an expression for the ionic conductance of a membrane nanopore that predicts scaling of the ionic conductance in thin membranes, and in thick ion-selective membranes, with fractional powers of the pore size and electrolyte concentration at low electrolyte concentrations. These results are important for understanding electrolyte transport in nanoporous membranes, particularly those where entrance effects are significant. Such membranes include those made from two-dimensional (2D) materials,\cite{sahuColloquiumIonicPhenomena2019} such as graphene\cite{rollingsIonSelectivityGraphene2016} and molybdenum disulfide,\cite{fengSinglelayerMoS2Nanopores2016} and those with low-friction internal surfaces,\cite{sisanEndNanochannels2011} such as carbon nanotubes.\cite{rankinEffectHydrodynamicSlip2016, Manghi2021CNTs} For example, our theory could help to explain scaling of the ionic conductance with fractional powers of the electrolyte concentration measured at low salt concentrations in boron nitride nanotube porins\cite{zhongwuIonTransportBNNT2024} and thin nanoporous membranes.\cite{chengGateableOsmoticPowerPolymer2023, liu2024hbn, shan2013graphene, VenkatesanGraphene2012} While we do not question the validity of previously proposed mechanisms, such as a salinity-dependent surface charge,\cite{Secchi2016, Biesheuvel2016, Manghi2018} interfacial slip,\cite{Manghi2018} or variable degrees of counter-ion binding at the surface,\cite{Uematsu2018CNTs} for explaining such fractional scaling in specific systems, our findings highlight that purely geometrical entrance effects can also be responsible for this phenomenon. Thus, the results in this work have significant implications for applications involving electric-field-driven transport of an electrolyte solution in a porous membrane, including reverse electrodialysis for energy generation, sensing and iontronics.

\section*{Supplementary material}

The supplementary material contains details of model parameters and variables; details of the system model; further details on the derivation of the theory in the Debye--H\"uckel regime; further details on the derivation of the electric-field-driven electric current for arbitrary ion--membrane interactions and a thin EDL; further details and supplementary results of finite-element method numerical simulations; further details of the fitted surface charge densities; \addition{quantification of the EDL width for an ion-selective membrane;} scaling laws for the advective contribution to the electric current (Debye--H\"uckel regime); and derivation of scaling of the ionic conductance with electrolyte concentration for a finite-length nanopore.

\section*{Acknowledgments}
This work was supported by the Australian Research Council under the Discovery Projects funding scheme (Grant No. DP210102155). H.C.M.B. acknowledges the support of an Australian Government Research Training Program Scholarship.

\section*{Author declarations}
\subsection*{Conflict of interest}
The authors have no conflicts to disclose.

\subsection*{Author Contributions}

\textbf{Holly C. M. Baldock}: Methodology (lead); Investigation (lead); Formal Analysis (lead); Writing -- original draft (lead); Writing -- review \& editing (supporting). \textbf{David M. Huang}: Conceptualization (lead); Methodology (supporting); Investigation (supporting); Formal Analysis (supporting); Funding acquisition (lead); Project administration (lead); Supervision (lead); Writing -- original draft (supporting); Writing -- review \& editing (lead).

\section*{Data availability}

The data that support the findings of this study are openly available in The University of Adelaide figshare repository at \url{https://doi.org/10.25909/29850083}, and in the Zenodo repositories at \url{https://doi.org/10.5281/zenodo.16758121},   \url{https://doi.org/10.5281/zenodo.16758170} and \url{https://doi.org/10.5281/zenodo.17216772}.

\section*{References}
\bibliography{2D_membrane_references}

\makeatletter\@input{2D_membrane_EDF-si-aux.tex}\makeatother

\end{document}



\title{Supplementary Material: Revisiting the access conductance of a nanopore in a charged membrane}
\author{Holly C. M. Baldock}
\author{David M. Huang}
\email{david.huang@adelaide.edu.au}
\affiliation{School of Physics, Chemistry and Earth Sciences, The University of Adelaide, SA 5005, Australia}

             
\maketitle 

\section{Model parameters and variables}
\label{ssec:parameters}

\begin{table}[!ht]
\caption{\label{stab:parameters}{Definitions of model parameters and variables.}}
\small
\renewcommand{\arraystretch}{0.85}
\begin{tabular*}{\columnwidth}{@{\extracolsep{\fill}} c l}
\hline
symbol & quantity \\
\hline
$e$ & elementary charge \\
$k_{\mathrm{B}}$ & Boltzmann constant \\
$T$ & temperature \\
$\epsilon$ & dielectric constant of the medium \\
$\epsilon_0$ & vacuum permittivity \\
$\eta$ & solution shear viscosity \\
$Z_i$ & valence of species $i$ \\
$D_i$ & diffusivity of species $i$ \\
$\mu_i$ & mobility of species $i$ \\
$a$ & pore radius \\
$L$ & membrane thickness \\
$\sigma$ & surface charge density \\
$\cinf^{(i)}$ & bulk concentration of species $i$ \\
$c_{\mathrm{s}}^{(i)}$ & concentration of species $i$ where the electric potential $\psi = 0$  \\
$\Delta \psi$ & applied potential difference between the reservoirs \\
$\lDinf$ & Debye screening length (for $\cinf$) \\
$l_{\mathrm{GC}}$ & Gouy--Chapman length \\
$\kappa_{\mathrm{b}}$ & bulk conductivity \\
$\lDu$ & Dukhin length of an ultrathin membrane \\
$\kappa_{\mathrm{s}}$ & surface conductance of an ultrathin membrane \\
$l_{\mathrm{Du}} ^{\infty}$ & Dukhin length of a planar channel \\
$\kappa_{\mathrm{s}}^{\infty}$ & surface conductivity near a planar wall \\
$\psi_0$ & equilibrium electric potential \\
$\psi_{\infty}$ & equilibrium electric potential at a planar surface \\
$\psi$ & electric potential \\
$c^{(i)}$ & concentration of species $i$ \\
$\rho$ & total ionic charge density \\
$p$ & pressure \\
$\boldsymbol{u}$ & solution velocity \\
$\boldsymbol{\phi}$ & stream function of flow velocity \\
$\boldsymbol{j}_i$ & flux density of species $i$ \\
\hline
\end{tabular*}
\end{table}

\sloppy

\section{Full details of system model}
\label{ssec:theo-full_model}

Assuming that the system can be described by continuum hydrodynamics, the governing equations for low-Reynolds-number steady-state transport of a dilute electrolyte solution are\cite{probstein1994}
\begin{align}
        \label{seq:poisson}
         \epsilon \epsilon_0 \nabla ^2 \psi + \rho & = 0, \\
        \label{seq:stokes}
        -\nabla p - \rho \nabla \psi + \eta \nabla ^2 \boldsymbol{u} & = 0,\\ 
         \label{seq:nernst-planck}
        \nabla \cdot \boldsymbol{j} _i = \nabla \cdot \left( c ^{(i)} \boldsymbol{u} - D_{i} \nabla c^{(i)} - Z_i e \mu _i c^{(i)} \nabla \psi \right) & =0,  \\
        \label{seq:continuity}
      \nabla \cdot \boldsymbol{u} & =0,
\end{align}
where $\psi$ is the electric potential, $\rho = e \sum_iZ_ic^{(i)}$ is the total ionic charge density, $\boldsymbol{u}$ is the solution velocity, $p$ is the pressure, and $\eta$ is the solution shear viscosity. Furthermore, $\boldsymbol{j}_i$, $c^{(i)}$, $Z_i$, $D_i$ and $\mu _i$ are the ion flux density, concentration, valence, diffusivity, and mobility, respectively, of species~$i$. We assume that $D_i$ and $\mu_i$ are related by the Einstein relation \cite{probstein1994} $\mu_i = D_i/(\kBT)$, where $k_{\mathrm{B}}$ is the Boltzmann constant and $T$ is the temperature. Moreover, $e$ is the elementary charge, $\epsilon$ is the dielectric constant of the medium (taken to be water here) and $\epsilon_0$ is the vacuum permittivity. We assume no slip of the fluid ($\boldsymbol{u} = 0$) and no penetration of the ions ($\boldsymbol{\hat{n}} \cdot \boldsymbol{j}_i = 0$) at the surface boundary (where $\boldsymbol{\hat{n}}$ is the unit vector normal to the surface). We consider fluid flow through a circular aperture of radius $a$ in an infinitesimally thin planar membrane induced by an applied gradient, as illustrated in Fig.~\ref{fig:schematic-2d_membrane} in the main paper. The circular aperture is modeled in oblate spheroidal cooordinates ($\zeta, \nu, \phi$), for which $0 \leq \zeta \leq 1$, $-\infty < \nu < \infty$, and $0 \leq \phi < 2\pi$,\cite{baldockCDF2025, rankinEntranceEffectsConcentrationgradientdriven2019, morse1953a} where $\zeta$ and $\nu$ can be related to a cylindrical coordinate system by $r = a \sqrt{(1 + \nu^2)(1 - \zeta^2 )}$ in the radial direction and $z = a\nu\zeta$ in the axial direction.

Here we consider flow of an electrolyte solution through the circular aperture shown in Fig.~\ref{fig:schematic-2d_membrane} in the main paper induced by a potential difference $\Delta \psi = \psi_{\mathrm{H}} - \psi_{\mathrm{L}}$ between the two sides of the membrane, where the concentration of the solution and pressure far from the pore is the same on both sides, i.e., $c_{\mathrm{H}} = c_{\mathrm{L}}$ and $p_{\mathrm{H}} = p_{\mathrm{L}}$, respectively. Analogous to Refs.~\citenum{baldockCDF2025} and \citenum{rankinEntranceEffectsConcentrationgradientdriven2019} for the similar case of concentration-gradient-driven flow in this geometry, we assume without loss of generality that the ion concentration in the presence of an applied gradient has the form
\begin{equation}
        \label{seq:conc}
        c^{(i)}(\zeta, \nu) = c_{\mathrm{s}}^{(i)}(\zeta, \nu) \exp{\left(-\frac{Z_i e \psi(\zeta, \nu)}{\kBT} \right)}, 
\end{equation}
where $c_{\mathrm{s}}^{(i)}$ is the concentration where the electric potential is zero and is to be determined. Substituting Eq.~\eqref{seq:conc} into Eq.~\eqref{seq:nernst-planck} and implementing the Einstein relation, the ion flux density of species~$i$ is
\begin{equation}
        \label{seq:ion_flux_dens}
        \boldsymbol{j} _i = \left( \boldsymbol{u} c_{\mathrm{s}}^{(i)} - D_i \nabla c_{\mathrm{s}} ^{(i)} \right) \exp{\left(-\frac{Z_i e \psi}{\kBT} \right)} ,
\end{equation}
where the first term is the advective ion flux density, while the electrophoretic and diffusive (electro-diffusive) ion flux densities are combined in the second term. Since $\psi \neq 0$ at the surface when there are electrostatic interactions between the ions and the membrane, and $\boldsymbol{\hat{n}} \cdot \boldsymbol{u} = 0$ at the surface, the zero flux boundary condition reduces to $\boldsymbol{\hat{n}} \cdot \boldsymbol{j}_i = \boldsymbol{\hat{n}} \cdot \nabla c_{\mathrm{s}}^{(i)} = 0$. For electric-field-driven flow through the pore geometry shown in Fig.~\ref{fig:schematic-2d_membrane}, the boundary conditions on the variables are $\psi \rightarrow \pm \frac{\Delta \psi}{2}$ in the upper and lower half planes, respectively, and $c^{(i)} \rightarrow \cinf^{(i)}$ at both boundaries. Substituting the boundary conditions $\psi \rightarrow \Delta \psi/2$ and $c^{(i)} \rightarrow \cinf^{(i)}$ for $\nu \rightarrow \infty$ into Eq.~\eqref{seq:conc} gives, for $\nu \rightarrow \infty$,
\begin{align}
    \label{seq:cs-boundary}
    \cinf^{(i)} \rightarrow c_{\mathrm{s}}^{(i)} \exp{\left(-\frac{Z_ie\Delta \psi}{2\kBT} \right)}.
\end{align}
Rearranging Eq.~\eqref{seq:cs-boundary} with respect to $c_{\mathrm{s}}^{(i)}$ gives, for $\nu \rightarrow \infty$,
\begin{align}
    \label{seq:cs-boundary-rearr}
    c_{\mathrm{s}}^{(i)}  \rightarrow  \cinf^{(i)} \exp{\left(\frac{Z_ie\Delta \psi}{2\kBT} \right)}.
\end{align}
In the linear--response regime, $Z_i e \Delta \psi \ll 2 \kBT$, such that we can approximate this boundary condition for $\nu \rightarrow \infty$ as
\begin{align}
     c_{\mathrm{s}}^{(i)}  \rightarrow  \cinf^{(i)} \left( 1 +  \frac{Z_ie\Delta \psi}{2\kBT} \right) .
\end{align}
As $\psi \rightarrow -\Delta \psi/2$ and $c^{(i)} \rightarrow \cinf^{(i)}$ for $\nu \rightarrow -\infty$, at this boundary
\begin{align}
     c_{\mathrm{s}}^{(i)}  \rightarrow  \cinf^{(i)} \left( 1 -  \frac{Z_ie\Delta \psi}{2\kBT} \right).
\end{align}
Thus, the boundary conditions on the virtual variable $c_{\mathrm{s}} ^{(i)}$ are $c_{\mathrm{s}} ^{(i)} \rightarrow \cinf^{(i)} \left(1\pm \frac{Z_i e \Delta \psi}{2 \kBT} \right)$ in the upper and lower half planes, respectively. 

Given the form of the ion flux density in Eq.~\eqref{seq:ion_flux_dens}, we can conveniently obtain general expressions for the electric-field-driven fluxes with respect to solutions to $c_{\mathrm{s}}^{(i)}$ in the linear--response regime without making any assumptions about the surface contributions to the electric potential or the magnitude of the non-equilibrium charge density. These expressions can be used to derive scaling laws for various limiting regimes of the width of the electric double layer and the strength of ion--membrane interactions. In the linear--response regime, we assume that each system variable can be represented by a perturbation expansion with respect to the equilibrium value, where $\varepsilon \ll 1$ is a dimensionless quantity that characterizes the perturbation due to the applied gradient. Hence, to first order
\begin{align}
    \label{seq:beta_velocity}
    \boldsymbol{u} & = \varepsilon \boldsymbol{u}_1, \\
    \label{seq:beta_cs}
    c_{\mathrm{s}}^{(i)}  & = c _{\mathrm{s}_0}^{(i)} + \varepsilon c_{\mathrm{s}_1}^{(i)}, \\
    \label{seq:beta_epot}
    \psi & = \psi _0 + \varepsilon \psi _1, \\
    \label{seq:beta_p}
    p & = p_0 + \varepsilon p_1,
\end{align}
where $c _{\mathrm{s}_0}^{(i)} = \cinf^{(i)}$ is constant and $\psi_0$ is an even function of $\nu$ for a uniformly charged membrane. Thus, we can use Eqs.~\eqref{seq:beta_velocity}--\eqref{seq:beta_p} to write the boundary conditions imposed on the electric potential and ion concentration far from the membrane as $\varepsilon \psi _1 \rightarrow \pm \frac{\Delta \psi}{2}$ as $\nu \rightarrow \pm \infty$, such that $\psi_0 \rightarrow 0$ and $ \varepsilon c_{\mathrm{s}_1} \rightarrow \pm \frac{Z_i e \Delta \psi}{2 \kBT}$ as $\nu \rightarrow \pm \infty$. Substituting Eqs.~\eqref{seq:beta_velocity}--\eqref{seq:beta_epot} into Eq.~\eqref{seq:ion_flux_dens} gives
\begin{align}
        \setcounter{equation}{14}
        \boldsymbol{j} _i & =  \left[ (\varepsilon\boldsymbol{u}_1)(\cinf^{(i)} + \varepsilon \bcs ^{(i)})  - D_i  \nabla (\cinf^{(i)}+ \varepsilon \bcs ^{(i)}) \right]\exp{\left(-\frac{Z_i e (\psi_0 + \varepsilon \psi_1)}{\kBT}\right)} \notag \\
        & \approx  \left[ (\varepsilon\boldsymbol{u}_1)(\cinf^{(i)} + \varepsilon \bcs ^{(i)})  - D_i  \nabla (\varepsilon \bcs ^{(i)}) \right] \left(1 - \frac{Z_i e \varepsilon \psi_1}{\kBT} \right)\exp{\left(-\frac{Z_i e \psi_0}{\kBT}\right)},
        \end{align}
where we have used the fact that $\cinf^{(i)}$ is constant, and $Z_i e (\varepsilon \psi_1) \ll {\kBT}$ in the linear--response regime. Truncating this equation at $O(\varepsilon)$, we can write the ion flux density of species $i$ to first order as
\begin{equation}
        \label{seq:ion_flux_dens-lin}
        \boldsymbol{j} _i = \left[ (\varepsilon\boldsymbol{u}_1)\cinf^{(i)}  - D_i  \nabla (\varepsilon \bcs ^{(i)}) \right]\exp{\left(-\frac{Z_i e \psi_0}{\kBT}\right)},
\end{equation}
which is the same as Eq.~\eqref{eq:ion_flux_dens-lin} in the main paper.


\section{Derivation of electric-field-driven fluxes for weak ion--membrane interactions (Debye--H\"uckel regime)}
\label{ssec:theo-DH}

Using Eq.~\eqref{eq:ion_flux_dens-lin} in the main paper, the electric current density $\boldsymbol{j}_{\mathrm{e}} = e\sum _i Z_i \boldsymbol{j}_i$ of a $Z$:$Z$ electrolyte (i.e. $Z_+ = -Z_- = Z$ and $\cinf^{+} = \cinf^{-} = \cinf$) is
\begin{align}
    \label{seq:elec_curr_dens}
    \boldsymbol{j} _\mathrm{e} = -Ze \left[ 2 \cinf (\varepsilon\boldsymbol{u}_1) \sinh \left(\frac{Ze \psi_0}{\kBT} \right) \right.
       &+ D_+  \nabla (\varepsilon c_{\mathrm{s}_1} ^+ ) \exp{\left(-\frac{Z e \psi_0}{\kBT}\right)}  \notag \\
       & - \left. D_- \nabla (\varepsilon c_{\mathrm{s}_1} ^- ) \exp{\left(\frac{Z e \psi_0}{\kBT}\right)} \right]
\end{align}
in the linear--response regime. Note that we can also write Eq.~\eqref{seq:elec_curr_dens} as
\begin{align}
    \label{seq:elec_curr_dens-exp}
    \boldsymbol{j} _\mathrm{e} = & -Ze \left\{ 2 \cinf (\varepsilon\boldsymbol{u}_1) \sinh \left(\frac{Ze \psi_0}{\kBT} \right) \right. \notag \\
       & \left. -\frac{1}{2} \nabla\left( \varepsilon c_{\mathrm{s}_1}^{+} + \varepsilon c_{\mathrm{s}_1}^{-} \right) \left[(\Dp)\sinh \left(\frac{Ze \psi_0}{\kBT} \right) - (\Dm)\cosh \left(\frac{Ze \psi_0}{\kBT} \right) \right] \right. \notag \\
        & \left. +\frac{1}{2} \nabla\left( \varepsilon c_{\mathrm{s}_1}^{+} - \varepsilon c_{\mathrm{s}_1}^{-} \right) \left[(\Dp)\cosh \left(\frac{Ze \psi_0}{\kBT} \right) - (\Dm)\sinh \left(\frac{Ze \psi_0}{\kBT} \right) \right] \right\} .
\end{align}
For electric-field-driven transport of a $Z$:$Z$ electrolyte, Eq.~\eqref{eq:cs-DH} in the main paper indicates that $\varepsilon c_{\mathrm{s}_1}^{+} = - \varepsilon c_{\mathrm{s}_1}^{-} = \varepsilon c_{\mathrm{s}_1}$, where
\begin{align}
    \label{seq:cs-DH}
    \varepsilon c_{\mathrm{s}_1} = \frac{\cinf Z e \Delta \psi}{\pi \kBT} \tan^{-1}{\nu}.
\end{align}
Note that Eq.~\eqref{seq:cs-DH} and Eq.~\eqref{eq:cs-DH} in the main paper may hold for electric potential energies that are larger than the thermal energy if, for example, $\varepsilon {c_{\mathrm{s}_1}^{(i)}}$ and $\psi_0$ are approximately orthogonal.\cite{rankinEntranceEffectsConcentrationgradientdriven2019} For example, if the tangential component of the electric field is small (i.e. the surface potential energy is approximately constant), then $\nabla ( {Z_i e \psi_0}/{(\kBT)} ) \cdot \nabla ({\varepsilon c_{\mathrm{s}_1}^{(i)}}/{\cinf^{(i)}} ) \approx 0$ at the membrane surface.
Since $\varepsilon c_{\mathrm{s}_1}^{+} - \varepsilon c_{\mathrm{s}_1}^{-} = 2 \varepsilon c_{\mathrm{s}_1}$ and $\varepsilon c_{\mathrm{s}_1}^{+} + \varepsilon c_{\mathrm{s}_1}^{-} = 0$, Eq.~\eqref{seq:elec_curr_dens-exp} reduces to
\begin{align}
    \label{seq:elec_curr_dens-rearr}
    \boldsymbol{j} _\mathrm{e} = & -Ze \left\{ 2 \cinf (\varepsilon\boldsymbol{u}_1) \sinh \left(\frac{Ze \psi_0}{\kBT} \right) \right. \notag \\
        & \left. + \, \, {\nabla} \left(\varepsilon c_{\mathrm{s}_1} \right) \left[(\Dp)\cosh \left(\frac{Ze \psi_0}{\kBT} \right) - (\Dm)\sinh \left(\frac{Ze \psi_0}{\kBT} \right) \right] \right\} .
\end{align} 
Since $\nabla = \left(\frac{1}{h_{\zeta}}\frac{\partial}{\partial \zeta}, \frac{1}{h_\nu} \frac{\partial}{\partial \nu} \right)$ in oblate-spheroidal coordinates where $ h_\zeta=a\sqrt{(\nu+\zeta^2)/(1-\zeta^2)}$ and $h_\zeta=a\sqrt{(\nu+\zeta^2)/(1+\nu^2)}$ are scale factors in the convention of Morse and Feshbach,\cite{morse1953a} the gradient of $\varepsilon c_{\mathrm{s}_1}$ is $\nabla (\varepsilon c_{\mathrm{s}_1}) = \left(0,  \frac{\cinf Z e \Delta \psi}{h_\nu \pi \kBT} \frac{1}{1+\nu^2} \right)$. Substituting this expression and Eq.~\eqref{seq:elec_curr_dens-rearr} into Eq.~\eqref{eq:elec_curr-int} in the main paper gives
\begin{align}
    \label{seq:elec_curr-lin-gen} 
    I = -\frac{a \eps (\Dp) \Delta \psi}{(\lDinf)^2} \int ^1 _0 \mathrm{d} \zeta \left[\cosh{\left(\frac{Ze \psi_0|_{\nu=0}}{\kBT}\right)} -\left( \frac{\Dm}{\Dp} \right) \sinh{\left(\frac{Ze \psi_0|_{\nu=0}}{\kBT} \right)} \right] \notag \\
     - 4 \pi Ze \cinf \int^1 _0 \mathrm{d} \zeta \left[h_\zeta h_\phi (\varepsilon u_{\nu_1}) \sinh\left(\frac{Ze \psi_0}{\kBT} \right) \right]_{\nu=0} ,
\end{align}
where $h_\phi= a\sqrt{(1 + \nu^2)(1-\zeta^2)}$ is a scale factor,\cite{morse1953a} $\varepsilon u_{\nu_1}$ is the first-order, normal (i.e. $\nu$-) component of the fluid velocity through the aperture, and
\begin{align}
\label{seq:debye_length}
    \lDinf = \sqrt{\frac{\eps \kBT}{2(Ze)^2 \cinf}}
\end{align}
is the Debye screening length of a $Z$:$Z$ electrolyte with bulk electrolyte concentration $\cinf$. Note that the final term in Eq.~\eqref{seq:elec_curr-lin-gen} is the advective contribution. Assuming that $|Z_ie\psi_0| \ll \kBT$ (Debye--H\"uckel regime) and that advection is negligible relative to electro-diffusion (low-Peclet-number flow) in the electric current density, 
Eq.~\eqref{seq:elec_curr-lin-gen} reduces to
\begin{align}
    \label{seq:elec_curr-DH} 
    I = -\frac{a \eps (\Dp) \Delta \psi}{(\lDinf)^2} \left[ 1 + \frac{1}{2}\int ^1 _0 \mathrm{d} \zeta \left(\frac{Ze \psi_0|_{\nu=0}}{\kBT}\right)^2 -\left( \frac{\Dm}{\Dp} \right) \int ^1 _0 \mathrm{d} \zeta {\left(\frac{Ze \psi_0|_{\nu=0}}{\kBT} \right)} \right],
\end{align}
where
\begin{align}
    \label{seq:epot-DH}
      \psi_0 =  \frac{\sigma a}{\eps} \left[ \int ^{\mathrm{\infty}} _0 \mathrm{d} s \, \frac{J_1(s) J_0(\hat{r} s)}{\sqrt{(a/\lDinf)^2 + s^2}} e^{-\hat{z}\sqrt{(a/\lDinf)^2 + s^2}} + \frac{e^{-(a/\lDinf) \hat{z}}}{(a/\lDinf)} \right]
\end{align}
is the equilibrium electric potential of an ultrathin membrane containing a circular pore in the Debye--H\"uckel regime, $\hat{z} = z/a$ amd $\hat{r} = r/a$ are the dimensionless cylindrical coordinates, and $J_0$ and $J_1$ are Bessel functions of the first kind.\cite{maoElectroosmoticFlowNanopore2014} Equation~\eqref{seq:elec_curr-DH} is similar to Eq.~\eqref{eq:elec_curr-DH-full} in the main paper, but Eq.~\eqref{seq:elec_curr-DH} does not assume that $D_+ \approx D_-$. 

Furthermore, the solute flux $J$ induced by an applied gradient across the membrane is
\begin{align}
    \label{seq:sol_flux-int}
    J & = \iint _S \mathrm{d} S \,  \boldsymbol{\hat{n}} \cdot \boldsymbol{j},
\end{align}
where $\boldsymbol{j} = \sum _i \boldsymbol{j}_i$ is the solute (or total ion) flux density, $\boldsymbol{\hat{n}} = \boldsymbol{\hat{\nu}}$ is the unit vector normal to the pore mouth (at $\nu =0$), and the surface integral is across the pore aperture. Using Eq.~\eqref{seq:ion_flux_dens-lin} (or, equivalently, Eq.~\eqref{eq:ion_flux_dens-lin} in the main paper), we can write the solute flux density of a $Z$:$Z$ electrolyte in the linear--response regime as
\begin{align}
    \label{seq:sol_flux-density_expand}
    \boldsymbol{j} = \, \, &  2 \cinf (\varepsilon\boldsymbol{u}_1) \cosh \left(\frac{Ze \psi_0}{\kBT} \right) \notag \\
       & \left. -\frac{1}{2} \nabla\left( \varepsilon c_{\mathrm{s}_1}^{+} + \varepsilon c_{\mathrm{s}_1}^{-} \right) \left[(\Dp)\cosh \left(\frac{Ze \psi_0}{\kBT} \right) - (\Dm)\sinh \left(\frac{Ze \psi_0}{\kBT} \right) \right] \right. \notag \\
        & +\frac{1}{2} \nabla\left( \varepsilon c_{\mathrm{s}_1}^{+} - \varepsilon c_{\mathrm{s}_1}^{-} \right) \left[(\Dp)\sinh \left(\frac{Ze \psi_0}{\kBT} \right) - (\Dm)\cosh \left(\frac{Ze \psi_0}{\kBT} \right) \right] .
\end{align}
Substituting Eq.~\eqref{seq:cs-DH}---where $\varepsilon c_{\mathrm{s}_1} ^{+} - \varepsilon c_{\mathrm{s}_1} ^{-} = 2\varepsilon c_{\mathrm{s}_1}$ and $\varepsilon c_{\mathrm{s}_1} ^{+} + \varepsilon c_{\mathrm{s}_1} ^{-} = 0$ for a $Z$:$Z$ electrolyte---into Eq.~\eqref{seq:sol_flux-density_expand} and substituting this expression into Eq.~\eqref{seq:sol_flux-int} gives
\begin{align}
    \label{seq:sol_flux-lin-gen} 
    J = \frac{a \eps (\Dp) \Delta \psi}{Ze (\lDinf)^2} \int ^1 _0 \mathrm{d} \zeta \left[ \sinh{\left(\frac{Ze \psi_0|_{\nu=0}}{\kBT}\right)} - \left(\frac{\Dm}{\Dp} \right) \cosh{\left(\frac{Ze \psi_0|_{\nu=0}}{\kBT} \right)} \right] 
    \notag \\ 
    + 4 \pi \cinf \int^1 _0 \mathrm{d} \zeta \left[h_\zeta h_\phi (\varepsilon u_{\nu_1}) \cosh\left(\frac{Ze \psi_0}{\kBT} \right) \right]_{\nu=0}.
\end{align}
Note that the final term in Eq.~\eqref{seq:sol_flux-lin-gen} is the advective contribution. Assuming that $|Z_ie\psi_0| \ll \kBT$, Eq.~\eqref{seq:sol_flux-lin-gen} reduces to
\begin{equation}
    \label{seq:sol_flux-lin-DH} 
    J = \frac{a \eps (\Dp) \Delta \psi}{Ze (\lDinf)^2} \left[ {\int ^1 _0 \mathrm{d} \zeta \, \frac{Ze \psi_0|_{\nu=0}}{\kBT}} - \left(\frac{\Dm}{\Dp} \right) \right]  + 4 \pi \cinf \int^1 _0 \mathrm{d} \zeta \left[h_\zeta h_\phi (\varepsilon u_{\nu_1}) \right]_{\nu=0}. 
\end{equation}
As the flow rate induced by an applied gradient across the membrane is
\begin{equation}
    \label{seq:flow_rate-int}
    Q = \iint _S \mathrm{d}S \, \boldsymbol{u} \cdot \boldsymbol{\hat{n}},
\end{equation}
then
\begin{equation}
    Q = \int^{2 \pi} _0 \mathrm{d} \phi \int ^1 _0  \mathrm{d} \zeta \, h_\zeta h_\phi \varepsilon u_{\nu_1} |_{\nu=0} = 2 \pi \int ^1 _0 \mathrm{d} \zeta \, h_\zeta h_\phi \varepsilon u_{\nu_1} |_{\nu=0}
    \label{seq:flow_rate-int-exp}
\end{equation}
is flow rate through an ultrathin membrane in the linear--response regime. Substituting Eq.~\eqref{seq:flow_rate-int-exp} into Eq.~\eqref{seq:sol_flux-lin-DH} gives
\begin{equation}
    \label{seq:sol_flux-DH-exp} 
    J = \frac{a \eps (\Dp) \Delta \psi}{Ze (\lDinf)^2} \left[ {\int ^1 _0 \mathrm{d} \zeta \, \frac{Ze \psi_0|_{\nu=0}}{\kBT}} - \left(\frac{\Dm}{\Dp} \right) \right]  + 2 \cinf Q,
\end{equation}
Thus, inserting the equation derived in Ref.~\citenum{maoElectroosmoticFlowNanopore2014} for the electroosmotic flow rate in the Debye--H\"uckel regime into Eq.~\eqref{seq:sol_flux-DH-exp} gives
\begin{align}
    \label{seq:sol_flux-DH-sub}
    J &= \frac{a \eps (\Dp) \Delta \psi}{Ze (\lDinf)^2} \left[ {\int ^1 _0 \mathrm{d} \zeta \, \frac{Ze \psi_0|_{\nu=0}}{\kBT}} - \left(\frac{\Dm}{\Dp} \right) \right] \notag \\ 
    & \quad + \frac{2a^3 \kBT (\eps)^2 \Delta \psi}{\pi (Ze)^2 \eta (\lDinf)^4} \int^1 _0 \mathrm{d}\zeta \, \zeta^2 \int^{\infty}_0 \mathrm{d} \nu \, \frac{\psi_0}{1 + \nu^2}
\end{align}
as the electric-field-driven solute flux in the Debye--H\"uckel regime. 

\subsubsection{Thick electric double layer}
\label{ssec:theo-DH-thickEDL}

When the width of the electric double layer is much larger than the pore radius ($\lDinf \gg a$), $\psi_0 \approx \sigma \lDinf/(\eps)$ in the Debye--H\"uckel regime. In this limit, Eq.~\eqref{seq:elec_curr-DH} reduces to
\begin{align}
     I \approx -\frac{a \eps (\Dp) \Delta \psi}{(\lDinf)^2}\left[1 -  2 \sgn(\sigma)\left(\frac{\Dm}{\Dp} \right)\left(\frac{\lDinf}{l_{\mathrm{GC}}} \right) + 2\left(\frac{\lDinf}{l_{\mathrm{GC}}} \right)^2 \right],
  \label{seq:elec_curr-DH-thickEDL}
\end{align}
where
\begin{align}
    \lGC = \frac{2 \eps \kBT}{Ze |\sigma|}
\end{align}
is the Gouy--Chapman length for a $Z$:$Z$ electrolyte. Assuming $|\Dm|/(\Dp) \ll l_{\mathrm{GC}}/(2\lDinf) + \lDinf/l_{\mathrm{GC}}$ (i.e. ignoring contribution due to difference in ion diffusivities) gives
\begin{align}
     I \approx -\frac{a \eps (\Dp) \Delta \psi}{(\lDinf)^2}\left[1 + 2\left(\frac{\lDinf}{l_{\mathrm{GC}}} \right)^2 \right] ,
     \label{seq:elec_curr-DH-thickEDL-reduce}
\end{align}
which is identical to Eq.~\eqref{eq:elec_curr-DH-thickEDL} in the main paper. For the case of a planar surface, $Ze \psi_\infty/(\kBT) = 2\sinh^{-1}(\lDinf/\lGC) \ll 1$ constitutes the Debye--H\"uckel regime;\cite{Herrero2024BoltzmannFormulae} this condition rearranges to $\lDinf / \lGC \ll 2\sinh(1/2) \approx 1$. For an ultrathin membrane, we expect that a similar condition to $\lGC \gg \lDinf$ constitutes the Debye--H\"uckel regime, for which the bulk contribution (first term) in Eq.~\eqref{seq:elec_curr-DH-thickEDL-reduce} dominates. Furthermore, substituting $\psi_0 \approx \sigma \lDinf/(\eps)$ into Eq.~\eqref{seq:sol_flux-DH-sub} gives (for $\lDinf \gg a$)
\begin{align}
    \label{seq:sol_flux-DH-thickEDL}
    J &\approx \frac{a \eps (\Dp) \Delta \psi}{Ze (\lDinf)^2} \left[2 \sgn(\sigma) \frac{\lDinf}{\lGC} - \left(\frac{\Dm}{\Dp} \right) \right] + \frac{a^3 \sigma \kBT \eps \Delta \psi}{3 (Ze)^2 \eta (\lDinf)^3} .
\end{align}
As the first term in Eq.~\eqref{seq:sol_flux-DH-thickEDL} is proportional to $a/\lDinf$ and the advective contribution is proportional to $(a/\lDinf)^3$---with both terms proportional to $\sigma$ (noting that $\lGC \propto 1/\sigma$)---the advective solute flux can be ignored in the Debye--H\"uckel regime when $\lDinf \gg a$.

\subsubsection{Thin electric double layer}
\label{ssec:theo-DH-thinEDL}

\begin{figure}[!ht]
\centering
\includegraphics[scale = 1, trim={0.195cm 0.0875cm 0.195cm 0.0875cm},clip]{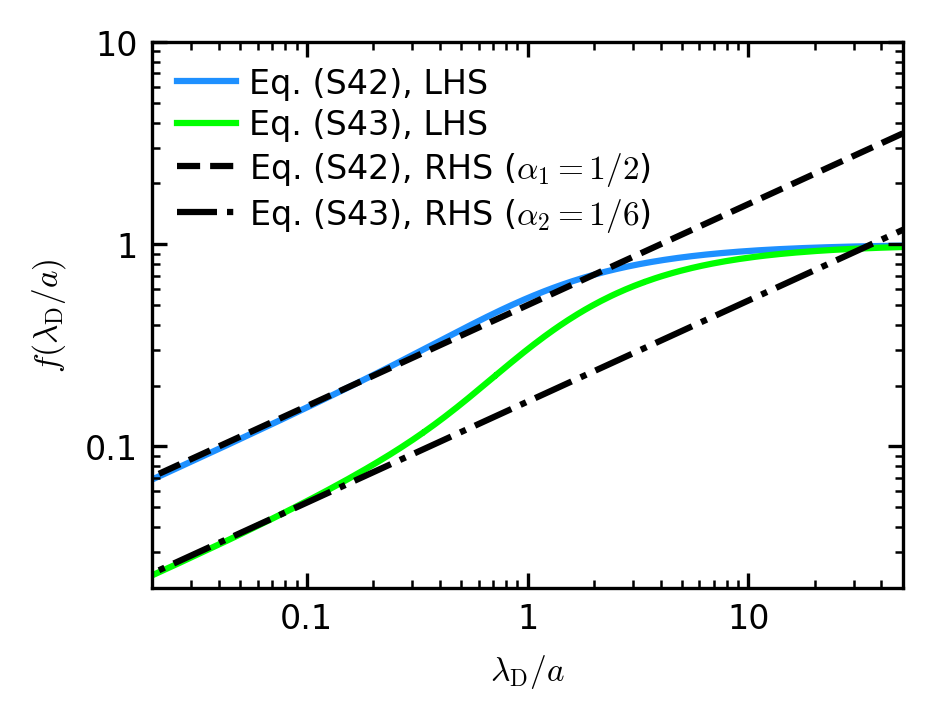}
\caption{\label{sfig:integrals-fits}Left-hand side (LHS) and right-hand side (RHS) of Eqs.~\eqref{seq:int_epot-DH-const} and \eqref{seq:int_epot-sq-DH-const} (i.e. $f(\lDinf/a)$) for $\alpha_1 = \frac{1}{2}$ and $\alpha_2 = \frac{1}{6}$, respectively, vs $\lDinf/a$.}
\end{figure}

Implementing the non-dimensionalization $\psi_0 = \frac{\sigma \lDinf}{\eps} \Tilde{\psi}_0 = 2 \sgn(\sigma)\frac{\kBT}{Ze}\frac{\lDinf}{\lGC} \Tilde{\psi}_0$ (such that $\tilde{\psi}_0 = 1$ at a planar surface), we can write 
\begin{equation}
    \label{seq:int_epot-DH-working}
     \int^1 _0 \mathrm{d} \zeta \, \frac{Ze \psi_0|_{\nu = 0}}{\kBT} = 2 \sgn(\sigma) \frac{\lDinf}{\lGC} \int^1 _0 \mathrm{d} \zeta \, \tilde{\psi}_0|_{\nu = 0} 
\end{equation}
and
\begin{equation}
    \label{seq:int_epot-sq-DH-working}
   \int^1 _0 \mathrm{d} \zeta \, {\left(\frac{Ze \psi_0|_{\nu = 0} }{\kBT} \right)}^2 = 4\left( \frac{\lDinf}{\lGC} \right)^2 \int^1 _0 \mathrm{d} \zeta \, \left(\tilde{\psi}_0|_{\nu = 0}\right)^2 ,
\end{equation}
where 
\begin{equation}
    \label{seq:epot-DH-dimensionless}
      \Tilde{\psi}_0 =  \frac{a}{\lDinf} \left[ \int ^{\mathrm{\infty}} _0 \mathrm{d} s \, \frac{J_1(s) J_0(\hat{r} s)}{\sqrt{(a/\lDinf)^2 + s^2}} e^{-\hat{z}\sqrt{(a/\lDinf)^2 + s^2}} + \frac{e^{-(a/\lDinf) \hat{z}}}{(a/\lDinf)} \right]
\end{equation}
is the dimensionless equilibrium electric potential near an ultrathin membrane in the Debye--H\"uckel regime. Using an analogous result in Ref.~\citenum{baldockCDF2025} for a different non-dimensionalization of $\psi_0$, it can be shown that 
\begin{equation}
    \label{seq:int_epot-DH-scaling}
     \int^1 _0 \mathrm{d} \zeta \, \tilde{\psi}_0|_{\nu = 0} \propto \left(\frac{\lDinf}{a} \right)^{\frac{1}{2}}
\end{equation}
and
\begin{equation}
    \label{seq:int_epot-sq-DH-scaling}
     \int^1 _0 \mathrm{d} \zeta \, \left( \tilde{\psi}_0|_{\nu = 0} \right)^2 \propto \left(\frac{\lDinf}{a} \right)^{\frac{1}{2}}
\end{equation}
for a thin electric double layer ($\lDinf \ll a$). Thus,
\begin{equation}
    \label{seq:int_epot-DH-const}
     \int^1 _0 \mathrm{d} \zeta \, \tilde{\psi}_0|_{\nu = 0} = \alpha_1 \left(\frac{\lDinf}{a} \right)^{\frac{1}{2}}
\end{equation}
and
\begin{equation}
    \label{seq:int_epot-sq-DH-const}
     \int^1 _0 \mathrm{d} \zeta \, \left( \tilde{\psi}_0|_{\nu = 0} \right)^2 = \alpha_2 \left(\frac{\lDinf}{a} \right)^{\frac{1}{2}},
\end{equation}
where $\alpha_1$ and $\alpha_2$ are constants, which we have determined by numerically calculating the integrals on the left-hand sides of Eqs.~\eqref{seq:int_epot-DH-const} and \eqref{seq:int_epot-sq-DH-const} for various values of $\lDinf/a$ and fitting them to their respective right-hand sides for $\lDinf \ll a$. For $\lDinf/a \leq 0.1$, we obtained $\alpha_1 = 0.488$ and $\alpha_2 = 0.165$, which we have simplified to $\alpha_1 \approx \frac{1}{2}$ and $\alpha_2 \approx \frac{1}{6}$. Figure~\ref{sfig:integrals-fits} compares the left- and right-hand sides of Eqs.~\eqref{seq:int_epot-DH-const} and \eqref{seq:int_epot-sq-DH-const} for $\alpha_1 = \frac{1}{2}$ and $\alpha_2 = \frac{1}{6}$, respectively. Substituting Eq.~\eqref{seq:int_epot-sq-DH-const} into Eq.~\eqref{seq:int_epot-sq-DH-working} and letting $\alpha_2 = \frac{1}{6}$ gives 
\begin{equation}
\label{seq:int_epot-sq-DH}
    \int^1 _0 \mathrm{d}\zeta \, \left( \frac{Ze \psi_0|_{\nu=0}}{k_{\mathrm{B}} T} \right) ^2 \approx \frac{2}{3} \left(\frac{\lDinf}{a} \right)^{\frac{1}{2}} \left(\frac{\lDinf}{\lGC} \right)^2 ,
\end{equation}
which is analogous to Eq.~\eqref{eq:int_epot-sq-DH} in the main paper. Similarly, substituting Eq.~\eqref{seq:int_epot-DH-const} into Eq.~\eqref{seq:int_epot-DH-working} and letting $\alpha_1 = \frac{1}{2}$ gives 
\begin{equation}
\label{seq:int_epot-DH-try}
    \int^1 _0 \mathrm{d}\zeta \, \frac{Ze \psi_0|_{\nu=0}}{k_{\mathrm{B}} T} \approx \mathrm{sgn} (\sigma) \left(\frac{\lDinf}{a} \right)^{\frac{1}{2}} \frac{\lDinf}{\lGC} .
\end{equation}
Note that Eqs.~\eqref{seq:int_epot-sq-DH} and \eqref{seq:int_epot-DH-try} with Eq.~\eqref{seq:elec_curr-DH} indicate that the electric-field-driven electric current and solute flux are both proportional to $a^{\frac{1}{2}}$ for $\lDinf \ll a$.

From Onsager's reciprocal relations,\cite{1931onsager} in the linear-response regime the electric current driven by a concentration difference applied across the membrane, $I^\mathrm{CDF}$, will obey the same scaling relationships as the solute flux driven by an electric potential difference,  $J^\mathrm{EDF}$. An identical expression to Eq.~\eqref{seq:int_epot-DH-scaling}  was used in Ref.~\citenum{baldockCDF2025} to derive the scaling $I_{\mathrm{CDF}} \propto a^{\frac{1}{2}}$, but fitting to simulation data in Ref.~\citenum{baldockCDF2025} yielded $I_{\mathrm{CDF}} \propto a^{\frac{1}{4}}$. As the concentration-gradient-driven electric current density is governed by the difference between cation and anion concentrations (when $D_+ \approx D_-$), it was suggested in Ref.~\citenum{baldockCDF2025} that small discrepancies in the predicted ion concentrations could significantly affect the electric current. Given that $\varepsilon c_{\mathrm{s}_1}^+ = - \varepsilon c_{\mathrm{s}_1}^-$ for electric-field-driven electrolyte transport---where $c^{(i)} = c_{\mathrm{s}}^{(i)} \exp \left(-\frac{Z_i e \psi}{\kBT} \right) \appropto \frac{Z_i e \Delta \psi}{\kBT} \cinf^{(i)} \exp \left(-\frac{Z_i e \psi}{\kBT} \right)$ in the theory and $Z_+ = Z = -Z_-$ for a $Z$:$Z$ electrolyte---the electric-field-driven solute flux density is governed by the difference between the cation and anion concentrations. Thus, the electric-field-driven solute flux may also be affected by small discrepancies in the predicted ion concentrations. Similarly to the case of the concentration-gradient-driven electric current in Ref.~\citenum{baldockCDF2025}, a discrepancy between the scaling relationships of the electric-field-driven solute flux in the theory and simulations could be expected. Consequently, we have checked whether the scaling relationships in Eqs.~\eqref{seq:int_epot-DH-scaling} and \eqref{seq:int_epot-DH-const} accurately capture those in the simulations before continuing with this derivation.

\begin{figure}[!ht]
    \centering
    \includegraphics{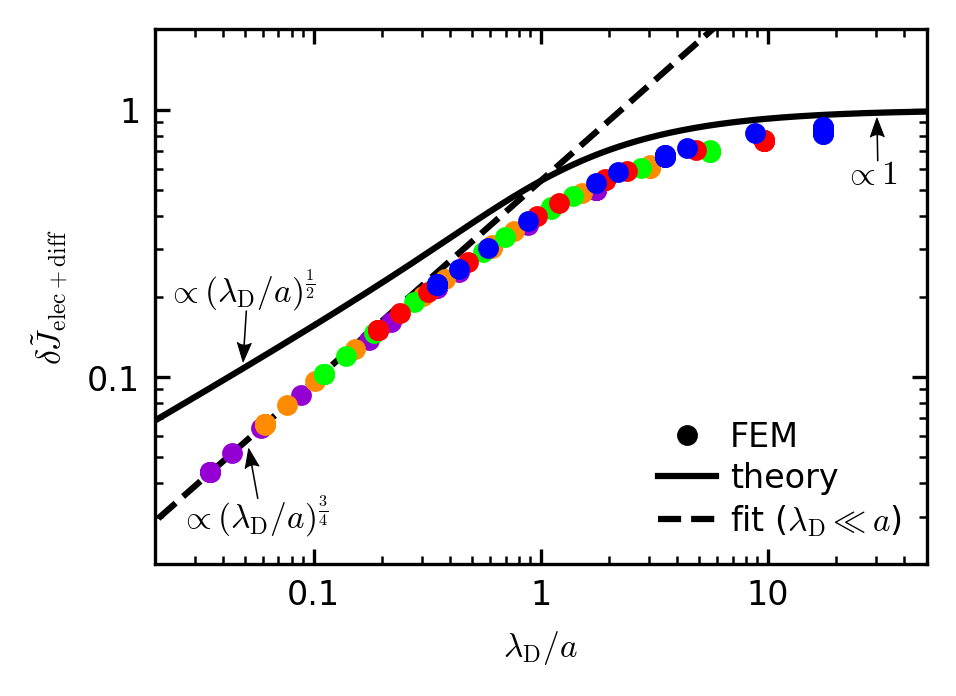}
    \caption{Dimensionless surface contribution to the electro-diffusive solute $\delta \tilde{J}_{\mathrm{elec+diff}}$ from FEM simulations for surface potential energies that are $< \kBT$ far from the pore mouth and bulk electrolyte concentration $\cinf = 0.3$ (blue), $1$ (red), $3$ (green), $10$ (orange) and $30$ (purple) mol~m$^{-3}$. For each bulk electrolyte concentration, surface charge densities where the surface potential energies are $< \kBT$ far from the pore mouth are shown. The solid line was Eq.~\eqref{seq:deltaJ-tilde} calculated numerically using Eq.~\eqref{seq:epot-DH-dimensionless}, while the dashed line is a linear fit of $\delta \tilde{J}_{\mathrm{elec+diff}}$ from simulations where $\lDinf/a \leq 0.1$ to $(\lDinf/a)^{\frac{3}{4}}$ ($\mathrm{coefficient} \approx 0.542$). Simulations with surface potential energies that are $< 0.1 \kBT$ far from the pore mouth have been excluded.}
    \label{sfig:deltaJ-tilde_universal}
\end{figure}

Firstly, Eq.~\eqref{seq:sol_flux-DH-exp} shows that the electro-diffusive and advective solute fluxes are separately conserved in the Debye--H\"uckel regime, so we take
\begin{equation}
    \label{seq:elecdiff_sol_flux-DH} 
    J_{\mathrm{elec + diff}} = \frac{a \eps (\Dp) \Delta \psi}{Ze(\lDinf)^2} \left[ {\int ^1 _0 \mathrm{d} \zeta \, \frac{Ze \psi_0|_{\nu=0}}{\kBT}} - \left(\frac{\Dm}{\Dp} \right) \right] 
\end{equation}
as the electro-diffusive solute flux. While the low-Péclet-number assumption may not always be accurate for the electric-field-driven solute flux, we can separate the electro-diffusive and advective solute fluxes in the Debye--H\"uckel regime. The bulk contribution to the solute flux (i.e. the solute flux at no surface charge) is $J^{(0)} = -\frac{a \eps (\Dm) \Delta \psi}{Ze (\lDinf)^2}$, such that the surface contribution to the solute flux can be given as $\delta J = J - J^{(0)}$. Subjecting the equilibrium electric potential in Eq.~\eqref{seq:elecdiff_sol_flux-DH} to the non-dimensionalization $\psi_0 = \frac{\sigma \lDinf}{\eps}\tilde{\psi}_0$, we can write the surface contribution to the electro-diffusive solute flux as
\begin{equation}
    \label{seq:elecdiff_sol_flux-surf-DH}  
    \delta J_{\mathrm{elec + diff}} = \frac{a \sigma (\Dp) \Delta \psi}{\kBT \lDinf} {\int ^1 _0 \mathrm{d} \zeta \, \tilde{\psi}_0}|_{\nu=0}.
\end{equation}
We subject the surface contribution to the electro-diffusive solute flux (in the Debye--H\"uckel regime) in Eq.~\eqref{seq:elecdiff_sol_flux-surf-DH} to the non-dimensionalization
\begin{equation}
    \label{seq:deltaJ-DH-dimensionless} 
    \delta J_{\mathrm{elec + diff}} = \frac{a \sigma (\Dp) \Delta \psi}{\kBT \lDinf} \delta \tilde{J}_{\mathrm{elec+diff}},
\end{equation}
where 
\begin{equation}
    \label{seq:deltaJ-tilde}
     \delta \tilde{J}_{\mathrm{elec+diff}} = {\int ^1 _0 \mathrm{d} \zeta \, \tilde{\psi}_0}|_{\nu=0}
\end{equation}
in the theory. Note that $\tilde{\psi}_0 \approx 1$ yields $\delta \tilde{J}_{\mathrm{elec+diff}} \approx 1$ for a thick electric double layer ($\lDinf \gg a$). We see that Eq.~\eqref{seq:deltaJ-tilde} is similar to the dimensionless surface contribution to the concentration-gradient-driven electric current used in Ref.~\citenum{baldockCDF2025}, although a different non-dimensionalization of $\psi_0$ is used here. We obtained $\delta J_{\mathrm{elec + diff}}$ from the simulations (more details on  the finite-element method (FEM) simulations are given in Sec.~\ref{sec:results} of the main paper and Sec.~\ref{ssec:FEM}) by subtracting the analytical expression for the bulk contribution to the solute flux from $J_{\mathrm{elec + diff}}$ calculated from the simulations, then compared these results with the analytical expression in Eq.~\eqref{seq:elecdiff_sol_flux-surf-DH} (or, equivalently, Eq.~\eqref{seq:deltaJ-tilde}). Figure~\ref{sfig:deltaJ-tilde_universal} shows that, for simulations where $Ze |\psi_{0}| < \kBT$ far from the pore mouth, the dimensionless surface contribution to the electro-diffusive solute flux collapses on to a single universal curve as a function of $\lDinf/a$. 

A linear fit of $\log(\delta \tilde{J}_{\mathrm{elec+diff}})$ from simulations where $\lDinf/a \leq 0.1$ to $\log(\lDinf/a)$ indicates that $\delta \tilde{J}_{\mathrm{elec+diff}} \propto (\lDinf/a)^{\frac{3}{4}}$, while a linear fit of $\delta \tilde{J}_{\mathrm{elec+diff}} \propto (\lDinf/a)^{\frac{3}{4}}$ from simulations where $\lDinf/a \leq 0.1$ to $(\lDinf/a)^{\frac{3}{4}}$ gave a coefficient of $0.542$ (i.e. $\delta \tilde{J}_{\mathrm{elec+diff}} \approx 0.542 \times (\lDinf/a)^{\frac{3}{4}}$ when $\lDinf \ll a$).  Such a result indicates that, in the simulations,
\begin{align}
    \label{seq:delJ_tilde-DH-thinEDL} 
     \delta \tilde{J}_{\mathrm{elec+diff}} \approx \frac{1}{2} \left(\frac{\lDinf}{a} \right)^{\frac{3}{4}},
\end{align}
where we have assigned the prefactor $0.542 \approx 0.5$ for simplicity. Re-dimensionalizing $\delta {J}_{\mathrm{elec+diff}}$ and adding the bulk contribution $J^{(0)}$ gives, when $\lDinf \ll a$,
\begin{equation}
\label{seq:elecdiff_sol_flux-DH-thinEDL}
     J_{\mathrm{elec + diff}} \approx \frac{\eps (\Dp) \Delta \psi}{Ze \lDinf} \left[\sgn(\sigma) \left(\frac{a}{\lDinf} \right)^{\frac{1}{4}} \frac{\lDinf}{\lGC} -  \frac{a}{\lDinf} \left(\frac{\Dm}{\Dp} \right)\right].
\end{equation}
Substituting the scaling relationships of the electroosmotic flow rate previously derived in Ref.~\citenum{baldockCDF2025} into the final term in Eq.~\eqref{seq:sol_flux-DH-sub} gives 
\begin{equation}
    \label{seq:adv_sol_flux-DH-thinEDL} 
    J_{\mathrm{adv}} \sim \frac{a \sigma \kBT \eps \Delta \psi}{(Ze)^2 \eta \lDinf},
\end{equation}
as the advective solute flux when $\lDinf \ll a$ (in the Debye--H\"uckel regime). Note that $J = {J}_{\mathrm{elec+diff}} + J_{\mathrm{adv}}$ gives the full expression for the electric-field-driven solute flux. As the first term in Eq.~\eqref{seq:elecdiff_sol_flux-DH-thinEDL} (i.e. $\delta {J}_{\mathrm{elec+diff}}$) is proportional to $(a/\lDinf)^{\frac{1}{4}}$ and the advective contribution is proportional to $a/\lDinf$---with both terms proportional to $\sigma$ (note that $\lGC \propto 1/\sigma$)---the advective solute flux is significant in the Debye--H\"uckel regime when $\lDinf \ll a$. 

Substituting Eq.~\eqref{seq:deltaJ-tilde} (i.e. the analytical expression for $\delta \tilde{J}_{\mathrm{elec+diff}}$) into Eq.~\eqref{seq:elec_curr-DH} gives
\begin{align}
    \label{seq:elec_curr-DH-sub} 
    I = -\frac{a \eps (\Dp) \Delta \psi}{(\lDinf)^2} \left[ 1 + \frac{1}{2}\int ^1 _0 \mathrm{d} \zeta \left(\frac{Ze \psi_0|_{\nu=0}}{\kBT}\right)^2 - 2\mathrm{sgn}(\sigma) \frac{\lDinf}{\lGC}\left( \frac{\Dm}{\Dp} \right) \delta \tilde{J}_{\mathrm{elec}+\mathrm{diff}} \right].
\end{align}
As the term in 
the electric-field-driven electric current density (see Eq.~\eqref{seq:elec_curr_dens-rearr})
with the prefactor $(\Dm)$ is also governed by the difference between cation and anion concentrations, we assume that Eq.~\eqref{seq:delJ_tilde-DH-thinEDL} accurately describes the scaling of this term with the pore radius and Debye length in the Debye--H\"uckel regime when $\lDinf \ll a$ (rather than the theory). Substituting Eqs.~\eqref{seq:int_epot-sq-DH} and \eqref{seq:delJ_tilde-DH-thinEDL} into Eq.~\eqref{seq:elec_curr-DH-sub} gives
\begin{align}
    I &\approx -\frac{\eps (\Dp) \Delta \psi}{\lDinf} \left[\frac{a}{\lDinf} -\sgn(\sigma) \left(\frac{a}{\lDinf} \right)^{\frac{1}{4}} \left(\frac{\Dm}{\Dp} \right) \frac{\lDinf}{\lGC} + \frac{1}{3} \left(\frac{a}{\lDinf} \right)^{\frac{1}{2}} \left( \frac{\lDinf}{\lGC} \right)^2 \right]  
    \label{seq:elec_curr-DH-thinEDL}
\end{align}
for $\lDinf \ll a$. Assuming that $|\Dm|/(\Dp) \ll a^{\frac{3}{4}} \lGC/(\lDinf)^{\frac{7}{4}} + a^{\frac{1}{4}} (\lDinf)^{\frac{3}{4}} /(3\lGC)$, we can ignore the contribution due to the difference in ion diffusitivies such that
\begin{equation}
    I \approx -\frac{\eps (\Dp) \Delta \psi}{\lDinf} \left[\frac{a}{\lDinf} + \frac{1}{3} \left(\frac{a}{\lDinf} \right)^{\frac{1}{2}} \left(\frac{\lDinf}{\lGC} \right)^2 \right]
    \label{seq:elec_curr-DH-thinEDL-reduc}
\end{equation}
for $\lDinf \ll a$, which is the same as Eq.~\eqref{eq:elec_curr-DH-thinEDL} in the main paper. 

Although deriving an analytical expression for the advective electric current in the Debye--H\"uckel regime is less straightforward than for the advective solute flux, we have verified that the advective contribution to the electric current is negligibly small inside the pore in the FEM simulations (see Sec.~\ref{ssec:FEM}). We have also derived scaling laws for the advective electric current---which is conserved in the Debye--H\"uckel regime---for a thick electric double layer, and fit scaling laws to the simulation data for a thin electric double layer, in Sec.~\ref{ssec:adv_elec_curr}. Using these scaling laws, we can predict that the advective electric current is negligibly small in the Debye--H\"uckel regime (see  Sec.~\ref{ssec:adv_elec_curr}).


\section{Derivation of electric current for a thin electric double layer (outside Debye--H\"uckel regime)}
\label{ssec:thinEDL-gen}

For a system that can be described using Eq.~\eqref{seq:poisson} and Eq.~\eqref{eq:ion_conservation-lin} of the main paper, we first consider electric-field-driven flow through a planar channel with channel length $L_z$ (in the $z$ direction), width $2 L_y$ (in the $y$ direction) and depth $L_x$ (in the $x$ direction). We assume that $L_z \gg 2 L_y \gg L_x$ to ignore entrance effects and the effect of the channel boundaries in the $x$ direction. In this case, the applied (and induced) gradients and flow velocity are tangential to the surface, such that $\varepsilon c_{\mathrm{s}_1}^{(i)} = \varepsilon c_{\mathrm{s}_1}^{(i)} (z)$, $\varepsilon \psi_1 = \varepsilon \psi_1 (z)$ and $\varepsilon u_{y_1} \approx 0$. On the other hand, the equilibrium electric potential decreases exponentially with distance from the surface, such that $\psi_0 = \psi_0(y)$. Thus, the solution to Eq.~\eqref{eq:ion_conservation-lin} in the main paper for this geometry, subject to the boundary conditions $\varepsilon \psi_1 = 0$ at $z=0$ and $\varepsilon \psi_1 = \Delta \psi$ at $z = L_z$ (i.e. $\varepsilon c_{\mathrm{s}_1}^{(i)} = 0$ at $z=0$ and $\varepsilon c_{\mathrm{s}_1}^{(i)} \rightarrow \cinf^{(i)} Z_i e \Delta \psi/(\kBT)$ at $z = L_z$), is
\begin{equation}
    \varepsilon c_{\mathrm{s}_1}^{(i)} = \cinf^{(i)} \frac{Z_i e}{\kBT} \frac{\Delta \psi}{L_z} z.
\end{equation}
Thus, the $y$-component of the ion flux density is zero and its $z$-component is
\begin{equation}
    j_z^{(i)} = \cinf^{(i)} \exp \left(-\frac{Z_i e \psi_0}{\kBT} \right) \left( \varepsilon u_{z_1} - D_i \frac{Z_i e}{\kBT} \frac{\Delta \psi}{L_z} \right),
\end{equation}
such that the $z$-component of the electric current density of a $Z$:$Z$ electrolyte is
\begin{align}
    \label{seq:elec_curr_dens-DH}
    j_{\mathrm{e}_z} = &-\frac{\eps}{2(\lDinf)^2} \frac{\Delta \psi}{L_z} (\Dp) \cosh \left(\frac{Ze \psi_0}{\kBT} \right) \notag \\ 
    &+  \left[\frac{\eps}{2(\lDinf)^2} \frac{\Delta \psi}{L_z} (\Dm) - 2 Ze \cinf \varepsilon u_{z_1} \right] \sinh \left(\frac{Ze \psi_0}{\kBT} \right).
\end{align}
Taking the channel walls to be at $y = 0$ and $2L_y$ and exploiting the symmetry around $y= L_y$, the electric current through the channel is
\begin{equation}
    \label{seq:elec_curr-int-plane}
    I = 2 L_x \int^{L_y} _0 \mathrm{d} y \, j_{\mathrm{e}_z},
\end{equation}
where substituting Eq.~\eqref{seq:elec_curr_dens-DH} into Eq.~\eqref{seq:elec_curr-int-plane} and ignoring advection gives
\begin{equation}
 \label{seq:elec_curr-plane}
    I = -\frac{\eps (\Dp) \Delta \psi}{(\lDinf)^2} \frac{L_x}{L_z}  \int^{L_y} _0 \mathrm{d} y \, \left[\cosh \left(\frac{Ze \psi_0}{\kBT} \right) - \left(\frac{\Dm}{\Dp} \right) \sinh\left(\frac{Ze \psi_0}{\kBT} \right) \right].
\end{equation}
For a non-overlapping electric double layer, the equilibrium electric potential in the channel is equivalent to that near an unbounded, charged plane at $y=0$, i.e.\cite{Herrero2024BoltzmannFormulae}
\begin{equation}
    \label{seq:epot-plane}
    \psi_0 = \frac{4 \kBT}{Ze} \sgn(\sigma) \tanh^{-1} \left[\gamma \exp \left(-\frac{y}{\lDinf} \right) \right],
\end{equation}
where
\begin{align}
\label{seq:gamma-plane}
    \gamma = \tanh \left[\frac{1}{2} \sinh^{-1} \left(\frac{\lDinf}{\lGC} \right) \right] =  -\frac{\lGC}{\lDinf} + \sqrt{\left(\frac{\lGC}{\lDinf}\right)^2+1}\,.
\end{align}
Substituting Eqs.~\eqref{seq:epot-plane} and \eqref{seq:gamma-plane} into Eq.~\eqref{seq:elec_curr-plane}, then evaluating the integral for $L_y/\lDinf \rightarrow \infty$ (i.e. non-overlapping electric double layer) gives
    \begin{equation}
    I \approx -\frac{L_x}{L_z} \left[2 L_y \kappa_{\mathrm{b}} + \kappa_{\mathrm{s}}^{\infty} - \frac{Ze \sigma}{2\kBT}(\Dm) \right]\Delta \psi ,
        \label{seq:elec_curr-plane-thinEDL}
\end{equation}
where 
\begin{align}
    \label{seq:bulk_conduct}
    \kappa_{\mathrm{b}} = \frac{\eps (\Dp)}{2(\lDinf)^2}
\end{align}
is the bulk conductivity of a $Z$:$Z$ electrolyte and 
\begin{align}
    \label{seq:surf_conduct-prelim}
    \kappa_\mathrm{s} ^{\infty} & = \frac{Ze(\Dp) |\sigma|}{2 \kBT} \gamma \\
    \label{seq:surf_conduct-plane}   
    & = \frac{Ze(\Dp) |\sigma|}{2 \kBT} \left[-\frac{\lGC}{\lDinf} + \sqrt{\left( \frac{\lGC}{\lDinf} \right)^2 + 1} \right]
\end{align}
is the surface conductivity (near a planar wall) of a $Z$:$Z$ electrolyte where $D_+ \approx D_-$ 
that appears in Ref.~\citenum{leeLargeElectricSizeSurfaceCondudction2012} and is analogous to Eq.~\eqref{eq:surf_conduct-plane} in the main paper. Ignoring the difference in ion diffusivities, Eq.~\eqref{seq:elec_curr-plane-thinEDL} reduces to
\begin{align}
     I \approx -\frac{L_x}{L_z} \left (2 L_y \kappa_{\mathrm{b}} + \kappa_{\mathrm{s}}^{\infty} \right) \Delta \psi .
\end{align}
The surface contribution to the electric current through a planar channel (i.e. $\delta I = I - I^{(0)}$, where $I^{(0)} = -2 \Delta \psi L_x L_y \kappa_{\mathrm{b}} /{L_z}$) for a non-overlapping electric double layer is 
\begin{align}
     \delta I \approx -\frac{L_x}{L_z} \kappa_{\mathrm{s}}^{\infty} \Delta \psi ,
\end{align}
which is just the product of the applied potential difference, the surface conductivity (near a planar wall) and a dimensionless prefactor that arises from the aspect ratio of the channel (i.e. ${L_x}/{L_z}$). Similarly, for the case of electric-field-driven electrolyte transport through a long cylindrical pore of length $L$ and radius $a$ where $L \gg 2a$, the electric current for a thin electric double layer and $D_+ \approx D_-$ is \cite{leeLargeElectricSizeSurfaceCondudction2012}
\begin{equation}
    I \approx -\frac{\pi a}{L} \left(a \kappa_{\mathrm{b}} + 2\kappa_{\mathrm{s}}^{\infty} \right) \Delta \psi,
    \label{seq:elec_curr-cyl-thinEDL}
\end{equation}
where 
\begin{equation}
    \delta I \approx -\frac{2\pi a}{L} \kappa_{\mathrm{s}}^{\infty} \Delta \psi
    \label{seq:elec_curr-cyl-thinEDL-surf}
\end{equation}
is the surface contribution to the electric current and $2a/L$ is the aspect ratio of the pore. As previously highlighted in Ref.~\citenum{leeLargeElectricSizeSurfaceCondudction2012}, the electric current depends on the aspect ratio of the pore rather than explicitly on its diameter when the bulk contribution is negligibly small.

For electric-field-driven electrolyte transport through a circular aperture of radius $a$ in an infinitesimally thin planar membrane (i.e. $L/a \rightarrow 0$), it is intuitive to assume that $\delta I$ does not depend on the pore radius as $\lDinf/a \rightarrow 0$, such that $\delta I \sim \kappa_{\mathrm{s}}^{\infty} \Delta \psi$. Such a result appears in the theory in Ref.~\citenum{leeLargeElectricSizeSurfaceCondudction2012} but differs to the scaling relationships in Eq.~\eqref{seq:elec_curr-DH-thinEDL-reduc}, which predict that $\delta I \propto a^{\frac{1}{2}}$ for a thin electric double layer in the Debye--H\"uckel regime. The unusual prefactor $(a /\lDinf)^{\frac{1}{2}}$ in Eq.~\eqref{seq:elec_curr-DH-thinEDL-reduc}, which is not seen in thicker membranes, arises from the geometrical interplay between the variables that describe the properties of the membrane and electrolyte solution  as the ions enter and exit the bulk. This phenomenon is most significant near the membrane surface and thus most relevant to a thin electric double layer. Note that the Debye length $\lDinf$ and the Gouy--Chapman length $\lGC$ are length scales that describe the properties of an electrolyte solution and a charged (planar) surface, respectively, while the pore radius controls the geometry of an ultrathin membrane. 

Substituting Eq.~\eqref{seq:bulk_conduct} into Eq.~\eqref{seq:elec_curr-DH-thinEDL-reduc}, the electric current in an ultrathin membrane for $Ze |\psi_0| \ll \kBT$ and $\lDinf \ll a$ can be written as 
\begin{equation}
    I \approx -\left[ 2a \kappa_{\mathrm{b}} + \frac{2}{3}\left(\frac{a}{\lDinf} \right)^{\frac{1}{2}} \frac{\eps (\Dp) \lDinf}{2 (\lGC)^2} \right] \Delta \psi .
   \label{seq:elec_curr-DH-rearr}
\end{equation}
When $\lGC \gg \lDinf$, which encompasses the Debye--H\"uckel regime, Eq.~\eqref{seq:gamma-plane} reduces to
\begin{equation}
\label{seq:gamma-plane-DH}
    \gamma \approx \frac{\lDinf}{2 \lGC}.
\end{equation}
Substituting Eq.~\eqref{seq:gamma-plane-DH} into Eq.~\eqref{seq:surf_conduct-prelim} gives
\begin{equation}
\label{seq:surf_conduct-plane-DH}
    \kappa_{\mathrm{s}}^{\infty} \approx \frac{Ze(\Dp) |\sigma|}{4 \kBT} \frac{\lDinf}{\lGC} = \frac{\eps (\Dp)\lDinf}{2 (\lGC)^2}
\end{equation}
as the surface conductivity near a planar wall for a $Z$:$Z$ and $D_+ \approx D_-$ electrolyte in the Debye--H\"uckel regime. Equation~\eqref{seq:surf_conduct-plane-DH} is analogous to Eq.~\eqref{eq:surf_conduct-plane-DH} in the main paper. Similarly to the approach in Ref.~\citenum{baldockCDF2025} for extending the theory of concentration-gradient-driven electrolyte transport through a two-dimensional membrane to outside the Debye--H\"uckel regime, we could substitute the factor $\eps (\Dp) \lDinf /(2(\lGC)^{2})$ in Eq.~\eqref{seq:elec_curr-DH-rearr} with $\kappa_{\mathrm{s}}^{\infty}$ in Eq.~\eqref{seq:surf_conduct-plane} to extend the theory in Sec.~\ref{ssec:theo-DH} to outside the Debye--H\"uckel regime. In this case, $2 (a/\lDinf)^{\frac{1}{2}}/3$ plays an analogous role to the aspect ratio of the pore in the theory for a planar channel. However, the assumption that $\delta I$ scales with a product of $(a/\lDinf)^{\frac{1}{2}}$ and a function of $\lDinf/\lGC$ that describes the properties of a planar surface (i.e. $\kappa_{\mathrm{s}}^{\infty}$) may be too simplistic, and could break down at very large magnitudes of the surface potential energy.

Substituting Eq.~\eqref{seq:surf_conduct-plane-DH} into Eq.~\eqref{seq:elec_curr-plane-thinEDL}, we can write the electric current through a planar channel for $Ze |\psi_0| \ll \kBT$ (Debye--H\"uckel regime) and a non-overlapping electric double layer, where $D_+ \approx D_-$, as
\begin{equation}
    I \approx -\frac{L_x}{L_z} \left[2 L_y \kappa_{\mathrm{b}} + \frac{\eps (\Dp)}{2 \lGC} \times \frac{\lDinf}{\lGC} \right] \Delta \psi.
   \label{seq:elec_curr-plane-DH}
\end{equation}
Next, we write the electric current in an ultrathin membrane for $Ze |\psi_0| \ll \kBT$ and $\lDinf \ll a$ (see Eqs.~\eqref{seq:elec_curr-DH-thinEDL-reduc} and \eqref{seq:elec_curr-DH-rearr}) as
\begin{equation}
    I \approx -\left[ 2a \kappa_{\mathrm{b}} + \frac{\eps (\Dp)}{2 \lGC} \times \frac{2 (a \lDinf)^{\frac{1}{2}}}{3 \lGC} \right] \Delta \psi . 
    \label{seq:elec_curr-DH-rearr-rep}
\end{equation}
Noting that the first term (bulk contribution) inside the square brackets in Eq.~\eqref{seq:elec_curr-DH-rearr-rep} is identical to that of Eq.~\eqref{seq:elec_curr-plane-DH}, we see that Eq.~\eqref{seq:elec_curr-plane-DH} contains the prefactor ${L_x}/{L_z}$---which is the aspect ratio of the channel---while Eq.~\eqref{seq:elec_curr-DH-rearr-rep} contains no such prefactor. While the final term (surface contribution) in Eq.~\eqref{seq:elec_curr-DH-rearr-rep} and in Eq.~\eqref{seq:elec_curr-plane-DH} both contain the prefactor ${\eps (\Dp)}/{(2 \lGC)}$, the final term in Eq.~\eqref{seq:elec_curr-plane-DH} contains the factor ${\lDinf}/{\lGC}$ and that in Eq.~\eqref{seq:elec_curr-DH-rearr-rep} contains the factor $2(a \lDinf)^{\frac{1}{2}}/(3 \lGC)$. 
We assume that the electric-field-driven electric current through an ultrathin membrane takes the form
\begin{equation}
    I \approx - \left(2 a \kappa_{\mathrm{b}} + \kappa_{\mathrm{s}} \right) \Delta \psi.
   \label{seq:elec_curr-gen}
\end{equation}
where
\begin{equation}
    \kappa_{\mathrm{s}} \approx \frac{\eps (\Dp)}{2 \lGC} \times \frac{2 (a \lDinf)^{\frac{1}{2}}}{3 \lGC}
       \label{seq:surf_conduct-DH}
\end{equation}
is the surface conductance of an ultrathin membrane for a $Z$:$Z$ and $D_+ \approx D_-$ electrolyte in the Debye--H\"uckel and thin electric-double-layer regime. Note that Eq.~\eqref{seq:elec_curr-gen} is the same as Eq.~\eqref{eq:elec_curr-gen} in the main paper, while Eq.~\eqref{seq:surf_conduct-DH} is analogous to Eq.~\eqref{eq:surf_conduct-DH} in the main paper. While Eqs.~\eqref{seq:elec_curr-gen} and \eqref{eq:elec_curr-gen} are similar to the theory in Ref.~\citenum{leeLargeElectricSizeSurfaceCondudction2012}, we have derived an expression for the surface conductance of an ultrathin membrane for a thin electric double layer that depends on the pore radius, rather than using the surface conductivity near a planar wall. Next, we work on extending Eq.~\eqref{seq:surf_conduct-DH} to outside the Debye--H\"uckel regime.

Rearranging Eq.~\eqref{seq:surf_conduct-plane-DH} gives
\begin{equation}
\label{seq:surf_conduct-DH-exp}
    \kappa_{\mathrm{s}}^{\infty} \approx \frac{\eps (\Dp)}{2 \lDinf} \left(\frac{\lDinf}{\lGC} \right)^2
\end{equation}
near a planar wall, and rearranging Eq.~\eqref{seq:surf_conduct-DH} gives
\begin{equation}
    \kappa_{\mathrm{s}} \approx \frac{\eps (\Dp)}{2 \lDinf} \times  \left(\sqrt{\frac{2}{3}} \frac{a^{\frac{1}{4}}(\lDinf)^{\frac{3}{4}}}{\lGC} \right)^2 
       \label{seq:surf_conduct-plane-DH-exp}
\end{equation}
in an ultrathin membrane, where Eqs.~\eqref{seq:surf_conduct-DH-exp} and \eqref{seq:surf_conduct-plane-DH-exp} both apply for $Ze|\psi_0| \ll \kBT$ and $\lDinf \ll a$. In Eq.~\eqref{seq:elec_curr-plane-DH}, 
$\lDinf/\lGC$ describes the interactions of the ions (in solution) with a charged, planar surface (i.e. $\psi_{\infty} \propto \lDinf/\lGC$ when $Ze|\psi_0| \ll \kBT$). Upon comparing Eqs.~\eqref{seq:surf_conduct-DH-exp} and \eqref{seq:surf_conduct-plane-DH-exp}, we assume that $(2/3)^{\frac{1}{2}} a^{\frac{1}{4}}(\lDinf)^{\frac{3}{4}}/\lGC$ similarly describes the influence of the interactions of the ions with the surface of a charged, ultrathin membrane (for $\lDinf \ll a$) on the surface conductance. Such a result can be explained by the geometrical interplay between $a$, $\lDinf$ and $\lGC$ that arises from the non-uniformity of the electric field lines outside the pore, which is not significant in a long planar channel ($L_z \gg L_x$) nor a long cylindrical pore ($L \gg 2a$). Outside the Debye--H\"uckel regime, we infer that, for an ultrathin membrane, $(2/3)^{\frac{1}{2}} a^{\frac{1}{4}}(\lDinf)^{\frac{3}{4}}/\lGC$ plays an analogous role to $\lDinf/\lGC$ in the theory of the electric-field-driven electric current through a planar channel for a thin electric double layer (i.e. in $\kappa_{\mathrm{s}}^{\infty}$). Given that we can use Eq.~\eqref{seq:surf_conduct-plane} to write
\begin{equation}
    \kappa_{\mathrm{s}}^{\infty} = \frac{\eps (\Dp)}{\lDinf} \times \frac{\lDinf}{\lGC} \left[ -\frac{\lGC}{\lDinf} + \sqrt{ \left(\frac{\lGC}{\lDinf}\right)^2 + 1} \right]
           \label{seq:surf_conduct-plane-rearr}
\end{equation}
near a planar wall for a thin electric double layer, we substitute all instances of $\lDinf/\lGC$ in Eq.~\eqref{seq:surf_conduct-plane-rearr} with $(2/3)^{\frac{1}{2}}a^{\frac{1}{4}}(\lDinf)^{\frac{3}{4}}/ \lGC$ (equivalent to substituting $1/\lGC$ with $(2/3)^{\frac{1}{2}} (a/\lDinf)^{\frac{1}{4}}/\lGC$) to yield 
\begin{equation}
    \kappa_{\mathrm{s}} \approx \frac{\eps (\Dp)}{\lDinf} \times \sqrt{\frac{2}{3}} \frac{a^{\frac{1}{4}}(\lDinf)^{\frac{3}{4}}}{\lGC} \left[ -\sqrt{\frac{3}{2}} \frac{\lGC}{a^{\frac{1}{4}}(\lDinf)^{\frac{3}{4}}} + \sqrt{ \left(\sqrt{\frac{3}{2}} \frac{\lGC}{a^{\frac{1}{4}}(\lDinf)^{\frac{3}{4}}} \right)^2 + 1} \right] .
    \label{seq:surf_conduct-exp}
\end{equation}
Thus, we can rearrange Eq.~\eqref{seq:surf_conduct-exp} to write the surface conductance of an ultrathin membrane for a thin electric double layer and any magnitude of the electric potential as
\begin{equation}
    \kappa_{\mathrm{s}} \approx \frac{Ze(\Dp) |\sigma|}{2 \kBT} \left[-\frac{\lGC}{\lDinf} + \sqrt{\left( \frac{\lGC}{\lDinf} \right)^2 + \frac{2}{3} \left( \frac{a}{\lDinf}\right)^{\frac{1}{2}}} \right],
           \label{seq:surf_conduct}
\end{equation}
which is analogous to Eq.~\eqref{eq:surf_conduct} in the main paper. Note that we have given Eq.~\eqref{seq:surf_conduct} in the convention of Ref.~\citenum{leeLargeElectricSizeSurfaceCondudction2012} here for ease of comparison with their theory, which just entails writing $\eps/\lGC$ in Eq.~\eqref{seq:surf_conduct} as $Ze |\sigma|/(2 \kBT)$. We can rearrange Eq.~\eqref{seq:elec_curr-gen} (using Eq.~\eqref{seq:surf_conduct}) to write the magnitude of the surface charge density as
\begin{align}
    |\sigma| \approx \frac{\sqrt{6} \eps \kBT}{Ze} \left(\frac{\lDinf}{a} \right)^{\frac{1}{4}} \left[ \left( - \frac{I/\Delta \psi + 2a \kappa_{\mathrm{b}}}{\eps(\Dp)} + \frac{1}{\lDinf} \right)^2 - \left(\frac{1}{\lDinf} \right)^{2} \right]^{\frac{1}{2}},
    \label{seq:surface-charge}
\end{align}
where $I$ is the measured electric current of an ultrathin membrane in this context. Given that the ionic conductance is $G = -I/\Delta \psi$, Eq.~\eqref{seq:surface-charge} could be convenient for calculating the surface charge density of an ultrathin membrane from the measured ionic conductance.

\newpage

\section{Finite-element method simulations}
\label{ssec:FEM}

\begin{figure}[b]
    \centering
    \includegraphics[scale = 3]{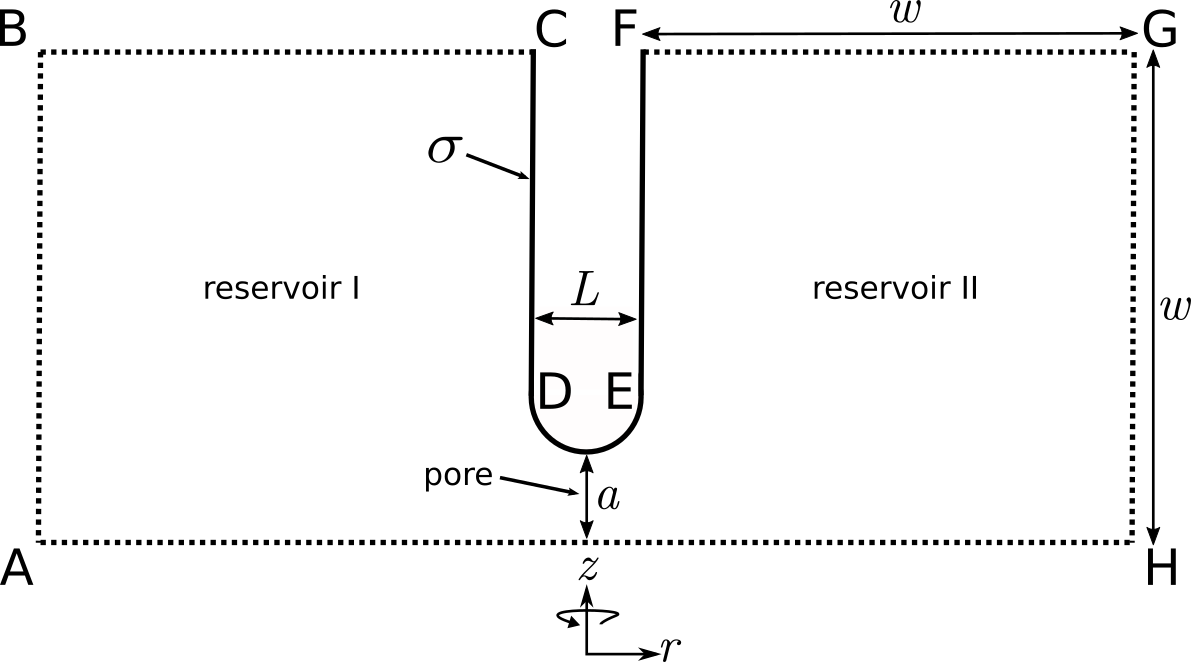}
    \caption[Computational domain]{Schematic of the two-dimensional axisymmetric computational domain used in the FEM simulations (not to scale). The geometry has rotational symmetry about the boundary AH, where the solid lines denote solid–liquid boundaries and the dashed lines denote liquid boundaries.}
    \label{sfig:schematic-FEM}
\end{figure}

The continuum hydrodynamic flow equations, given in Eqs.~\eqref{seq:poisson}--\eqref{seq:continuity}, were solved using finite-element method (FEM) simulations with version 4.3a of COMSOL Multiphysics\cite{comsol4.3a} for a thin planar membrane of thickness $L$ containing a circular aperture of radius $a$ connecting two large cylindrical fluid reservoirs (Fig.~\ref{sfig:schematic-FEM}). A fully coupled solver, which is a damped version of Newton's method, and the PARDISO direct solver were used to solve the equations, where the damping option used to achieve convergence was "Automatic highly nonlinear (Newton)". A thickness adjustment factor of $10$ was used for the boundary layer mesh at the solid--liquid boundaries. Table~\ref{stab:boundaries-FEM} lists the boundary conditions used to solve the equations. We verified that the measured flow rate, solute flux and electric current did not change significantly with a finer mesh, a larger reservoir size or higher discretization orders for the most extreme parameter combinations shown in Table~\ref{stab:boundaries-FEM} (discrepancy of $< 5\%$ for flow rate, solute flux and electric current at $\Delta \psi = 70$ mV). 


\begin{table}[!ht]
\centering
\caption[Boundary conditions for FEM simulations without slip]{\label{stab:boundaries-FEM}Boundary conditions used to solve the continuum hydrodynamic flow equations in the FEM simulations with no slip at the membrane surface, where $\boldsymbol{\hat{n}}$ is the unit normal to the surface and $i$ labels the ion type.}
\renewcommand{\arraystretch}{0.9}
\begin{tabular*}{.75\columnwidth}{@{\extracolsep{\fill}} l l}
\hline
boundary & conditions \\
\hline
AH      & $\boldsymbol{\hat{n}} \cdot \nabla c^{(i)} = \boldsymbol{\hat{n}} \cdot \boldsymbol{u} = \boldsymbol{\hat{n}} \cdot \nabla \boldsymbol{u} = 0$ \\
AB      & $c^{(i)} = c^{(i)}_{\mathrm{H}} = \cinf$, $p = p_{\mathrm{H}} = 0$, $\psi = \psi_{\mathrm{H}} = \Delta \psi$  \\
GH          & $c^{(i)} = c^{(i)}_{\mathrm{L}} = \cinf$, $p = p_{\mathrm{L}} = 0$, $\psi = \psi_{\mathrm{L}} = 0$ \\
BC and FG         &  $\boldsymbol{\hat{n}} \cdot \boldsymbol{j}_i = \boldsymbol{\hat{n}} \cdot \boldsymbol{u} = \boldsymbol{\hat{n}} \cdot \nabla \boldsymbol{u} = 0$ \\
CD, DE, and EF         & $\boldsymbol{\hat{n}} \cdot \boldsymbol{j}_i = \boldsymbol{u} = 0$ \\ 
\hline
\end{tabular*}
\end{table}

\begin{figure}[!ht]
 \centering
    \includegraphics[scale = 1, trim={0.175cm 0.0875cm 0.175cm 0.0875cm},clip]{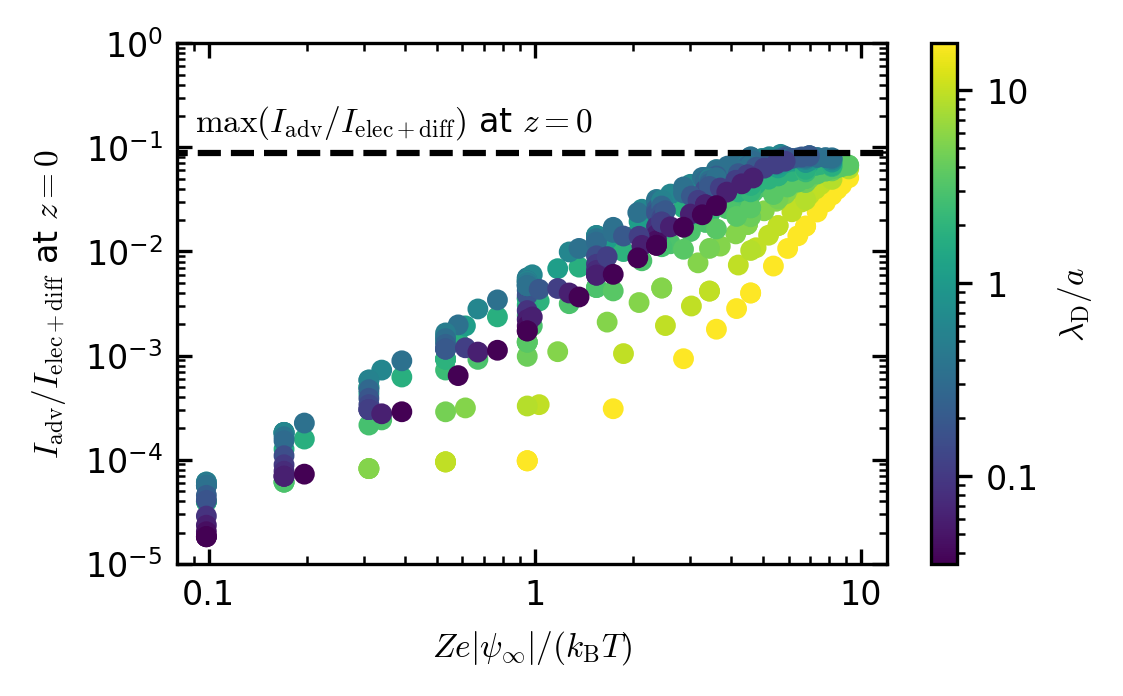}
    \caption{\label{sfig:Iadv-contr}Advective electric current, $I_{\mathrm{adv}}$, over the electro-diffusive electric current, $I_{\mathrm{elec+diff}}$, at $z = 0$ vs the magnitude of the equilibrium potential energy at a planar surface, $Ze|\psi_{\infty}|$, relative to the thermal energy, $\kBT$ (symbols). The color map indicates variations in the Debye length $\lDinf$ over the pore radius $a$. The dashed line indicates the maximum $I_{\mathrm{adv}}/I_{\mathrm{elec+diff}}$ at $z=0$.}
\end{figure}

A surface charge density $\sigma$ was applied to the boundaries DC, EF and DE. The boundaries AB, BC, FG and GH were of width $\mathrm{max}(12(l^{\infty}_{\mathrm{Du}} + 2a), 75\lDinf)$, where $\lDinf$ and $\lDuinf = \kappa_{\mathrm{s}}^{\infty}/\kappa_{\mathrm{b}}$ are the Debye screening length and Dukhin length of a planar channel, respectively, and $a$ is the pore radius. The membrane thickness $L$ was no larger than $1/5$ of the pore radius, and less than $1/9$ of the Debye length. The membrane surface between points D and E in Fig.~\ref{sfig:schematic-FEM} was given a finite radius of curvature of $L/2$. A boundary layer mesh was used at all solid--liquid boundaries, with 5 boundary layers and a boundary layer stretching factor (mesh element growth rate) of $1.2$ . A predefined ("Normal") element size was used in the simulation domain and the maximum element size at the boundary between points D and E in Fig.~\ref{sfig:schematic-FEM} was $2L$. The maximum element size between the points D and C, and E and F, was $\lDinf/5$. Cubic and quadratic discretizations were used for the solute concentration and electric potential, respectively, while second-order elements were used for the velocity components and linear elements were used for the pressure field in the Stokes equation. Table~\ref{stab:variables-FEM} lists the parameters used in the simulations; however, simulations with a surface charge density of $-100$ mC m$^{-2}$ were carried out only for pore radii up to $5$~nm. Although the smallest Gouy--Chapman length was $\lesssim 2 L$, we verified that halving the membrane thickness did not change the measured fluxes significantly (discrepancy of $< 5\%$ for the flow rate, solute flux and electric current at $\Delta \psi = 70$ mV). The ratios of the advective to the electro-diffusive contributions to the electric current inside the pore (i.e. $z=0$ cutline) in all simulations where $|\sigma| \geq 1$ mC m$^{-2}$ are shown in Fig.~\ref{sfig:Iadv-contr}.

\begin{table}[!ht]
\centering
\caption[Parameters for FEM simulations]{\label{stab:variables-FEM}Parameters used in the FEM simulations, where the ion mobilities were calculated using the Einstein relation $\mu _i = D_i /\kBT$ with ion diffusivities $D_i$ chosen to be those for KCl. \cite{grayAIPHandbook1972} 
}
\renewcommand{\arraystretch}{0.9}
    \begin{tabular*}{0.85\columnwidth}{@{\extracolsep{\fill}} l c c c}

\hline
quantity & symbol & unit & value \\
\hline
ion diffusivity ($+$ve)    & $D_+$      & m$^2$ s$^{-1}$ & $1.960 \times 10^{-9}$ \\
ion diffusivity ($-$ve)    & $D_-$      & m$^2$ s$^{-1}$ & $2.030 \times 10^{-9}$ \\
ion valence ($+$ve)        & $Z_+$      & --              & $1$ \\
ion valence ($-$ve)        & $Z_-$      & --              & $-1$ \\
membrane thickness         & $L$        & nm              & $0.2$ \\ 
pore radius                & $a$        & nm              & $1$--$50$ 
\\
bulk electrolyte concentration & $\cinf$    & mol~m$^{-3}$    & $0.3$--$30$ 
\\
surface charge density     & $\sigma$   & mC~m$^{-2}$     & $-(0.1$--$100)$ 
\\
potential difference       & $\Delta \psi$    &~mV        & $10$--$70$ \\
solution density           & $\rho_{\mathrm{w}}$          & g cm$^{-3}$ & 1 \\
dielectric constant        & $\epsilon$ & --              & $78.46$ \\ 
temperature                & $T$        & K               & $298$ \\
\hline
\end{tabular*}
\end{table}

\clearpage

\subsection{Fitted surface charge densities}
\label{ssec:fitted_surface-charge}

\begin{table}[htbp]
\caption{Fits of Eq.~\eqref{eq:acc_resist} (using Eq.~\eqref{eq:surf_conduct}) and Eq.~\eqref{eq:lee-access_conductance} (using Eq.~\eqref{eq:surf_conduct-plane}) for $\alpha=2$ and $\beta = 1/2$, $1$, $2$ and $4$ to the simulation data in Fig.~\ref{fig:acc_resist-vs-radius-cave-thinEDL}(b); $\sigma$ is the surface charge density in the simulations, $|\hat{\sigma}|$ is the fitted magnitude of the surface charge density, $\mathrm{RE} = |\hat{\sigma}|/|\sigma| - 1$ is the relative error, $\mathrm{SE}$ is the standard error from the fit, and $R^2$ is the coefficient of determination calculated from the theoretical ionic conductances (using $\hat{\sigma}$) and the ionic conductances from simulations. To weight each data point equally, we fitted the theoretical ionic conductance over the simulation ionic conductance as a function of electrolyte concentration to a constant value of $1$. \label{stab:fitted-surface_charge}}
\centering
\renewcommand{\arraystretch}{0.75}
\begin{tabular*}{\columnwidth}{@{\extracolsep{\fill}} ll c c c c c}
\hline
theory &&
{$\sigma$ (mC m$^{-2})$} & 
{$|\hat{\sigma}|$ (mC m$^{-2}$)} & 
{$\mathrm{RE}$} & 
{$\mathrm{SE}$ (mC m$^{-2}$)} & 
$R^2$ \\
\hline
this work && $-1$ & $0.726$ & $-0.274$ & $0.265$ & $0.974$ \\
\multirow{4}{*}{Lee at al.} 
&$\beta = 1/2$ & $-1$ & $0.607$ & $-0.393$ & $0.206$ & $0.974$ \\
&$\beta = 1$ & $-1$ & $0.427$ & $-0.573$ & $0.143$ & $0.974$ \\
&$\beta = 2$ & $-1$ & $0.301$ & $-0.699$ & $0.100$ & $0.974$ \\
&$\beta = 4$ & $-1$ & $0.213$ & $-0.787$ & $0.071$ & $0.974$ \\
\hline
this work && $-10$ & $9.83$ & $-0.017$ & $0.398$ & $0.983$ \\
\multirow{4}{*}{Lee at al.} 
&$\beta = 1/2$ & $-10$ & $11.8$ & $0.178$ & $0.571$ & $0.989$\\
&$\beta = 1$ & $-10$ & $6.96$ & $-0.304$ & $0.476$ & $0.984$ \\
&$\beta = 2$ & $-10$ & $4.29$ & $-0.571$ & $0.364$ & $0.973$ \\
&$\beta = 4$ & $-10$ & $2.74$ & $-0.726$ & $0.262$ & $0.962$ \\
\hline
this work  && $-30$ & $29.9$ & $-0.004$ & $0.462$ & $0.988$ \\
\multirow{4}{*}{Lee at al.} 
&$\beta = 1/2$ & $-30$ & $46.2$ & $0.542$ & $2.55$ & $0.879$ \\
&$\beta = 1$ & $-30$ & $25.2$ & $-0.161$ & $1.95$ & $0.691$ \\
&$\beta = 2$ & $-30$ & $14.2$ & $-0.528$ & $1.46$ & $0.343$ \\
&$\beta = 4$ & $-30$ & $8.23$ & $-0.726$ & $1.05$ & $-0.121$ \\
\hline
this work && $-60$ & $59.5$ & $-0.009$ & $0.486$ & $0.991$ \\
\multirow{4}{*}{Lee at al.} 
&$\beta = 1/2$ & $-60$ & $103$ & $0.722$ & $0.217$ & $0.598$ \\
&$\beta = 1$ & $-60$ & $54.3$ & $-0.095$ & $0.141$ & $0.302$ \\
&$\beta = 2$ & $-60$ & $29.4$ & $-0.511$ & $0.0966$ & $-0.307$ \\
&$\beta = 4$ & $-60$ & $16.3$ & $0.728$ & $0.0671$ & $-1.33$ \\
\hline
this work && $-100$ & $98.6$ & $-0.014$ & $0.663$ & $0.994$ \\
\multirow{4}{*}{Lee at al.}  
&$\beta = 1/2$ & $-100$ & $182$ & $0.818$ & $13.2$ & $0.375$ \\
&$\beta = 1$ & $-100$ & $93.8$ & $-0.062$ & $7.80$ & $0.119$ \\
&$\beta = 2$ & $-100$ & $49.5$ & $-0.505$ & $4.94$ & $-0.419$ \\
&$\beta = 4$ & $-100$ & $26.9$ & $-0.731$ & $3.29$ & $-1.44$ \\
\hline
\end{tabular*}
\end{table}

\begin{figure}[!ht]
 \centering
    \includegraphics[scale = 1]{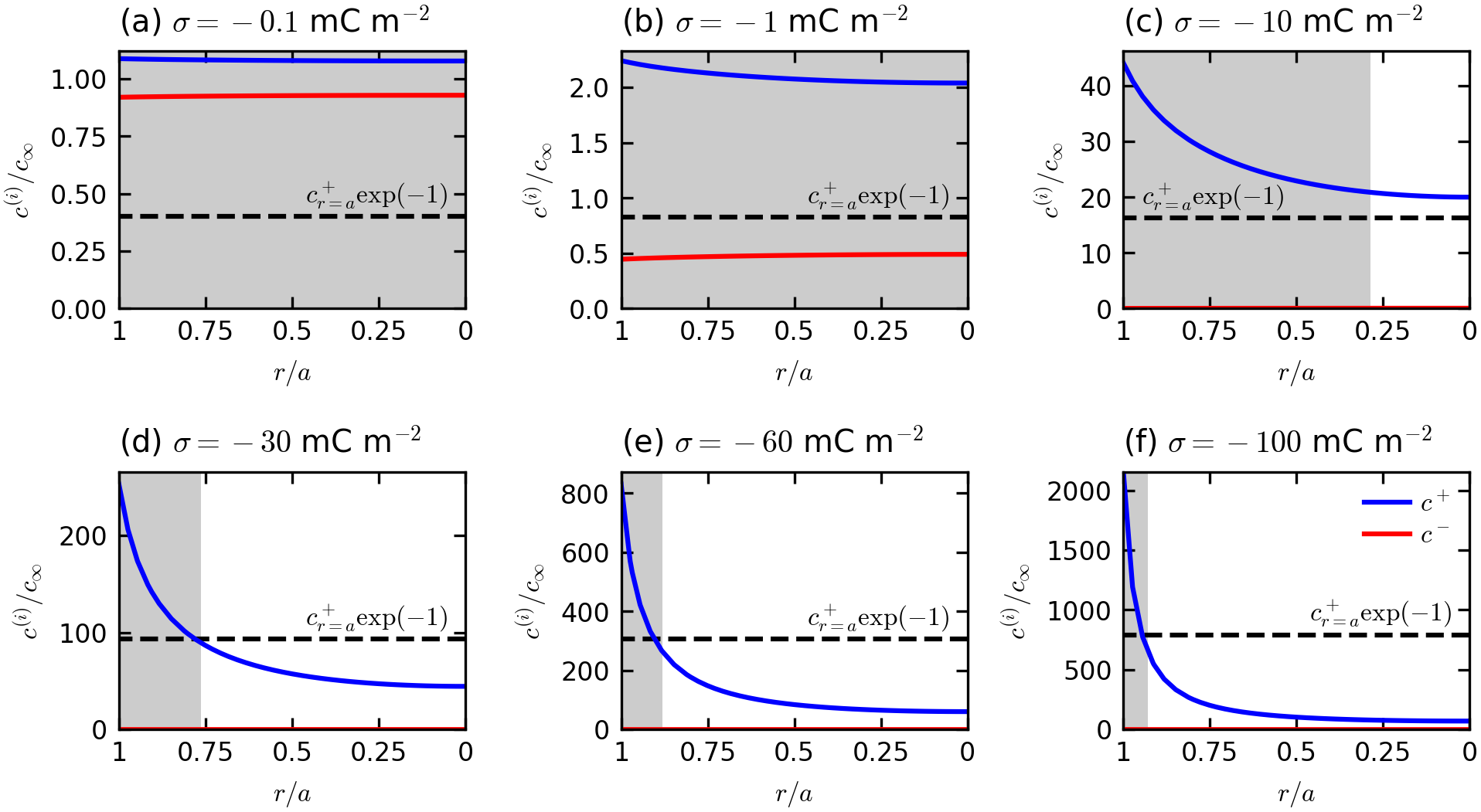}
        \caption{
        \label{sfig:EDL_width}
        \addition{Equilibrium ion concentration profiles inside the pore (i.e. $z=0$) relative to the bulk electrolyte concentration $\cinf$ vs the radial coordinate relative to the pore radius $a$ from FEM simulations with $\cinf = 0.3$~mol~m$^{-3}$ and $a = 5$~nm, which corresponds to a Debye length $\lDinf \approx 3.5a$. Surface charge densities $\sigma$ of (a) $-0.1$~mC~m$^{-2}$, (b) $-1$~mC~m$^{-2}$, (c) $-10$~mC~m$^{-2}$, (d) $-30$~mC~m$^{-2}$, (e) $-60$~mC~m$^{-2}$ and (f) $-100$~mC~m$^{-2}$ are shown, where the boundary between shaded and unshaded regions indicates a distance of one Gouy--Chapman length $\lGC$ from the pore edge; note that $\lGC > a$ in (a) and (b). Blue solid lines are the counter-ion concentrations ($c^+$), red solid lines are the co-ion concentrations ($c^-$), while horizontal dashed lines indicate $1/e$ of the counter-ion concentration at the pore edge.} 
        }
 \end{figure}

\addition{
\subsection{EDL width for ion-selective membrane}
\label{ssec:EDL_width}

To verify that the Gouy--Chapman length quantifies the EDL width inside the pore of an ion-selective ultrathin membrane, we have plotted equilibrium ion concentration profiles from FEM simulations for a bulk electrolyte concentration of $0.3$~mol~m$^{-3}$, pore radius of $5$~nm and a range of surface charge densities with magnitudes of $0.1$--$100$~mC~m$^{-2}$ in Fig.~\ref{sfig:EDL_width}. For all the parameters considered in Fig.~\ref{sfig:EDL_width}, the Debye length is $3.5$ times the pore radius, while the Dukhin length (using the theory for an ultrathin membrane) is $0.0008a$, $0.08a$, $6a$, $26a$, $57a$ and $100a$ in Fig.~\ref{sfig:EDL_width}(a), (b), (c), (d), (e) and (f), respectively. Thus, the condition for an ion-selective membrane in Eq.~\eqref{eq:selec_pore-cond} in the main paper is comfortably satisfied by the parameters considered in Fig.~\ref{sfig:EDL_width}(d)--(f). When Eq.~\eqref{eq:selec_pore-cond} is satisfied, the counter-ion concentration decays to $1/e$ of its pore-edge value within one Gouy--Chapman length of the pore edge (see. Fig.~\ref{sfig:EDL_width}(d)--(f)), indicating that the Gouy--Chapman length quantifies the EDL width. As the Gouy--Chapman length is much smaller than the pore radius under the conditions considered in Fig.~\ref{sfig:EDL_width}(d)--(f), the theory of the surface conductance for a thin EDL in Eq.~\eqref{eq:surf_conduct} in the main paper holds. While Fig.~\ref{sfig:EDL_width}(a) and (b) depict significant EDL overlap (i.e. similar surface and centerline ion concentrations), the bulk conductance dominates under these conditions and the thin EDL assumption becomes irrelevant to the accuracy of the theory for the ionic conductance.}

\clearpage

\section{Access electrical resistance: Theory vs simulations}
\label{ssec:acc_resist}

\subsection{Weak ion--membrane interactions}
\label{ssec:acc_resist-DH}

\begin{figure}[!b]
 \centering
 \begin{minipage}{.5\textwidth}
\begin{flushleft}
    \includegraphics[scale = 1, trim={0.175cm 0.0875cm 0.175cm 0.0875cm},clip]{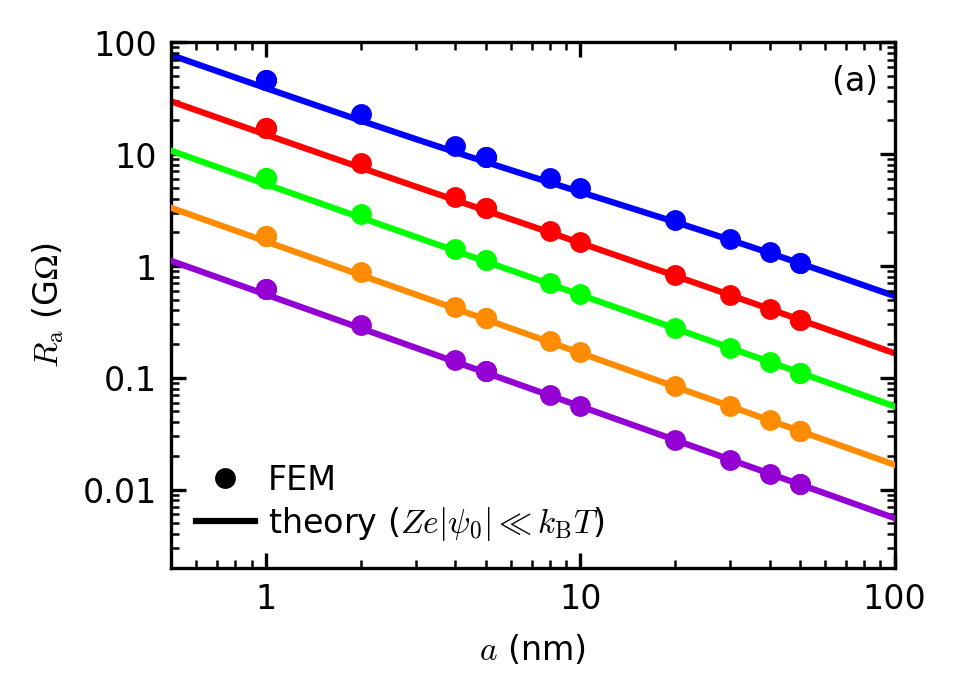}
\end{flushleft}
  \end{minipage}%
 \begin{minipage}{.5\textwidth}
   \begin{flushright}
    \includegraphics[scale = 1, trim={0.175cm 0.0875cm 0.175cm 0.0875cm},clip]{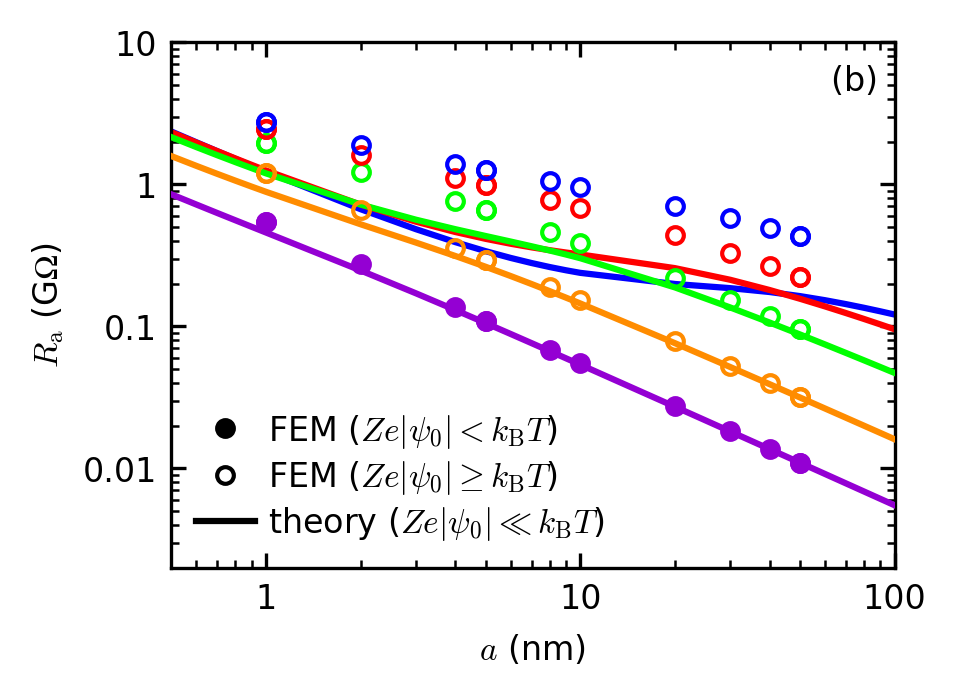}
  \end{flushright}
  \end{minipage}
    \caption{\label{sfig:acc_resist-vs-radius-DH}Access electrical resistance, $R_{\mathrm{a}}$, vs pore radius $a$ for surface charge densities of (a)~$-1$~mC~m$^{-2}$ and (b)~$-10$~mC~m$^{-2}$, and bulk electrolyte concentrations $\cinf$ of $0.3$ (blue), $1$ (red), $3$ (green), $10$ (orange) and $30$ (purple)~mol~m$^{-3}$. Filled symbols are FEM simulations corresponding to surface potential energies that are $< \kBT$ and unfilled symbols are FEM simulations corresponding to surface potential energies that are $\geq \kBT$ far from the pore mouth. Solid lines are the theory (in the Debye--H\"uckel regime) obtained from Eq.~\eqref{seq:elec_curr-DH}.}
 \end{figure}

\begin{figure}[!ht]
 \centering
 \begin{minipage}{.5\textwidth}
\begin{flushleft}
    \includegraphics[scale = 1, trim={0.175cm 0.0875cm 0.175cm 0.0875cm},clip]{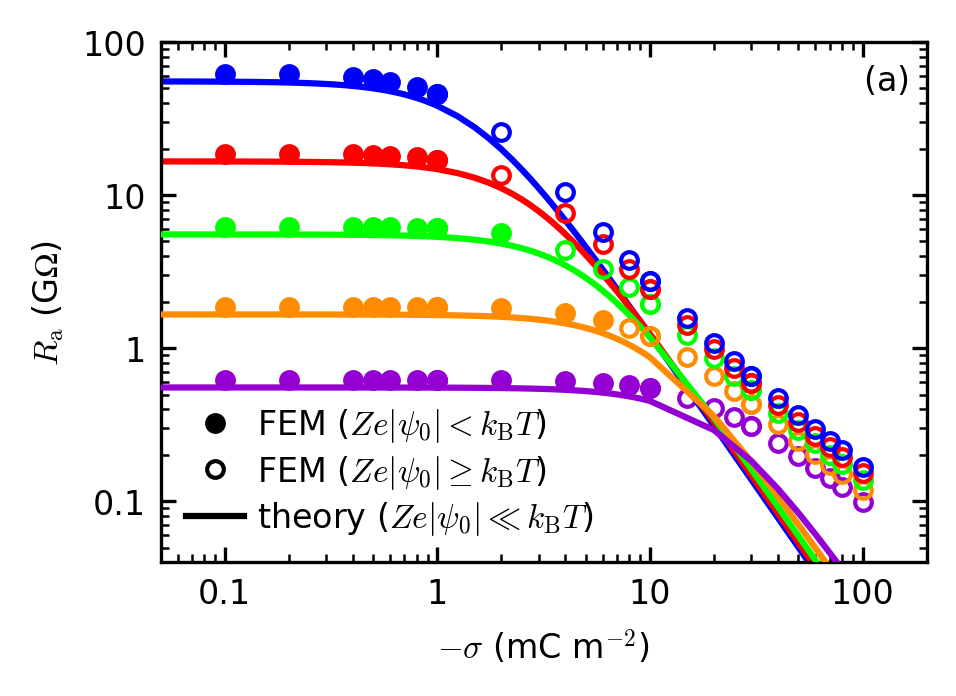}
\end{flushleft}
  \end{minipage}%
 \begin{minipage}{.5\textwidth}
   \begin{flushright}
    \includegraphics[scale = 1, trim={0.175cm 0.0875cm 0.175cm 0.0875cm},clip]{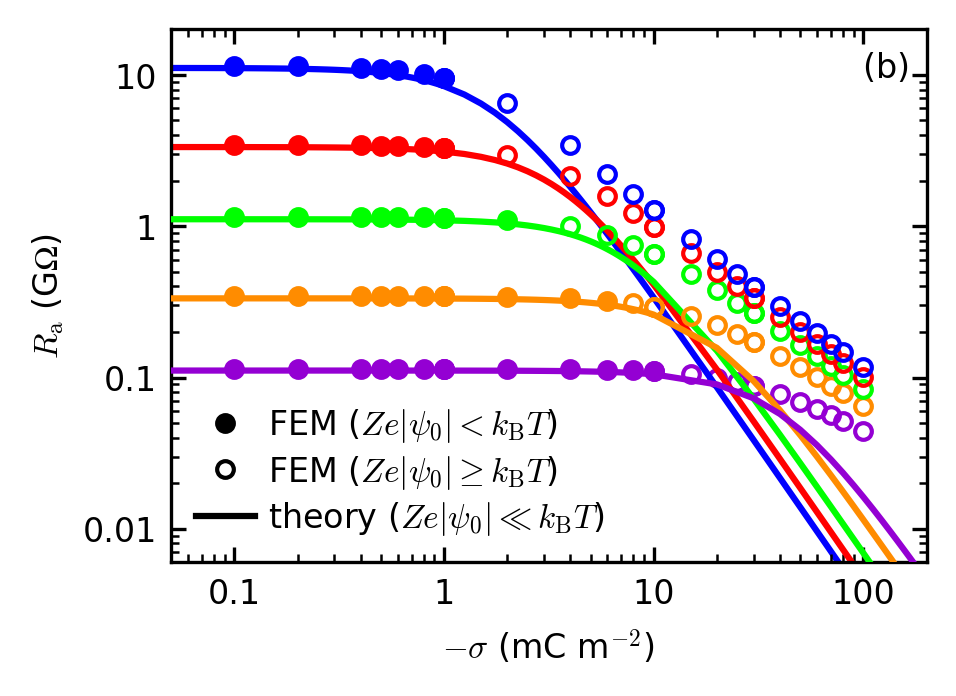}
  \end{flushright}
  \end{minipage}
    \caption{\label{sfig:acc_resist-vs-sigma-DH}Access electrical resistance, $R_{\mathrm{a}}$, vs surface charge density $\sigma$ for pore radii of (a)~$1$~nm and (b)~$5$~nm, and bulk electrolyte concentration $\cinf$ of $0.3$ (blue), $1$ (red), $3$ (green), $10$ (orange) and $30$ (purple)~mol~m$^{-3}$. Filled symbols are FEM simulations corresponding to surface potential energies that are $< \kBT$ and unfilled symbols are FEM simulations corresponding to surface potential energies that are $\geq \kBT$ far from the pore mouth. Solid lines are the theory (in the Debye--H\"uckel regime) obtained from Eq.~\eqref{seq:elec_curr-DH}.}
 \end{figure}

Assuming the electric-field-driven access electrical resistance of each pore end is $R_{\mathrm{a}} = \Delta \psi/(2I)$, where $I$ is the electric-field-driven electric current and $\Delta \psi$ is the potential difference applied across the membrane, we can substitute Eq.~\eqref{seq:elec_curr-DH} into this expression to predict the access electrical resistance for weak ion--membrane interactions (Debye--H\"uckel regime) and all $\lDinf/a$. Figure~\ref{sfig:acc_resist-vs-radius-DH} compares the theoretical access electrical resistance calculated from Eq.~\eqref{seq:elec_curr-DH} with the simulations as the pore radius $a$ is varied, while $\sigma = -1,-10$~mC~m$^{-2}$ and $\cinf = 0.3,1,3,10,30$~mol~m$^{-3}$ are both fixed. On the other hand, Fig.~\ref{sfig:acc_resist-vs-sigma-DH} compares the theoretical access electrical resistance in the Debye--H\"uckel regime with the simulations as the surface charge density $\sigma$ is varied, while $a = 1,5$~nm and $\cinf = 0.3,1,3,10,30$~mol~m$^{-3}$ are both fixed. Figures~\ref{sfig:acc_resist-vs-radius-DH} and \ref{sfig:acc_resist-vs-sigma-DH} indicate that Eq.~\eqref{seq:elec_curr-DH} is accurate when the magnitude of the surface potential energy far from the pore mouth, $Ze|\psi_{\infty}|$,  is smaller than the thermal energy $\kBT$, where $Ze|\psi_{\infty}| = 2\kBT \sinh^{-1}(\lDinf/\lGC)$ is the magnitude of the equilibrium surface potential energy of a plane (i.e. far from the pore mouth of an ultrathin membrane). When $Ze|\psi_{\infty}| > \kBT$, Eq.~\eqref{seq:elec_curr-DH} begins to break down.

\subsection{Arbitrary ion--membrane interactions and a thin electric double layer}
\label{ssec:access-resistance-thinEDL}

\begin{figure}[!ht]
 \centering
 \begin{minipage}{.5\textwidth}
\begin{flushleft}
    \includegraphics[scale = 1, trim={0.175cm 0.0875cm 0.175cm 0.0875cm},clip]{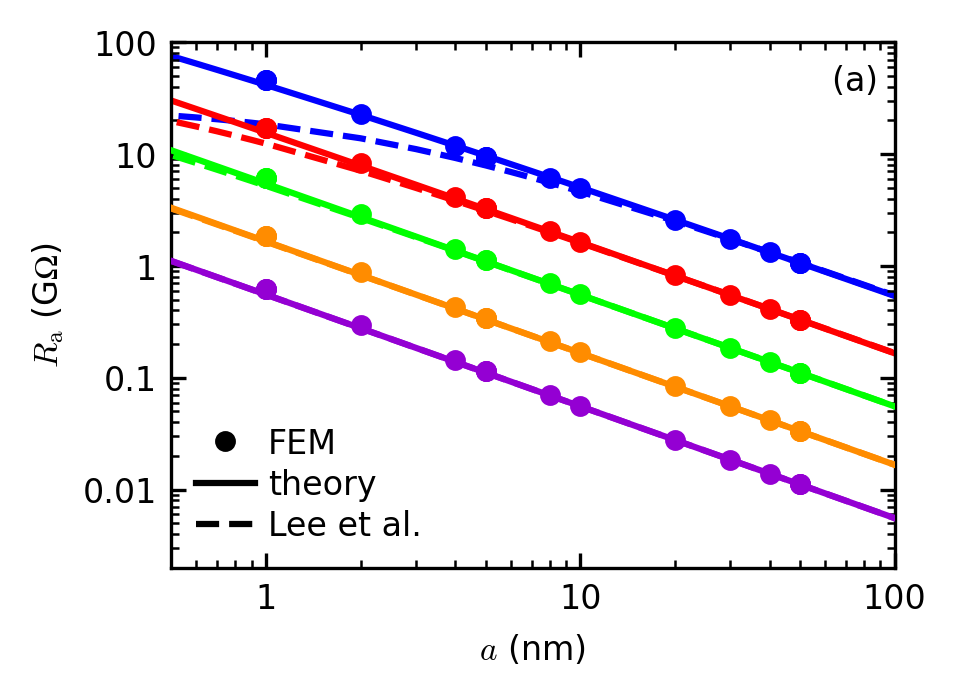}
    \includegraphics[scale = 1, trim={0.175cm 0.0875cm 0.175cm 0.0875cm},clip]{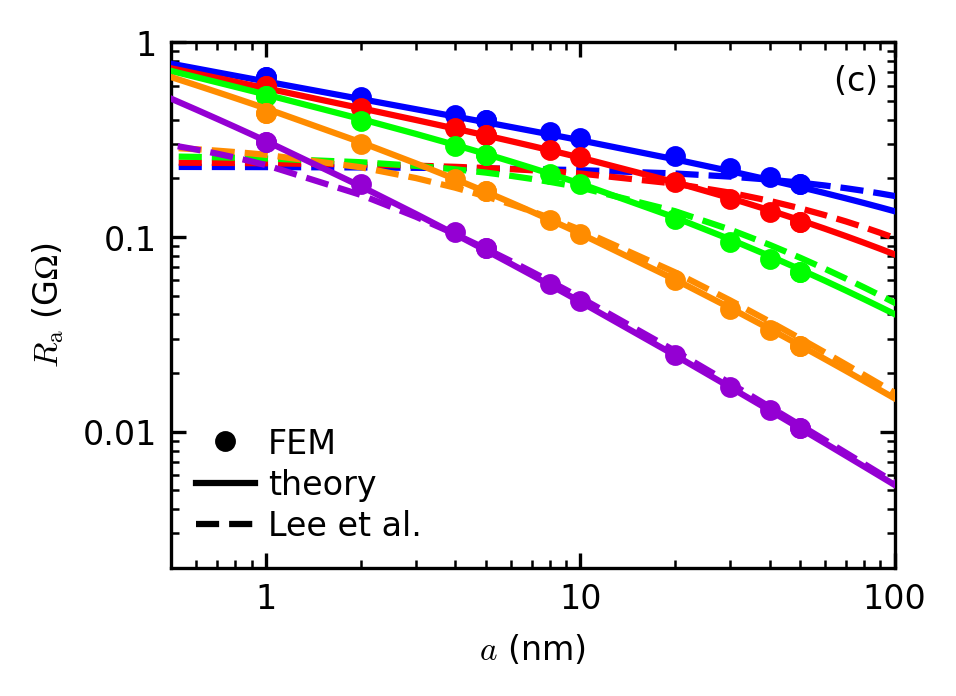}
\end{flushleft}
  \end{minipage}%
 \begin{minipage}{.5\textwidth}
   \begin{flushright}
    \includegraphics[scale = 1, trim={0.175cm 0.0875cm 0.175cm 0.0875cm},clip]{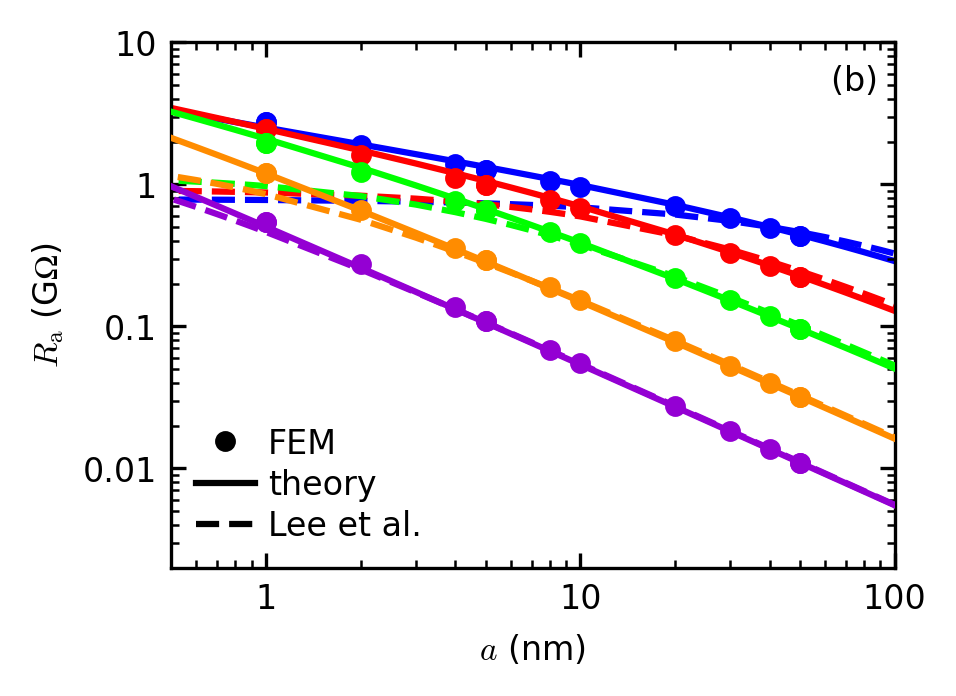}
    \includegraphics[scale = 1, trim={0.175cm 0.0875cm 0.175cm 0.0875cm},clip]{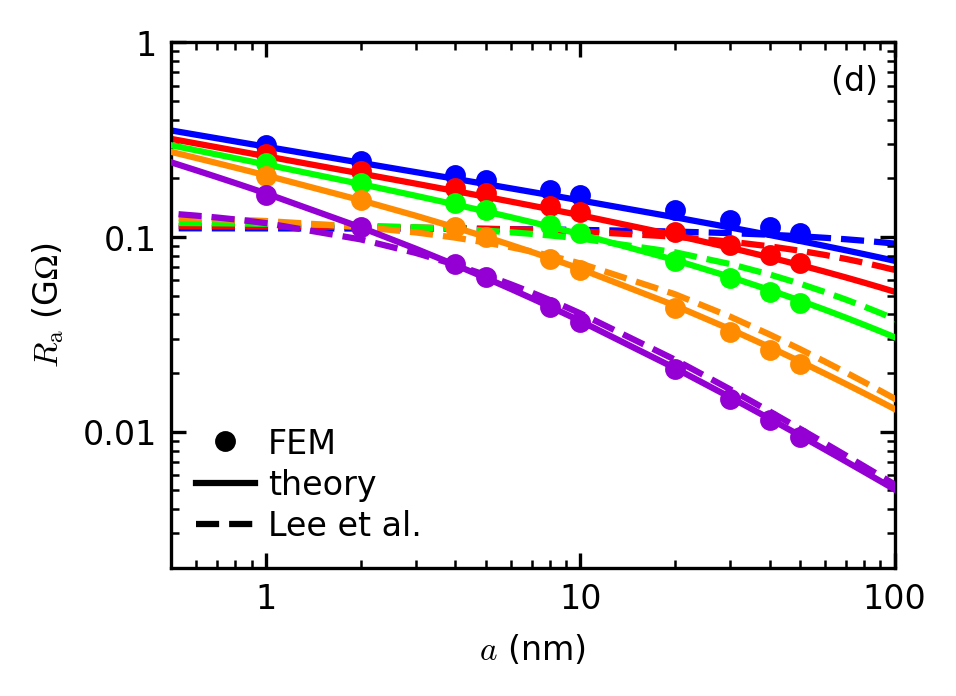}
  \end{flushright}
  \end{minipage}
      \caption{\label{sfig:acc_resist-vs-radius-thinEDL}Access electrical resistance, $R_{\mathrm{a}}$, vs pore radius $a$ for surface charge densities $\sigma$ of (a)~$-1$~mC~m$^{-2}$, (b) $-10$~mC~m$^{-2}$, (c) $-30$~mC~m$^{-2}$ and (d) $-60$~mC~m$^{-2}$, and bulk electrolyte concentrations $\cinf$ of $0.3$ (blue), $1$ (red), $3$ (green), $10$ (orange) and $30$ (purple)~mol~m$^{-3}$. Symbols are FEM simulations, solid lines are the theory derived in this work (see Eqs.~\eqref{eq:surf_conduct} and \eqref{eq:acc_resist} in the main paper), and dashed lines are the theory in Ref.~\citenum{leeLargeElectricSizeSurfaceCondudction2012} (see Eqs.~\eqref{eq:surf_conduct-plane} and \eqref{eq:lee-access_conductance} in the main paper) for $\alpha = \beta = 2$.}
 \end{figure}

\begin{figure}[!ht]
 \centering
 \begin{minipage}{.5\textwidth}
\begin{flushleft}
    \includegraphics[scale = 1, trim={0.175cm 0.0875cm 0.175cm 0.0875cm},clip]{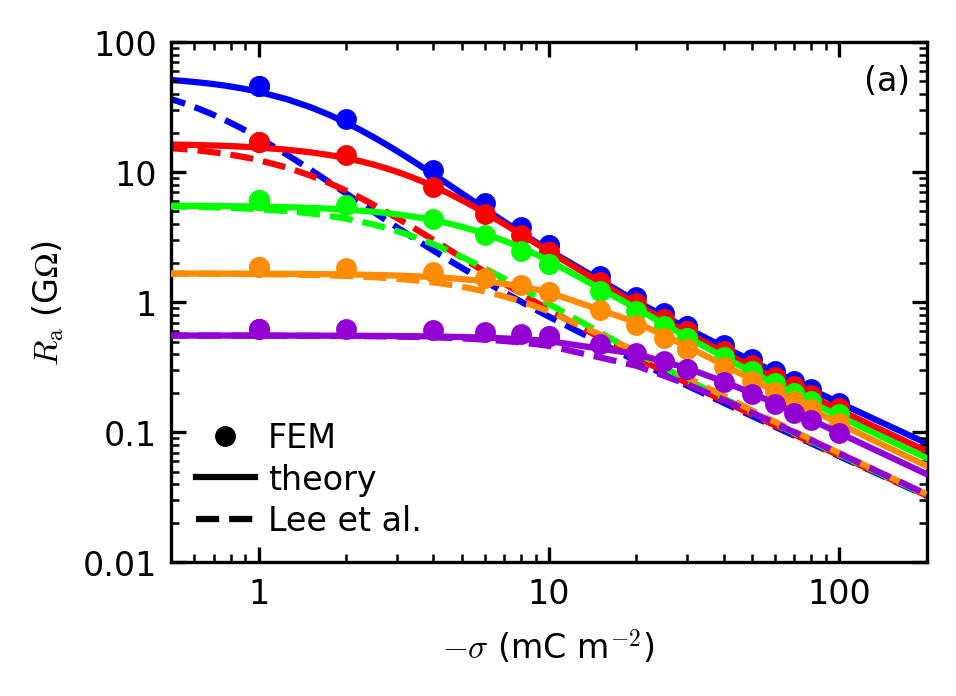}
\end{flushleft}
  \end{minipage}%
 \begin{minipage}{.5\textwidth}
   \begin{flushright}
    \includegraphics[scale = 1, trim={0.175cm 0.0875cm 0.175cm 0.0875cm},clip]{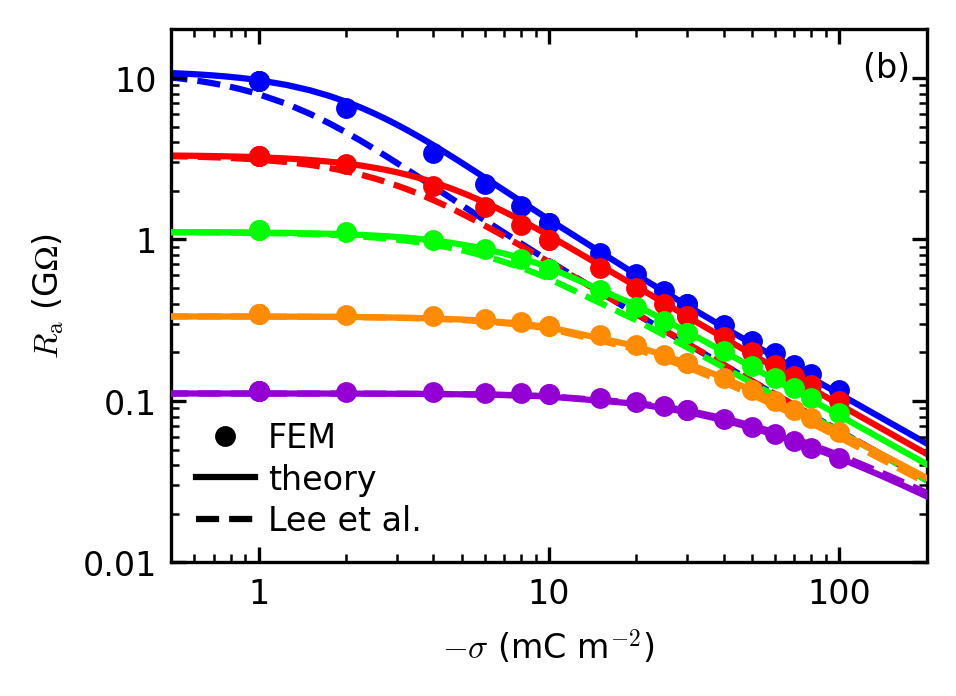}
  \end{flushright}
  \end{minipage}
    \caption{\label{sfig:acc_resist-vs-sigma-thinEDL}Access electrical resistance, $R_{\mathrm{a}}$, vs surface charge density $\sigma$ for pore radii of (a)~$1$~nm, and (b)~$5$~nm, and bulk electrolyte concentrations $\cinf$ of $0.3$ (blue), $1$ (red), $3$ (green), $10$ (orange) and $30$ (purple)~mol~m$^{-3}$. Symbols are FEM simulations, solid lines are the theory derived in this work (see Eqs.~\eqref{eq:surf_conduct} and \eqref{eq:acc_resist} in the main paper), and dashed lines are the theory in Ref.~\citenum{leeLargeElectricSizeSurfaceCondudction2012} (see Eqs.~\eqref{eq:surf_conduct-plane} and \eqref{eq:lee-access_conductance} in the main paper) for $\alpha = \beta = 2$.}
 \end{figure}

\begin{figure}[!ht]
 \centering
  \centering
 \begin{minipage}{.5\textwidth}
\begin{flushleft}
    \includegraphics[scale = 1, trim={0.175cm 0.0875cm 0.175cm 0.0875cm},clip]{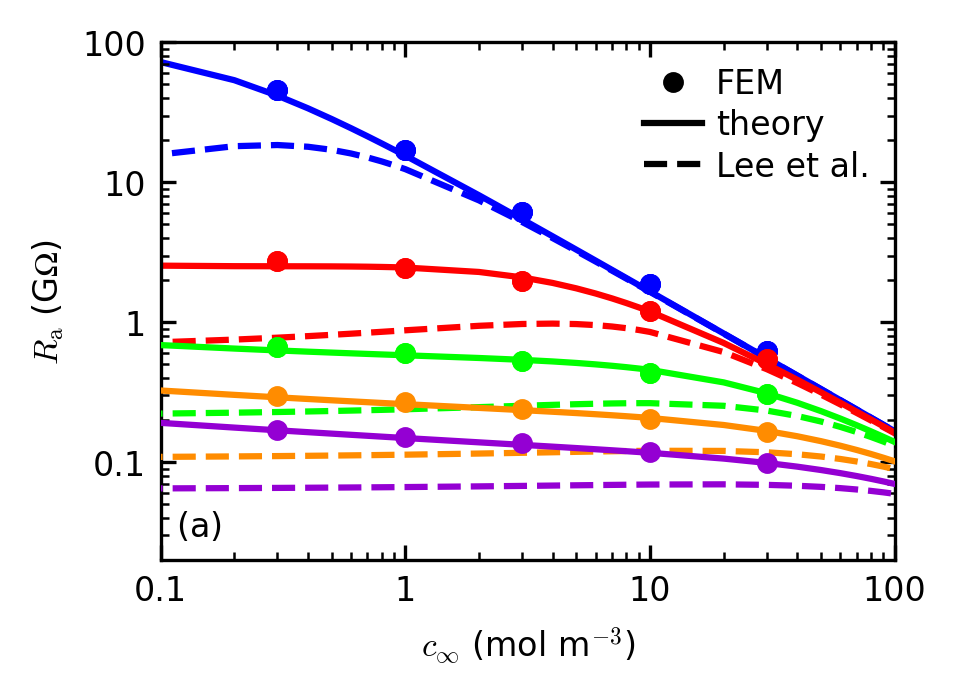}
\end{flushleft}
  \end{minipage}%
 \begin{minipage}{.5\textwidth}
   \begin{flushright}
    \includegraphics[scale = 1, trim={0.175cm 0.0875cm 0.175cm 0.0875cm},clip]{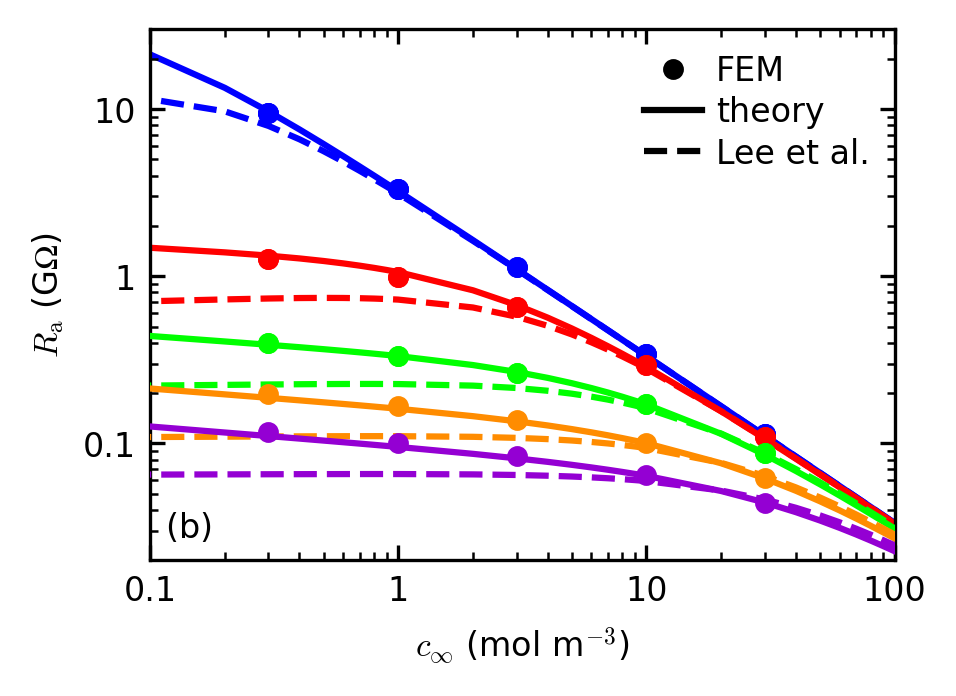}
  \end{flushright}
  \end{minipage}
    \caption{\label{sfig:acc_resist-vs-cave-thinEDL}Access electrical resistance, $R_{\mathrm{a}}$, vs bulk electrolyte concentration $\cinf$ for pore radii of (a)~$1$~nm, and (b)~$5$~nm, and surface charge densities $\sigma$ of $-1$ (blue), $-10$ (red), $-30$ (green), $-60$ (orange) and $-80$ (purple)~mC~m$^{-2}$. Symbols are FEM simulations, solid lines are the theory derived in this work (see Eqs.~\eqref{eq:surf_conduct} and \eqref{eq:acc_resist} in the main paper), and dashed lines are the theory in Ref.~\citenum{leeLargeElectricSizeSurfaceCondudction2012} (see Eqs.~\eqref{eq:surf_conduct-plane} and \eqref{eq:lee-access_conductance} in the main paper) for $\alpha = \beta = 2$.}
 \end{figure}

\begin{figure}[!ht]
 \centering
 \begin{minipage}{.5\textwidth}
\begin{flushleft}
    \includegraphics[scale = 1, trim={0.175cm 0.0875cm 0.175cm 0.0875cm},clip]{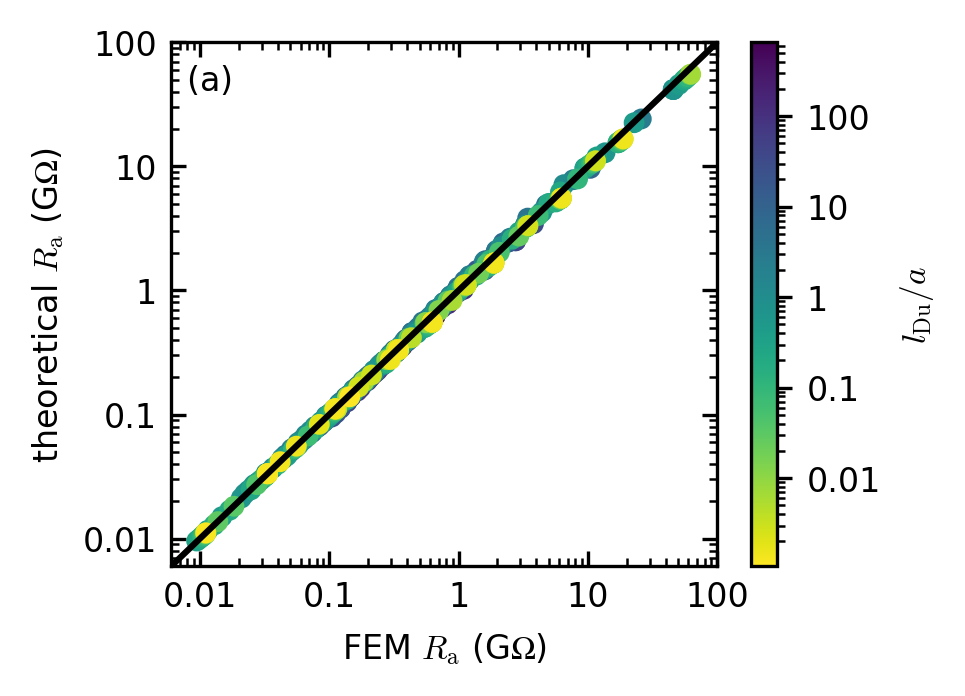}
\end{flushleft}
  \end{minipage}%
 \begin{minipage}{.5\textwidth}
   \begin{flushright}
    \includegraphics[scale = 1, trim={0.175cm 0.0875cm 0.175cm 0.0875cm},clip]{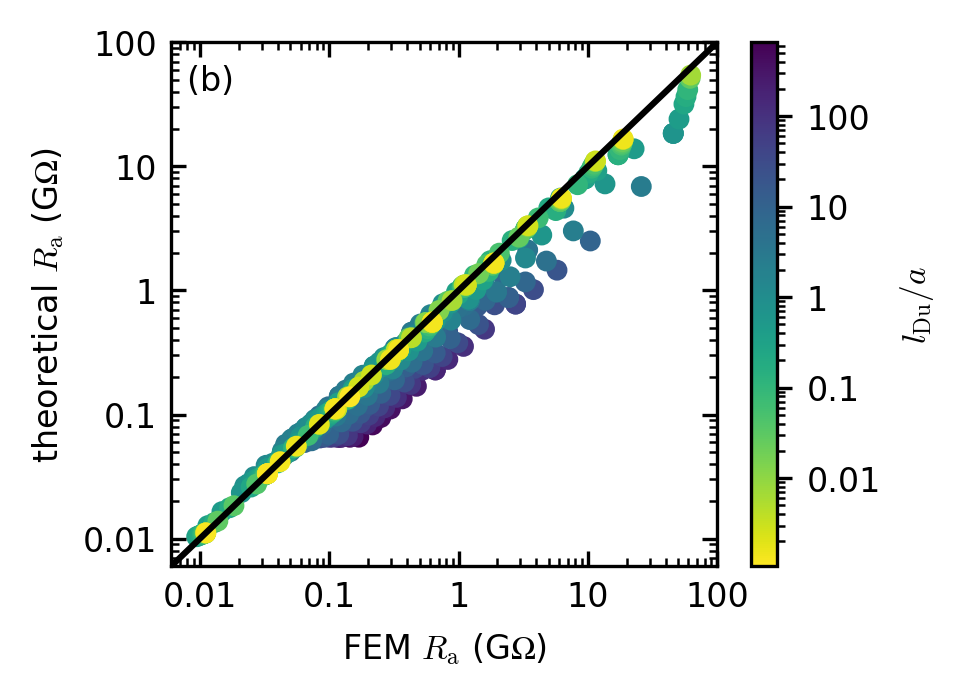}
  \end{flushright}
  \end{minipage}
    \caption{\label{sfig:agreement_lines}Theoretical access electrical resistance (symbols) for the theory (a) derived in this work and (b) in Ref.~\citenum{leeLargeElectricSizeSurfaceCondudction2012} for $\alpha = \beta = 2$, vs the access electrical resistance from FEM simulations for $\lDu \geq 0.001$, where $\lDu$ is the Dukhin length (of an ultrathin membrane) and $a$ is the pore radius. 
    The color map depicts variations $\lDu/a$ and the solid line indicates perfect agreement between the theory and simulations. }
 \end{figure}

Figure~\ref{sfig:acc_resist-vs-radius-thinEDL} compares the expressions for the access electrical resistance of each pore end, $R_\mathrm{a} = 1/G_\mathrm{a} \approx 1/(2 G_{L/(2a) \rightarrow 0})$, for the access conductance derived in this work (see Eqs.~\eqref{eq:surf_conduct} and \eqref{eq:acc_resist} in the main paper) and in Ref.~\citenum{leeLargeElectricSizeSurfaceCondudction2012} (see Eqs.~\eqref{eq:surf_conduct-plane} and \eqref{eq:lee-access_conductance} in the main paper) for $\alpha = \beta = 2$ with the simulations for a range of pore radii and bulk electrolyte concentrations, and surface charge densities of $-1$, $-10$, $-30$ and $-60$~mC~m$^{-2}$. Figure~\ref{sfig:acc_resist-vs-sigma-thinEDL} compares the expressions for the access electrical resistance 
derived in this work
and in Ref.~\citenum{leeLargeElectricSizeSurfaceCondudction2012} (for $\alpha = \beta = 2$) with the simulations for a range of surface charge densities and bulk electrolyte concentrations, and pore radii of $1$ and $5$~nm. 
Furthermore, Fig.~\ref{sfig:acc_resist-vs-cave-thinEDL} compares the expressions for the access electrical resistance derived in this work and in Ref.~\citenum{leeLargeElectricSizeSurfaceCondudction2012} (for $\alpha = \beta = 2$) with the simulations for a range of bulk electrolyte concentrations and surface charge densities, and pore radii of $1$ and $5$~nm. Note that Figs.~\ref{sfig:acc_resist-vs-radius-thinEDL}(d) and \ref{sfig:acc_resist-vs-cave-thinEDL}(a) are analogous to Fig.~\ref{fig:acc_resist-vs-radius-cave-thinEDL} in the main paper, just with the resistance plotted instead of the conductance. Figure~\ref{sfig:agreement_lines} compares the theoretical access electrical resistance in the theories derived in this work and in Ref.~\citenum{leeLargeElectricSizeSurfaceCondudction2012} (for $\alpha = \beta = 2$) with the access electrical resistance from simulations, where the color map in Fig.~\ref{sfig:agreement_lines} indicates variations in $\lDu/a$.


\clearpage

\subsection{Charged pore in uncharged membrane}
\label{ssec:access-resistance-uncharged-membrane}

\begin{figure}[!ht]
 \centering
  \centering
 \begin{minipage}{.5\textwidth}
\begin{flushleft}
    \includegraphics[scale = 1, trim={0.175cm 0.0875cm 0.175cm 0.0875cm},clip]{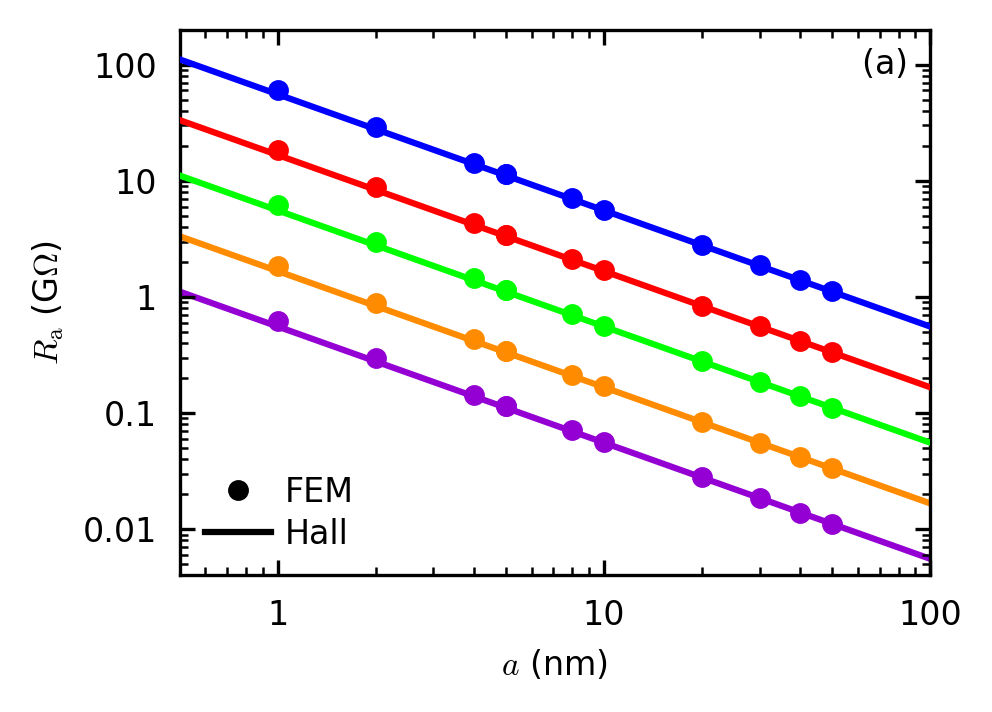}
\end{flushleft}
  \end{minipage}%
 \begin{minipage}{.5\textwidth}
   \begin{flushright}
    \includegraphics[scale = 1, trim={0.175cm 0.0875cm 0.175cm 0.0875cm},clip]{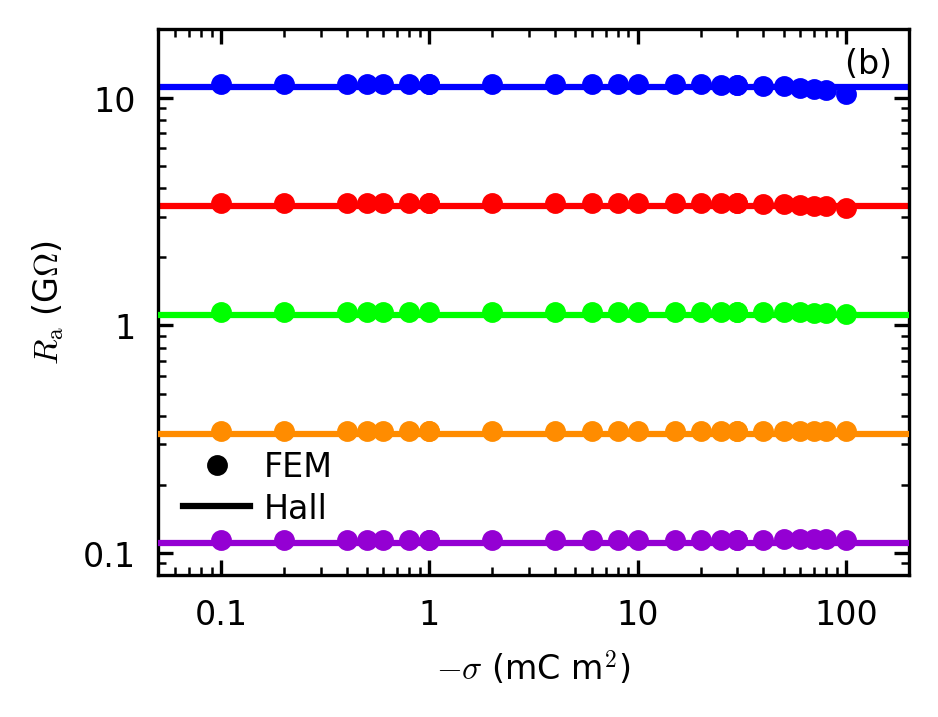}
  \end{flushright}
  \end{minipage}
    \caption{
    \label{sfig:acc_resist-2D-ring} 
    Access electrical resistance, $R_\mathrm{a}$, vs (a) pore radius $a$ for a surface charge density of $\sigma = -30$~mC~m$^{-2}$, and (b) $\sigma$ for $a=5$~nm, and bulk electrolyte concentrations of $\cinf = 0.3$ (blue), $1$ (red), $3$ (green), $10$ (orange), and $30$ (purple)~mol~m$^{-3}$. Symbols are the access resistances calculated from FEM simulations of an ultrathin membrane with charge at the pore edge only; and solid lines are Hall's theory for an uncharged pore, $R_{\mathrm{a}} = 1/(4a\kappa_{\mathrm{b}})$.\cite{hallAccessResistance1975}}
 \end{figure}

To study the effects of charging the surface only at the pore edge of the 0.2-nm-thick membrane described in Sec.~\ref{ssec:FEM}, surface charge was applied to the curved edge inside the pore and to a 0.1-nm-wide planar region (i.e. width equal to the curvature radius) of the membrane surface at the pore entrance. Simulations of transport of electrolyte solutions with bulk electrolyte concentrations $\cinf$ of $0.3$, $1$, $3$, $10$ and $30$~mol~m$^{-3}$ through pores with radii $a$ ranging from $1$ to $50$~nm were carried out for a surface charge density $\sigma$ of $-30$~mC~m$^{-2}$. Simulations of electrolyte transport through $5$-nm-radius pores with surface charge densities ranging from $0.1$ to $100$~mC~m$^{-2}$ were also performed. Note that for simulations of $5$-nm-radius pores, parameters corresponding to a Dukhin length near a planar wall, $\lDuinf$, of up to $\approx 1700$~nm (i.e. $\lDuinf/a \approx 340$) and a Dukhin length of a uniformly charged ultrathin membrane, $\lDu$, of up to $\approx 1000$~nm (i.e. $\lDu/a \approx 200$) were considered. Figure~\ref{sfig:acc_resist-2D-ring} compares the access electrical resistances calculated from FEM simulations of various pore radii and surface charge densities with Hall's theory for an uncharged pore,\cite{hallAccessResistance1975} and shows that this theory accurately quantifies the FEM access electrical resistances. In simulations of $5$-nm-radius pores (see Fig.~\ref{sfig:acc_resist-2D-ring}(b)), increasing the surface charge density magnitude from $0.1$~mC~m$^{-2}$ (i.e. $\lDuinf/a$ and $\lDu/a$ both $< 0.01$ for all $\cinf$) to $100$~mC~m$^{-2}$ results in a maximum reduction of $6 \%$ in the access electrical resistance. For a uniformly charged membrane containing a pore of the same size (where $0.3 \leq \cinf \leq 30$~mol~m$^{-3}$), the reduction in the electrical access resistance is up to two orders of magnitudes as $|\sigma|$ is increased from $0.1$ to $100$~mC~m$^{-2}$, as can be observed in Fig.~\ref{sfig:acc_resist-vs-sigma-thinEDL}(b).

\section{Scaling laws for advective electric current (Debye--H\"uckel regime)}
\label{ssec:adv_elec_curr}

Substituting Eq.~\eqref{seq:ion_flux_dens-lin} into Eq.~\eqref{eq:elec_curr-int} in the main paper gives
\begin{align}
         I =  e \int _0 ^{2 \pi} \, \mathrm{d} \phi \int _0 ^1 \, \mathrm{d} \zeta \left[Z_+ h_{\zeta} h_{\phi} \left(\varepsilon u_{\nu_1} \cinf^{+}   - \frac{D_+}{h_\nu} \frac{\partial (\varepsilon c_{\mathrm{s}_1} ^{+})}{\partial \nu} \right) \exp{\left(-\frac{Z_+ e \psi _0}{\kBT}\right)} \right. \nonumber \\
             + \left.  Z_- h_{\zeta} h_{\phi} \left( \varepsilon u_{\nu_1} \cinf^{-}  - \frac{D_-}{h_\nu} \frac{\partial (\varepsilon c_{\mathrm{s}_1} ^{-})}{\partial \nu} \right) \exp{\left(-\frac{Z_- e \psi _0}{\kBT}\right)}\right]_{\nu = 0} ,
\label{seq:elec_curr-lin-int-exp}
\end{align}
where 
\begin{equation}
    h_\zeta = a\sqrt{\frac{\nu^2 + \zeta^2}{1 - \zeta^2}} , 
\end{equation}
\begin{equation}
    h_\nu = a\sqrt{\frac{\nu^2 + \zeta^2}{1 + \nu^2}}  , 
\end{equation}
and 
\begin{equation}
    h_\phi = a\sqrt{(1+\nu^2)(1-\zeta^2)}    
\end{equation}
are the scale factors of the oblate spheroidal coordinates (in the convention in Ref.~\citenum{morse1953a}) and $\varepsilon u_{\nu_1}$ is the first-order, normal (i.e. $\nu$-) component of the fluid velocity through the aperture. Note that Eq.~\eqref{seq:elec_curr-lin-int-exp} is analogous to Eq.~\eqref{seq:elec_curr-lin-gen} in Sec.~\ref{ssec:theo-DH}. As Eq.~\eqref{eq:cs-DH} in the main paper predicts that the fluid velocity does not affect $\varepsilon c_{\mathrm{s}_1}^{(i)}$, then the advective and electro-diffusive contributions to the electric current are separately conserved when Eq.~\eqref{eq:cs-DH} is valid. Substituting Eq.~\eqref{eq:cs-DH} in the main paper into Eq.~\eqref{seq:elec_curr-lin-int-exp} and taking the advective contribution (i.e. terms containing $\varepsilon u_{\nu_1}$) gives the advective electric current as
\begin{equation}
    \label{seq:adv_elec_curr} 
    I_{\mathrm{adv}} = -\frac{2 \pi \eps \kBT}{Ze (\lDinf)^2} \int ^1 _0 \mathrm{d} \zeta \left[h_\zeta h_\phi \varepsilon u_{\nu_1} \sinh \left(\frac{Ze \psi_0}{\kBT} \right) \right]_{\nu=0}.
\end{equation}
Assuming that $Ze|\psi_0|  \ll \kBT$, Eq.~\eqref{seq:adv_elec_curr} reduces to
\begin{equation}
    \label{seq:adv_elec_curr-DH} 
     I_{\mathrm{adv}} = -\frac{2 \pi \eps}{(\lDinf)^2} \int ^1 _0 \mathrm{d} \zeta \left[h_\zeta h_\phi \varepsilon u_{\nu_1} \psi_0 \right]_{\nu=0},
\end{equation}
which does not have an analytical solution. However, we can determine approximate expressions for $I_{\mathrm{adv}}$ in the thin and thick electric-double-layer limits.

\subsubsection{Thick electric double layer}
\label{ssec:adv_elec_curr-thickEDL}

When the width of the electric double layer is much larger than the pore radius ($\lDinf \gg a$), we can use the approximation $\psi_0 \approx \sigma \lDinf/(\eps)$. In this case, Eq.~\eqref{seq:adv_elec_curr-DH} reduces to
\begin{equation}
     I_{\mathrm{adv}} \approx -\frac{2 \pi \sigma}{\lDinf} \int ^1 _0 \mathrm{d} \zeta \, h_\zeta h_\phi \varepsilon u_{\nu_1} |_{\nu=0}.
     \label{seq:adv_elec_curr-DH-thickEDL} 
\end{equation}
The flow rate induced by an applied gradient across the membrane is
\begin{equation}
    Q = \iint _S \mathrm{d}S \, \boldsymbol{u} \cdot \boldsymbol{\hat{n}},
    \label{seq:flow_rate-exp-rep}
\end{equation}
such that
\begin{equation}
    Q = \int^{2 \pi} _0 \mathrm{d} \phi \int ^1 _0  \mathrm{d} \zeta \, h_\zeta h_\phi \varepsilon u_{\nu_1} |_{\nu=0} = 2 \pi \int ^1 _0 \mathrm{d} \zeta \, h_\zeta h_\phi \varepsilon u_{\nu_1} |_{\nu=0}
    \label{seq:flow_rate-int-exp-rep}
\end{equation}
gives the flow rate through an ultrathin membrane in the linear--response regime. Substituting the left-hand-side of Eq.~\eqref{seq:flow_rate-int-exp-rep} into Eq.~\eqref{seq:adv_elec_curr-DH-thickEDL}, where $Q \approx a^3 \sigma \Delta \psi/(3 \eta \lDinf)$ when $Ze|\psi_0| \ll \kBT$ and $\lDinf \gg a$,\cite{maoElectroosmoticFlowNanopore2014} we can write the advective electric current  as
\begin{equation}
     I_{\mathrm{adv}} \approx -\frac{\sigma}{\lDinf} Q \approx -\frac{a^3 \sigma^2 \Delta \psi}{3 \eta (\lDinf)^2}.
     \label{seq:adv_elec_curr-DH-thickEDL-sub} 
\end{equation}

Summing Eq.~\eqref{seq:adv_elec_curr-DH-thickEDL-sub} and Eq.~\eqref{eq:elec_curr-DH-thickEDL}
in the main paper, which is the electro-diffusive electric current $I_{\mathrm{elec+diff}}$, and rearranging this expression gives
\begin{align}
     I \approx -\frac{a \eps \Delta \psi}{(\lDinf)^2}\left\{(\Dp) + 2\left(\frac{\lDinf}{\lGC} \right)^2 \left[ (\Dp) + \frac{2 \eps}{3 \eta} \left(\frac{\kBT}{Ze} \right)^2 \left(\frac{a}{ \lDinf} \right)^2 \right] \right\}.
     \label{seq:gen_elec_curr-DH-thickEDL}
\end{align}
As we assumed that $\lDinf \gg a$ (i.e. $a/\lDinf \rightarrow 0$) to derive Eq.~\eqref{seq:gen_elec_curr-DH-thickEDL}, the advective contribution to the electric current is negligibly small in this regime.

\subsubsection{Thin electric double layer}
\label{ssec:adv_elec_curr-thinEDL}

\begin{figure}[!ht]
    \centering
    \includegraphics[]{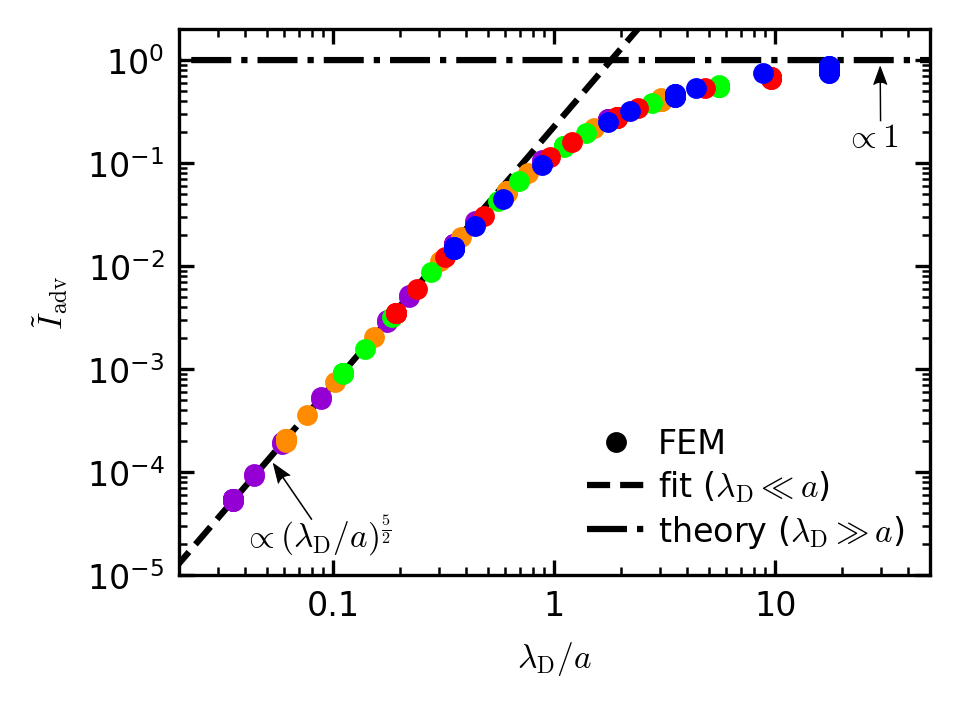}
    \caption{Dimensionless advective electric current $\tilde{I}_{\mathrm{adv}}$ vs $\lDinf/a$ from FEM simulations for surface potential energies that are $< \kBT$ far from the pore mouth and bulk electrolyte concentrations $\cinf$ of $0.3$ (blue), $1$ (red), $3$ (green), $10$ (orange) and $30$ (purple) mol~m$^{-3}$. The dashed line is a linear fit of $\tilde{I}_{\mathrm{adv}}$ from simulations where $\lDinf/a \leq 0.1$ to $(\lDinf/a)^{\frac{5}{2}}$ ($\mathrm{coefficient} \approx 0.227$) and the dash-dotted line is $\tilde{I}_{\mathrm{adv}} = 1$.}
    \label{seq:adv_elec_curr-universal}
\end{figure}

Using the theory of the advective electric current for a thick electric double layer (Eq.~\eqref{seq:adv_elec_curr-DH-thickEDL-sub}), 
we subject the advective electric current to the following non-dimensionalization
\begin{equation}
       I_{\mathrm{adv}} = -\frac{a^3 \sigma^2 \Delta \psi}{3 \eta (\lDinf)^2} \tilde{I}_{\mathrm{adv}},
       \label{seq:adv_elec_curr-dimensionless}
\end{equation}
where $\tilde{I}_{\mathrm{adv}} \approx 1$ when $\lDinf \gg a$. Figure~\ref{seq:adv_elec_curr-universal} shows that for simulations where $Ze |\psi_{0}| < \kBT$ far from the pore mouth, the dimensionless advective current collapses on to a single universal curve as a function of $\lDinf/a$. A linear fit of $\log(\tilde{I}_{\mathrm{adv}})$ from simulations (see Sec.~\ref{ssec:FEM} for further details) where $\lDinf/a \leq 0.1$ to $\log(\lDinf/a)$ indicates that $\tilde{I}_{\mathrm{adv}} \propto (\lDinf/a)^{\frac{5}{2}}$, while a linear fit of $\tilde{I}_{\mathrm{adv}}$ from simulations where $\lDinf/a \leq 0.1$ to $(\lDinf/a)^{\frac{5}{2}}$ gives a coefficient of $0.227$ (i.e. $\tilde{I}_{\mathrm{adv}} \approx 0.227 \times (\lDinf/a)^{\frac{5}{2}}$ when $\lDinf \ll a$). Re-dimensionalizing $\tilde{I}_{\mathrm{adv}}$ thus gives the scaling relationship, when $\lDinf \ll a$,
\begin{equation}
    I_{\mathrm{adv}} \sim \frac{(a \lDinf)^{\frac{1}{2}} \sigma^2 \Delta \psi}{\eta}.
    \label{seq:adv_elec_curr-thinEDL}
\end{equation} 

Summing Eq.~\eqref{seq:adv_elec_curr-thinEDL} and Eq.~\eqref{eq:elec_curr-DH-thinEDL} in the main paper
gives
\begin{align}
    I \approx -\frac{a \eps \Delta \psi}{(\lDinf)^2}\left\{(\Dp) + \frac{1}{3} \left(\frac{\lDinf}{\lGC} \right)^2 \left(\frac{\lDinf}{a} \right)^{\frac{1}{2}} \left[(\Dp) + \frac{\alpha_3 \eps}{\eta} \left(\frac{\kBT}{Ze} \right)^2\right] \right\}.
\label{seq:gen_elec_curr-thinEDL}
\end{align}
where $\alpha_3$ is a constant prefactor that is $\approx 4 \times 0.227 \approx 0.908$ (see Eq.~\eqref{seq:adv_elec_curr-dimensionless}).
The advective electric current is most significant when the bulk contribution (i.e. electric current at no surface charge) is relatively small, where $I^{0} = -a \eps (\Dp) \Delta \psi/(\lDinf)^2$ is the bulk contribution to the electric current. Thus, Eq.~\eqref{seq:gen_elec_curr-thinEDL} indicates that 
\begin{equation}
    \frac{I_{\mathrm{adv}}}{I_{\mathrm{elec+diff}}} \rightarrow \frac{\alpha_3 (\kBT)^2 \eps}{(Ze)^2 (\Dp) \eta}
    \label{seq:Iadv-vs-Iconv-thinEDL}
\end{equation}
is the maximum predicted ratio of advection to diffusion in the Debye--H\"uckel regime when $\lDinf \ll a$. Since the right-hand-side of Eq.~\eqref{seq:Iadv-vs-Iconv-thinEDL} $\approx 0.117 \ll 1$ for potassium chloride under ambient conditions (see Table~\ref{stab:variables-FEM}), the advective electric current is small compared with the electro-diffusive electric current when $\lDinf \ll a$. Since $I^{0}$ is large compared to all other terms in Eq.~\eqref{seq:gen_elec_curr-thinEDL} when $\lDinf \ll a$, ${I_{\mathrm{adv}}}/{I_{\mathrm{elec+diff}}}$ is likely to be much smaller than the upper limit predicted by Eq.~\eqref{seq:Iadv-vs-Iconv-thinEDL}. Note that the relationship in Eq.~\eqref{seq:Iadv-vs-Iconv-thinEDL} only holds when the electro-diffusive and advective electric current are separately conserved and exhibit analogous scaling relationships with $\lDinf/\lGC$ and $a/\lDinf$. These assumptions may break down outside of the Debye--H\"uckel regime and thin electric-double-layer limit. 

\section{Derivation of scaling of ionic conductance with electrolyte concentration (finite-thickness membrane)}
\label{ssec:chi}

Substituting Eqs.~\eqref{eq:bulk_conduct} and \eqref{eq:surf_conduct} in the main paper into the expression for the ionic conductance of an ultrathin membrane (i.e. $G_{L/(2a) \rightarrow 0} =2a \kappa_{\mathrm{b}} + \kappa_{\mathrm{s}}$) gives
\begin{align}
    G_{L/(2a) \rightarrow 0} &= \frac{a \eps (\Dp)}{(\lDinf)^2} + \frac{\eps(\Dp)}{\lGC} \left[-\frac{\lGC}{\lDinf} + \sqrt{\left(\frac{\lGC}{\lDinf} \right)^2 + \frac{2}{3} \left(\frac{a}{\lDinf} \right)^{\frac{1}{2}}} \right]
        \label{seq:conduct-ultrathin-raw} \\
    & = \frac{\eps(\Dp)}{a} \left[\left(\frac{a}{\lDinf} \right)^2  - \frac{a}{\lDinf} + \sqrt{\left(\frac{a}{\lDinf} \right)^2 + \frac{2}{3} \left( \frac{a}{\lDinf} \right)^\frac{1}{2} \left(\frac{a}{\lGC} \right)^2 } \right],
    \label{seq:conduct-ultrathin}
\end{align}
where we have rearranged Eq.~\eqref{seq:conduct-ultrathin-raw} to express $G_{L/(2a) \rightarrow 0}$ in terms of $a/\lGC$ and $a/\lDinf$. Similarly, substituting Eqs.~\eqref{eq:bulk_conduct} and \eqref{eq:surf_conduct-plane} in the main paper into the expression for the ionic conductance of a long cylindrical pore (i.e. $G_{\mathrm{p}} =\pi a (a\kappa_{\mathrm{b}} + 2\kappa_{\mathrm{s}}^{\infty})/L$ for a thin electric double layer)\cite{leeLargeElectricSizeSurfaceCondudction2012} gives
\begin{align}
    G_{\mathrm{p}} &= \frac{\pi a}{L}\left \{\frac{a \eps (\Dp)}{2 (\lDinf)^2} + \frac{2\eps(\Dp)}{\lGC} \left[-\frac{\lGC}{\lDinf} + \sqrt{\left(\frac{\lGC}{\lDinf} \right)^2 + 1} \right] \right \}
    \label{seq:conduct-thick-raw} \\ 
    & = \frac{\pi \eps(\Dp)}{2 L} \left[\left(\frac{a}{\lDinf} \right)^2  - \frac{4a}{\lDinf} + \sqrt{\left(\frac{4 a}{\lDinf} \right)^2 +\left(\frac{4 a}{\lGC} \right)^2 } \right].
    \label{seq:conduct-thick}
\end{align}
Assuming that the ionic conductance of a membrane of finite thickness can be approximated as $G = 1/(R_{\mathrm{p}} + 2R_{\mathrm{a}})$,\cite{leeLargeElectricSizeSurfaceCondudction2012} where $R_{\mathrm{a}} \approx 1/(2 G_{L/(2a) \rightarrow 0})$, we can write
\begin{equation}
    \frac{1}{G} \approx \frac{1}{G_{L/(2a) \rightarrow 0}} + \frac{1}{G_{\mathrm{p}}}.
    \label{seq:conduct-finite}
\end{equation}
By substituting Eqs.~\eqref{seq:conduct-ultrathin} and \eqref{seq:conduct-thick} into Eq.~\eqref{seq:conduct-finite}, and implementing the non-dimensionalization
\begin{equation}
    G = \frac{\eps(\Dp)}{a} \hat{G}
    \label{seq:conduct-non-dim}
\end{equation}
into this expression, we can write 
\begin{align}
    \frac{1}{\hat{G}} = \left[\left(\frac{a}{\lDinf} \right)^2  - \frac{a}{\lDinf} + \sqrt{\left(\frac{a}{\lDinf} \right)^2 + \frac{2}{3} \left( \frac{a}{\lDinf} \right)^\frac{1}{2} \left(\frac{a}{\lGC} \right)^2 } \right]^{-1} \notag \\ 
    + \frac{2L}{\pi a} \left[\left(\frac{a}{\lDinf} \right)^2  - \frac{4a}{\lDinf} + \sqrt{\left(\frac{4 a}{\lDinf} \right)^2 +\left(\frac{4 a}{\lGC} \right)^2 } \right]^{-1} .
    \label{seq:conduct-dim-less}
\end{align}
As shown in Eq.~\eqref{seq:conduct-dim-less}, the dimensionless ionic conductance can be quantified using the dimensionless variables ${\lDinf}/{a}$, ${\lGC}/{a}$ and ${L}/{a}$, where only ${\lDinf}/{a}$ depends on the electrolyte concentration.

Suppose we seek to approximate the scaling of the ionic conductance with electrolyte concentration using a single term, i.e. $G \propto (\cinf)^{\chi}$---at this stage, $\chi$ is to be determined. As $a/\lDinf$ is the only variable in the dimensionless ionic conductance that depends on the electrolyte concentration (where $\lDinf \propto (\cinf)^{-\frac{1}{2}}$), the relationship $G \propto (\cinf)^{\chi}$ is analogous to
\begin{equation}
    \hat{G} = \alpha_4\left(\frac{a}{\lDinf}\right)^{2\chi},
    \label{seq:conduct-simple_conc}
\end{equation}
where $\chi$ is the power law scaling of the ionic conductance with electrolyte concentration and $\alpha_4$ is a fitting parameter that does not depend on electrolyte concentration. Taking the logarithm of Eq.~\eqref{seq:conduct-simple_conc} gives
\begin{equation}
    \log(\hat{G}) = 2\chi \log\left(\frac{a}{\lDinf}\right) + \log(\alpha_4).
    \label{seq:conduct-log_conc}
\end{equation}
Taking the derivative of $\log(\hat{G})$ with respect to $\log\left({a}/{\lDinf}\right)$ and rearranging for $\chi$ gives
\begin{equation}
    \chi = \frac{1}{2} \frac{\partial \log(\hat{G})}{\partial \log(a/\lDinf)},
    \label{seq:conduct-theta}
\end{equation}
such that solving Eq.~\eqref{seq:conduct-theta} with respect to $\chi$ gives the power law scaling of the ionic conductance with electrolyte concentration at a given $L/a$, $a/\lDinf$ and $a/\lGC$. Note that Eq.~\eqref{seq:conduct-theta} is identical to Eq.~\eqref{eq:scaling_derivative} in the main paper. 

To determine the value of $\chi$ predicted by our model of a 
membrane pore as the electrolyte concentration is varied, we first calculated the values of $\hat{G}$ from Eq.~\eqref{seq:conduct-dim-less} at constant $L/a$ and $a/\lGC$ with varying $a/\lDinf$ on a uniform grid of $\log(a/\lDinf)$. Then, we estimated the derivative of $\log(\hat{G})$ with respect to $\log(a/\lDinf)$ at each value of $a/\lDinf$ using finite differences, where we used the sixth-order central finite difference. Halving the derivative at each point gave the value of $\chi$ at each $a/\lDinf$. Outside of the limits where the theory for the ionic conductance reduces to an expression that scales with a power law of the electrolyte concentration (for example, $G \propto \cinf$ when the bulk conductance is much larger than the surface conductance), $\chi$ can be viewed as the apparent scaling with electrolyte concentration that arises from a fitting of the ionic conductance with a power law of the electrolyte concentration. 

\bibliography{2D_membrane_references}